\let\accentvec\vec
\documentclass[10pt,a4paper,english]{article}
\providecommand{\tabularnewline}{\\}

\let\vec\accentvec
\usepackage[hyphens]{url}
\usepackage{ded}
\usepackage[all]{xy}
\usepackage{mathptmx}
\usepackage{amssymb}
\usepackage{color}
\usepackage{graphicx}
\usepackage{subfigure}
\usepackage{array}
\usepackage{hyperref}
\usepackage{algorithmic}
\usepackage{pifont}
\usepackage{MnSymbol}
\usepackage{marvosym}
\usepackage[english]{babel}
\usepackage{amsthm}
\usepackage{rotating}

\theoremstyle{plain}
\newtheorem{thm}{\protect\theoremname}
  \theoremstyle{definition}
  \newtheorem{example}[thm]{\protect\examplename}
  \theoremstyle{definition}
  \newtheorem{definition}[thm]{\protect\definitionname}

\def\avdashc#1{\ensuremath{\textbf{(\ensuremath{\vdash}#1)}}}
\def\axvdashc#1#2#3#4{\ianc{#2}{#3}{#4\avdashc{#1}}}

\def\ctxm#1#2#3{#1|^{#2}_{#3}}
\def\ctx#1#2{\ctxm{*}{#1}{#2}}
\def\ctxmg#1#2{\ctxm{#1}{#2}{\cdot}}
\def\proveseq#1#2#3#4{(#1\doteq#2)\Rightarrow(#3\doteq#4)}

\def\mathignore#1{\color{white}#1\color{black}}

\def\ioi{\mathcal{O}}
\def\cioi{\mathsf{O}^c}

\def\objl{context}
\def\pimodel{(\mathsf{O}^c,\mathcal{O},\relation,\sigma,\tau,\{\psi_1,\ldots,\psi_k\})}
\def\imodel{(\mathcal{L}^c,\mathcal{L},\mathsf{E}^c,\mathcal{E},\relation,\sigma,\tau,\{\psi_1,\ldots,\psi_k\})}

\usepackage{babel}
  \providecommand{\definitionname}{Definition}
  \providecommand{\examplename}{Example}
\providecommand{\theoremname}{Theorem}

\def\relation{\ensuremath{\Leftrightarrow}}

\def\related#1#2{\ensuremath{#1 \Leftrightarrow #2}}
\def\linkrel{\ensuremath{\leftrightarrow}}
\def\link#1{\ensuremath{\leftrightarrow_{#1}}}

\def\linkable#1#2#3{\ensuremath{#1 \link{#3} #2}}

\def\transmission#1#2#3{\ensuremath{{#1}\mapsto{#2}:{#3}}}
\def\transmissionP#1#2{\ensuremath{{#1}\mapsto{#2}}}

  \newcommand{\tick}{\includegraphics[height=2.5mm]{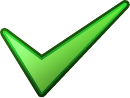}}
 \newcommand{\cross}{\includegraphics[height=2.5mm]{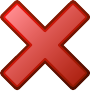}}
 \newcommand{\gtick}{\includegraphics[height=2.5mm]{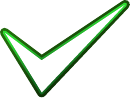}}
\newcommand{\gcross}{\includegraphics[height=2.5mm]{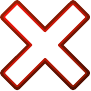}}

\begin{document}

\title{
Data Minimisation in Communication Protocols:\\
A Formal Analysis Framework and Application to Identity Management
}
\author{Meilof Veeningen \and Benne de Weger \and Nicola Zannone}


\maketitle

\begin{abstract}
With the growing amount of personal information exchanged over the Internet, privacy is becoming more and more a concern for users.
One of the key principles in protecting privacy is data minimisation.
This principle requires that only the minimum amount of information necessary to accomplish a certain goal is collected and processed.
``Privacy-enhancing'' communication protocols have been proposed to guarantee data minimisation in a wide range of applications.
However, currently there is no satisfactory way to assess and compare the privacy they offer in a precise way: existing analyses are either too informal and high-level, or specific for one particular system.
In this work, we propose a general formal framework to analyse and compare communication protocols with respect to privacy by data minimisation.
Privacy requirements are formalised independent of a particular protocol in terms of the knowledge of (coalitions of) actors in a three-layer model of personal information.
These requirements are then verified automatically for particular protocols by computing this knowledge from a description of their communication.
We validate our framework in an identity management (IdM) case study.
As IdM systems are used more and more to satisfy the increasing need for reliable on-line identification and authentication, privacy is becoming an increasingly critical issue.
We use our framework to analyse and compare four identity management systems.
Finally, we discuss the completeness and (re)usability of the proposed framework.

\end{abstract}

\section{Introduction}

As more and more personal information is exchanged over the Internet by businesses and other organisations, privacy risks are becoming a major concern.
For instance, e-health and identity management systems deal with large amounts of personal information.
There have been numerous reports of information from such systems being used for secondary purposes \cite{States.[2012]Prescriptiondrugdata}, or being stolen and abused by third parties \cite{20112011Costof}.
Legislation (e.g., EU Directive 95/46/EC, HIPAA) attempts to reduce these risks by requiring such systems to satisfy the \emph{data minimisation} principle.
That is, systems have to be designed to ensure that actors in such systems collect and store only the minimal amount of personal information needed to fulfil their task.
This means limiting the amount of shared personal information, but also limiting the use of identifiers that different actors can use to correlate their views on a data subject \cite{Hansen200435}.

One important approach to achieve privacy by data minimisation is the use of \emph{privacy-enhancing} communication protocols \cite{Hansen200435,Troncoso2011Designandanalysis}.
Such protocols use cryptographic primitives to ensure that participants learn as little information as possible, and that they have as little ability as possible to correlate information from different sources.
Privacy-enhancing protocols have been proposed for a wide range of applications: e.g.,~smart metering~\cite{Rial2011Privacy-preservingsmartmetering}, e-voting \cite{Fujioka1993PracticalSecretVoting}, and electronic toll collection \cite{Dahl2011FormalAnalysisof}.

Understanding the privacy differences between privacy-enhancing protocols designed for the same purpose is important, e.g., for system designers who want to use privacy-enhancing protocols, or for system administrators who want to select what system to use.
However, it is typically not straightforward to obtain such an understanding.
One reason is that privacy-enhancing protocols typically combine (advanced) cryptographic primitives in subtle ways; also, typical scenarios involve multiple actors which may collude in different coalitions to correlate their views on data subjects.
Existing comparisons in areas such as e-health \cite{DataProtectionCommissionerDataprotectionguidelines,Tinabo2009Anonymisationvs.Pseudonymisation:} or identity management \cite{EU2003IdentityManagementSystems,Hoepman2008ComparingIdentityManagement} are performed in an informal and high-level (and thus, possibly subjective) way, and thus their privacy assessments do not offer much insight into differences between systems and the reasons behind them.
On the other hand, proposals for privacy-enhancing systems typically assess the privacy of their own solution using terminology and criteria specific to the setting at hand, making it hard to compare different systems.
Thus, we need a practical way to compare different systems that is precise and verifiable, yet application-independent; and that provides sufficient detail for real insight into the privacy differences that exist between systems.

Formal methods provide the machinery to perform such a comparison.
Over the years, formal methods, e.g., the applied pi calculus \cite{Abadi2001Mobilevaluesnew}, have arisen as an important tool to analyse security of communication in IT systems  \cite{Abadi2001Mobilevaluesnew,Burrows1990logicofauthentication,Meadows2003Formalmethodscryptographic,Paulson1998inductiveapproachto}.
The idea is to express communication protocols in a suitable formalism, and then verify whether such a model of the protocol satisfies, e.g., authentication properties \cite{Burrows1990logicofauthentication} or secrecy properties \cite{Bella1998KerberosVersionIV:}.
Secrecy, in particular, expresses one aspect of privacy; namely, whether a certain piece of information is known by some party in a protocol.
However, it leaves unanswered a question which is equally important for privacy-enhancing protocols; namely, whether a certain piece of information can be linked to its corresponding data subject (who, in general, might not be a direct participant in the communication under analysis).

Recently, several research efforts have focused on the analysis of privacy properties using the applied pi calculus and related techniques \cite{Bruso2010FormalVerificationof,Dahl2011FormalAnalysisof,Delaune2009Verifyingprivacy-typeproperties,JannikDreier2011FormalTaxonomyof,Suriadi2010Strengtheningandformally}, in application domains such as electronic toll collection~\cite{Dahl2011FormalAnalysisof}, e-voting~\cite{Delaune2009Verifyingprivacy-typeproperties,JannikDreier2011FormalTaxonomyof}, and RFID systems~\cite{Bruso2010FormalVerificationof}.
While this approach has delivered considerable successes, several issues inhibit its use for our purposes, namely, practical and accessible high-level privacy analysis.
First, in many cases, properties are defined and verified specific to their respective settings or protocols \cite{Dahl2011FormalAnalysisof,Delaune2009Verifyingprivacy-typeproperties,JannikDreier2011FormalTaxonomyof}.
General definitions for the common privacy property of linkability exist \cite{Arapinis2010AnalysingUnlinkabilityand}, but they are aimed towards linking messages to their senders (whereas data minimisation concerns linking of information to its data subject) and defined with respect to an outside attacker (whereas data minimisation concerns the knowledge of actors or coalitions inside the system).
Second, such methods require considerable manual work for each property to be verified, in many cases including particular assumptions on the model to make computation feasible.
Third, analysis results are not summarised in a comprehensive and intuitive way, necessitating substantial manual review.

In our previous works \cite{Veeningen2010Modelingidentity,Veeningen2011FormalPrivacyAnalysis}, we have introduced building blocks for high-level privacy analysis of protocols to exchange personal information.
We introduced a three-layer model that captures the knowledge of personal information held by different (coalitions of) communicating parties \cite{Veeningen2010Modelingidentity,Veeningen2011FormalPrivacyAnalysis}.
The model captures the context in which pieces of information has been observed, as well as the contents they have.
We showed how relevant privacy requirements can be expressed as properties of items in this model.
We also showed how this model is determined from observations of communication between the different parties.
However, the model of \cite{Veeningen2010Modelingidentity,Veeningen2011FormalPrivacyAnalysis} only captures communication that uses a limited set of cryptographic primitives; moreover, it does not offer an implementation of the analysis method; and finally, it does not discuss in general what kinds of privacy requirements can be verified, or how to perform a privacy comparison in practice.

In this work, we combine our previous building blocks into a general framework for privacy comparison of communication protocols, and we apply the framework in an identity management case study.
Specifically, our contributions are as follows:
\begin{itemize}
 \item We present a framework to compare communication protocols with respect to privacy by data minimisation. Our framework gives precise, verifiable results with enough detail to obtain insight into privacy differences;
 \item We extend our previous formal method \cite{Veeningen2010Modelingidentity,Veeningen2011FormalPrivacyAnalysis} for the analysis of knowledge of personal information to cover additional primitives and cryptographic protocols (specifically, zero-knowledge proofs and issuing protocols for anonymous credentials);
 \item We provide an implementation of the formal method in Prolog to automate part of the comparison;
 \item We validate our framework by analysing and comparing four identity management systems: we show that a range of relevant privacy requirements can be captured by our model, and use our framework to formally analyse the identity management systems with respect to these requirements.
\end{itemize}

\begin{figure}[tb]
 \centering\includegraphics[scale=0.45]{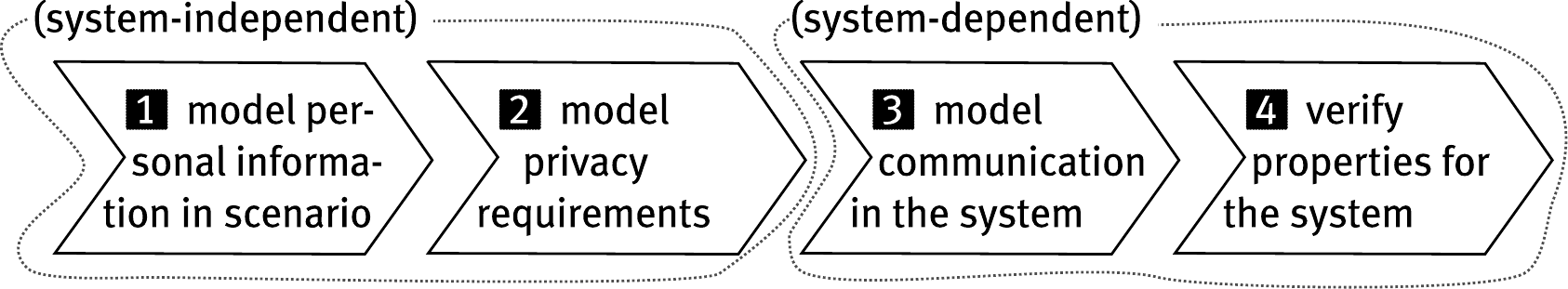}
 \caption{\label{fig:steps}Steps of our privacy comparison framework}
\end{figure}
Our privacy comparison framework consists of four steps, shown in Figure~\ref{fig:steps}.
The first two steps are to model the scenario and its requirements.
We introduce two formalisms: the \emph{Personal Information (PI) Model} ($\S$\ref{subsec:pi-model}) to model different types of personal information and their relations; and the \emph{view} of an actor to describe the partial knowledge about this information that this actor has at one point in time ($\S$\ref{subsec:view}).
The \emph{first step} of our method comprises modelling all personal information using a PI model, and modelling the initial knowledge of each actor as a view on that PI model.
This means modelling the personal information as used in the protocol instances in the scenario; however, it also means modelling other knowledge of personal information held initially by the actors.
This way, we can assess how links can be established between the knowledge learned from the protocol instances and the initial knowledge.
The \emph{second step} is to model data minimisation requirements, i.e., which personal information should become known or remain unknown to which actors in the system.
These requirements are phrased as properties of the views of actors after communication has taken place.
These first two steps are performed independently from the particular systems to be analysed.

The \emph{third step} is then to model the exchange of information in the information systems.
For this, we need to model the evolution of actor knowledge in such systems due to the exchange of messages.
We extend the PI model into an \emph{information model} that also includes messages using cryptographic primitives, and the non-personal information they may contain.
We express the messages that an actor has exchanged at a certain point in time using the notion of a \emph{knowledge base} on that information model ($\S$\ref{subsec:kb}).
We define a procedure to determine an actor's view from his knowledge base ($\S$\ref{subsec:deduction}), and present an algorithm that implements it ($\S$\ref{subsec:prolog}).
Finally, we introduce states to formalise the knowledge of all actors in the system at one point in time, and traces to capture a series of communications that transforms one state into another ($\S$\ref{sec:traces}).

The \emph{fourth step} is to verify which systems satisfy which requirements.
This step is performed automatically using our Prolog implementation\footnote{The tool and formal model of our case study are available at \url{www.mobiman.me/downloads/}.}.
Given a PI model, set of formalised requirements, initial state and trace, this tool first determines the state of the system after communication; then uses our formal procedure to compute the corresponding views of the actors in the system, and finally determines which requirements hold in these views.

We validate our framework by applying it to an identity management case study.
Identity management (IdM) systems~\cite{Sommer2008PRIMEArchitectureV3,SampoKellomaki(editor)2010,shibboleth} offer reliable on-line identification and authentication to service providers by outsourcing these tasks to ``identity providers''.
Identity providers endorse information about their users, and provide means for authenticating a user in a service provision.
To organisations, identity providers offer reduced cost for obtaining reliable user information; to users, they offer increased convenience by letting them reuse authentication credentials.
The amount of personal information exchanged in such systems makes privacy a critical issue; this is reflected by the large number of privacy-enhancing IdM systems that have been proposed \cite{Bangerter2004CryptographicFrameworkControlled,Chadwick2009AttributeAggregationin,Vossaert2010User-centricidentitymanagement}.
However, while high-level sketches of privacy issues~\cite{Alpar2011IdentityCrisis,Bhargav-Spantzel2007PrivacyRequirementsin,Hansen200435,Landau2009AchievingPrivacyin} and comparisons of systems~\cite{EU2003IdentityManagementSystems,Hoepman2008ComparingIdentityManagement} exist, no comprehensive set of relevant privacy requirements for IdM systems has been proposed, nor do there exist precise formal comparisons.
We demonstrate that our framework can be used to perform such a comparison.
In Section~\ref{sec:case-study}, we present our case study: we introduce IdM, discuss privacy requirements for IdM systems, and introduce four IdM systems.
In Section~\ref{sec:formal-case-study}, we use our framework to formally compare the privacy offered by these four IdM systems with respect to the requirements introduced above, and discuss the results.

Finally, we discuss the completeness and (re)usability of our framework ($\S$\ref{sec:discussion}).
We conclude the paper by discussing related work ($\S$\ref{sec:related-work}), drawing conclusions, and pointing to interesting directions for future work ($\S$\ref{sec:conclusion}).

\section{A Model for Knowledge of Personal Information}\label{sec:pi-model}

In this section, we present the Personal Information (PI) Model and actor views.
A \emph{PI model} ($\S$\ref{subsec:pi-model}) describes personal information in an information system at a certain point in time; the \emph{view} of an actor involved in the system on this PI model ($\S$\ref{subsec:view}) captures the knowledge about this information held by that actor.
Privacy requirements ($\S$\ref{subsec:model-reqs}) are modelled as properties of items from these views.
The PI model is used in step 1 of our framework (Figure~\ref{fig:steps}) to model personal information, and it is the basis for the model of communication in step 3.
Views are used in step 1 to express initial knowledge of actors; in step 2 to model requirements; and in step 4 to compare actual knowledge to these requirements.
Our model is based on two main assumptions:
\begin{itemize}
 \item Discrete information --- 
There is a finite set of pieces of personal information that each belong to a particular data subject.
In particular, we allow knowledge about finitely many boolean predicates on personal information (e.g., ``Alice's age is below $k$'' for some particular value $k$).
Each piece of information has a well-defined contents. (However, different pieces of information may have the same contents.)
 \item Discrete knowledge ---
Actors may or may not be able to learn these pieces of information; and they may or may not be able to learn that these pieces of information are about the same data subject.
In both cases, we do not allow uncertainty: either an actor knows a piece of information or a link, or he does not.
\end{itemize}
The above abstractions are common in the protocol verification literature~\cite{Meadows2003Formalmethodscryptographic}, and simplify both the specification of requirements and the modelling of protocols.
We discuss approaches that do not make these abstractions in Section~\ref{sec:related-work}.

\subsection{Personal Information Model}\label{subsec:pi-model}

A Personal Information (PI) model describes all personal information present in an information system at a certain point in time.

\subsubsection{Personal Information}

A piece of personal information in the PI model is a \emph{specific} string that has a \emph{specific} meaning as personal information about a \emph{specific} person, e.g., ``the age of Alice''.
We distinguish between two types of digital personal information: \emph{identifiers} and \emph{data items}.
Identifiers are unique within the system; for data items, this is not necessarily the case.
The sets of identifiers and data items are denoted $\mathcal{I}$ and $\mathcal{D}$, respectively.
We express that pieces of personal information satisfy certain fixed boolean \emph{properties} relevant to the application domain by a set $\{\psi_1,\ldots,\psi_k\}$ of partial functions $\mathcal{I}\cup\mathcal{D}\to\mathcal{D}$ that assign properties to some of the identifiers and data items.
For instance, suppose $\psi_j$ represents the property that an age is over 60.
If $\psi_j(d)$ is defined, i.e., $d$ has an image under partial function $\psi_j$, then $d$ represents the age of a data subject who is over 60 and $\psi_j(d)$ represents the fact that this data subject has an age over 60.
If $\psi_j(d)$ is not defined, i.e., $d$ does not have an image under partial function $\psi_j$, then either $d$ does not represent an age, or it represents an age below 60.
The set $\mathcal{E}$ of \emph{entities} models the real-world persons whom the considered information is about.
Elements of the set $\mathcal{O}:=\mathcal{E}\cup\mathcal{I}\cup\mathcal{D}$ are called \emph{items of interest}.
The link between information and its subject is captured by the \emph{related} relation, an equivalence relation on $\mathcal{O}$ denoted \relation.
Namely, given two items of interest $o_1,o_2\in\mathcal{O}$, $\related{o_{1}}{o_{2}}$ means that $o_1$ and $o_2$ are information about the same person.

These concepts, however, are insufficient to model all privacy aspects of communication protocols that we are interested in.
First, it is relevant to know whether different pieces of information have the same contents or not.
For instance, Alice's age may be the same as Bob's, and Alice's age may be the same as Alice's apartment number.
Whether this is the case influences what information can be determined from cryptographic primitives: for instance, an actor can determine a piece of information from its cryptographic hash if he knows another piece of information with the same contents.
Second, it is relevant to know how an actor obtained a piece of information.
We assumed that actors combine pieces of information that they know belong to the same data subject.
However, if an actor learns the same piece of information (e.g.,~``the age of Alice'') several times (e.g., in two protocol instances with different session identifiers), he may not know that it is the same information.
Thus, to represent the knowledge of this actor, we need to distinguish between these two ``representations'' of the information.

\subsubsection{Three-Layer Model}

Because of the need to distinguish different instances of the same piece of information, as well as to reason about message contents, we introduce a three-layer representation of personal information.
The representation consists of the \emph{context layer}, \emph{information layer}, and \emph{contents layer}.
At the information layer, as described above, the information itself is represented, e.g., ``Alice's city''.
At the context layer, information is described in terms of the context in which it has been observed, e.g., ``the city of the user in protocol instance \#1''.
At the contents layer, information is described in terms of the strings actually transmitted in a protocol, e.g., ``Eindhoven''.
Actor knowledge is modelled using the context layer and reasoned about using the contents layer; the information layer is used to specify privacy requirements or visualise analysis results~\cite{Veeningen2012FormalModellingof}.

At the {\objl} layer, we model the \emph{context} in which an actor knows pieces of information.
A context is an item $\ctx{\eta}{k}$, where $\eta$ is called the \emph{domain}, and $k$ is called the $\emph{profile}$.
A domain is any separate digital ``place'' where personal information is stored or transmitted, e.g., a database or an instance of a communication protocol.
A profile represents a particular data subject in a domain, e.g., an entry about one person in a database, or a logical role in a protocol instance (however, different profiles in a domain may still represent the same data subject, e.g., duplicate entries in a database).

In such a context, pieces of information are represented by \emph{variables}.
A variable describes the type of information in the context of that domain, e.g. ``session identifier'' or ``age attribute''.
Namely, the piece of information with variable $v$ in context $\ctx{\eta}{k}$ is denoted $\ctxm{v}{\eta}{k}$.
Context-layer representations of entities, identifiers, and data items are modelled by \emph{context entities} $\mathsf{E}^c$, \emph{context identities} $\mathsf{I}^c$, and \emph{context data items} $\mathsf{D}^c$, respectively.
The set $\mathsf{O}^c:=\mathsf{E}^c\cup\mathsf{I}^c\cup\mathsf{D}^c$ is the set of all \emph{context personal items}.
The unique context entity in context $\ctxm{*}{\eta}{k}$ is denoted $\ctxm{ds}{\eta}{k}$.
Properties of identifiers and data items are modelled at the context layer by extending the partial functions $\psi_i$ above.

We represent personal information at the contents layer as elements from an arbitrary set $\mathfrak{C}$ of \emph{message contents}.
In fact, for our purposes the exact representation is not relevant; it suffices to know which pieces of information have the same contents, and which do not.

Apart from these three descriptions of pieces of personal information, the PI model also defines mappings between the three layers.
Namely, it defines a mapping $\sigma$ from the {\objl} layer to the information layer; and a mapping $\tau$ from the information layer to the contents layer.
Properties of $\sigma$ and $\tau$ reflect characteristics of the different pieces of information, as shown below. 

Formally, a PI model is defined as follows:

\begin{definition}\label{def:pi-model}
 A \emph{Personal Information (PI) Model} is a tuple
$$\pimodel$$
so that:
\begin{itemize}
 \item $\mathsf{O}^c$ is a set of \emph{context personal items}, partitioned into $\mathsf{O}^c=\mathsf{E}^c\cup\mathsf{I}^c\cup\mathsf{D}^c$. Here, $\mathsf{E}^c$ are \emph{context entities} $\ctxm{ds}{\kappa}{n}$ with arbitrary domain $\kappa$ and profile $n$; $\mathsf{I}^c$ and $\mathsf{D}^c$ are \emph{context identifiers} and \emph{context data items} $\ctxm{v}{\kappa}{n}$ with arbitrary variable $v$, domain $\kappa$ and profile $n$, respectively;

 \item $\mathcal{O}$ is a set of \emph{items of interest}, partitioned into $\mathcal{O}=\mathcal{E}\cup\mathcal{I}\cup\mathcal{D}$. Here, $\mathcal{E}$ are \emph{entities}; $\mathcal{I}$ are \emph{identifiers}; and $\mathcal{D}$ are \emph{data items};

 \item $\relation\subset\mathcal{O}\times\mathcal{O}$ is the  \emph{related} relation on $\ioi$: an equivalence relation so that every item of interest is related to exactly one entity;

 \item $\sigma$ is a map $\cioi\to\ioi$ so that $\sigma(\mathsf{E}^c)\subset\mathcal{E}$; $\sigma(\mathsf{I}^c)\subset\mathcal{I}$; and $\sigma(\mathsf{D}^c)\subset\mathcal{D}$; and $\related{\sigma(\ctxm{x}{\eta}{k})}{\sigma(\ctxm{y}{\eta}{k})}$ for any $\ctxm{x}{\eta}{k}$, $\ctxm{y}{\eta}{k}$;

 \item $\tau$ is a map $\mathcal{I}\cup\mathcal{D}\to\mathfrak{C}$ so that $\forall i_1,i_2\in\mathcal{I}$: if $\tau(i_1)=\tau(i_2)$, then $i_1=i_2$;

\item $\{\psi_1,\ldots,\psi_n\}$ are partial functions $\psi_i:\mathsf{I}^c\cup\mathsf{D}^c\to\mathsf{D}^c$, $\mathcal{I}\cup\mathcal{D}\to\mathcal{D}$ so that:
1)~$\psi_i(o)$ is related to $o\in\mathcal{I}\cup\mathcal{D}$ whenever defined;
2)~$\tau(\psi_i(o_1))=\tau(\psi_i(o_2))$ whenever defined for some $o_1,o_2\in\mathcal{I}\cup\mathcal{D}$;
3)~$\psi_i(\mathsf{o})$ has the same context as $\mathsf{o}\in\mathsf{I}^c\cup\mathsf{D}^c$ whenever defined;
4)~$\sigma(\psi_i(\mathsf{o}))=\psi_i(\sigma(\mathsf{o}))$ for every $\mathsf{o}\in\mathsf{I}^c\cup\mathsf{D}^c$ for which $\psi_i(\mathsf{o})$ is defined.
\end{itemize}
\end{definition}
The first two bullets of the definition define information at the {\objl} and information layers, respectively; the third bullet defines personal relations at the information layer.
The fourth and fifth bullet define the mapping between the three layers: we demand that the contents of identifiers are unique.
The sixth bullet introduces properties both at the context and information layers.
Properties at the information layer preserve relation $\relation$ (1) and have contents independent from the item they are a property of (2); properties at the context layer preserve context (3) and are consistent with the information-layer properties (4).

We introduce notation for context personal items $\ctxm{x}{\eta}{k}$, $\ctxm{y}{\chi}{l}$ representing the same information or contents.
If $\sigma(\ctxm{x}{\eta}{k})=\sigma(\ctxm{y}{\chi}{l})$, then we write $\ctxm{x}{\eta}{k}\equiv\ctxm{y}{\chi}{l}$ and we call $\ctxm{x}{\eta}{k}$ and $\ctxm{y}{\chi}{l}$ \emph{equivalent}.
If $\tau(\sigma(\ctxm{x}{\eta}{k}))=\tau(\sigma(\ctxm{y}{\chi}{l}))$, then we write $\ctxm{x}{\eta}{k}\doteq\ctxm{y}{\chi}{l}$ and we call them \emph{content equivalent}.
Clearly, equivalence implies content equivalence.

The next example shows a PI model as used in step 1 of our analysis framework, i.e., to model all personal information present in a particular scenario.

\begin{example}\label{exa:pi-model}
\begin{figure}
 \centering\includegraphics[scale=0.6]{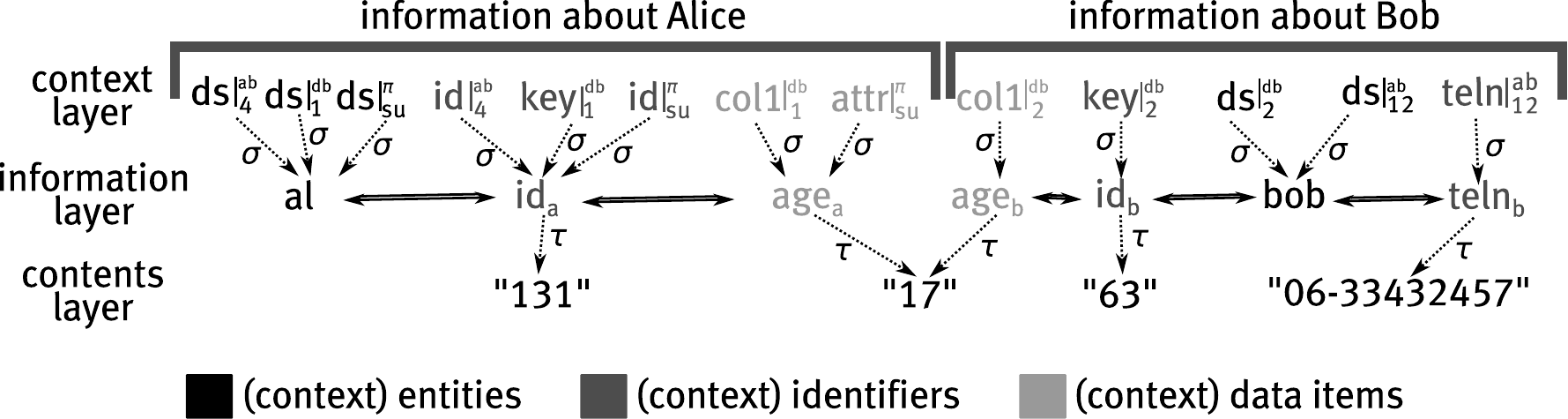}
 \caption{Personal Information Model of Example~\ref{exa:pi-model}\label{fig:exa-pi-model}}
\end{figure}
Figure~\ref{fig:exa-pi-model} shows a PI model representing personal information about two entities, Alice ($al\in\mathcal{E}$) and Bob ($bob\in\mathcal{E}$), in a simple scenario.
Recall that a PI model is used to express all personal information in a scenario, regardless of which protocols are used; regardless of who knows the information, and also including other information that it may be linked to by the actors involved.
In this scenario, a client and a server exchange information about Alice.
Namely, the server has a database with personal information about different entities; the server and client engage in a protocol to exchange information about Alice; and the client combines the results with her address book.

At the information layer of this PI model, Alice has identifier $id_a$ and an age $age_a$; Bob has identifier $id_b$, age $age_b$, and telephone number $teln_b$.
Alice and Bob happen to have the same age, so $\tau(age_a)=\tau(age_b)$; the other pieces of information have distinct contents.
(This example does not consider attribute properties.)

At the context layer of this PI model, the personal information in this scenario is modelled as follows:
\begin{itemize}
\item domain $db$ (database held by the server): Each profile $k\in\{1,2\}$ in this domain represents a database entry consisting of database key $\ctxm{key}{db}{k}$ and column value $\ctxm{col1}{db}{k}$.
As shown in the figure, the keys and column values map to the data subjects' identifiers and ages, respectively.
The data subject of profile $k$ is represented by context entity $\ctxm{ds}{db}{k}$.
\item domain $ab$ (address book of the client): Each profile $k\in\{4,12\}$ in this domain represents an entry in the address book. 
The fourth entry of the address book contains an identifier $\ctxm{id}{ab}{4}$; the 12th entry contains a telephone number $\ctxm{teln}{ab}{12}$.
\item domain $\pi$ (protocol instance): The client and server engage in an instance $\pi$ of a protocol in which identifier $\ctxm{id}{\pi}{su}$ and attribute $\ctxm{attr}{\pi}{su}$ are exchanged about data subject $su$; in this case, the subject is Alice and the attribute is her age.
\end{itemize}
(In a full analysis using our framework, we would also model the client and server as entities.
This allows us to reason about knowledge about their involvement in this scenario.
For simplicity, we omit them here.)
\qed
\end{example}

\subsection{Views: Actor Knowledge}\label{subsec:view}

Each actor in an information system has partial knowledge about the personal information in that system.
Our framework analyses privacy by data minimisation by verifying that this partial knowledge satisfies certain requirements.
We model actors as a finite set $\mathcal{A}$.
We require each actor to be an entity in the PI model, i.e., $\mathcal{A}\subset\mathcal{E}$.
The knowledge of an actor at some point in time consists of knowledge of some pieces of personal information, and knowledge that some of these pieces of information are about the same person.
We capture this knowledge as a \emph{view} on the PI model, consisting of a set of \objl-layer items and an equivalence relation on them:

\begin{definition}\label{def:view}
Let $M=\pimodel$ be a PI Model.
A \emph{view} on $M$ is a tuple $V=(\mathsf{O}_*,\linkrel_*)$ such that:
\begin{itemize}
 \item $\mathsf{O}_*\subset\mathsf{O}^c$ is the set of \emph{detectable} items;
 \item $\linkrel_*\subset\mathsf{O}_*\times\mathsf{O}_*$ is the \emph{associability} relation: an equivalence relation on $\mathsf{O}_*$.
\end{itemize}
\end{definition}
The view of actor $a\in\mathcal{A}$, determined in step 4 of our framework, is denoted $V_a=(\mathsf{O}_a,\linkrel_a)$.
From a privacy perspective, we are not just interested in the views of single actors $a\in\mathcal{A}$, but also in the views of coalitions $A\subset\mathcal{A}$.
Such a view represents knowledge of personal information when the actors in the coalition combine all information (e.g., databases, communication transcripts) they have.
The view of coalition $A\subset\mathcal{A}$ after communication is denoted $V_A=(\mathsf{O}_A,\linkrel_A)$.
It contains at least the knowledge of each individual actor in the coalition.

We next show an example of the views of actors and coalitions, as they may be obtained in step four of our framework when analysing a particular communication protocol.

\begin{example}\label{exa:view}
\begin{figure}[tb]
 \centering\includegraphics[scale=0.6]{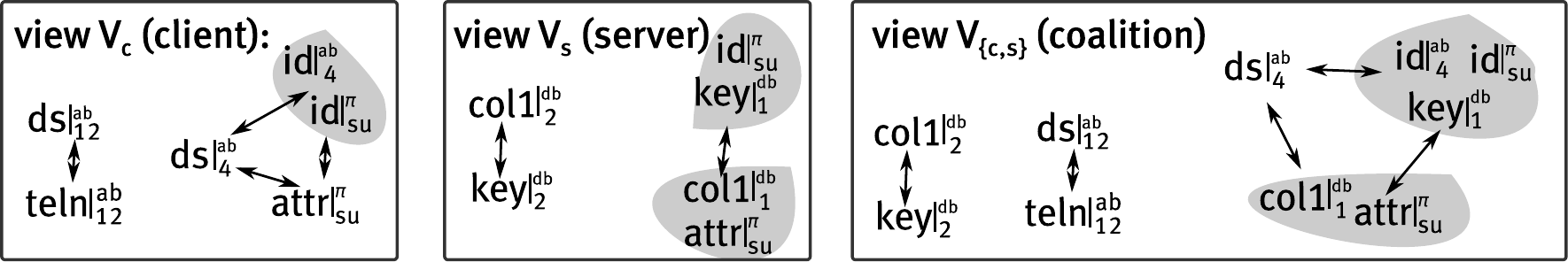}
 \caption{
Views of actors $c$ and $s$ and coalition $\{c,s\}$ in a scenario (Example~\ref{exa:view}).
Detectable context personal items are shown; grey areas are sets of equivalent items.
Associations are represented by arrows; for simplicity, they are shown up to equivalence.
\label{fig:exa-view}
}
\end{figure}

 Consider the PI model $M$ from Example~\ref{exa:pi-model}.
 The actors in this information system are the client and server, i.e., $\mathcal{A}=\{c,s\}$.
 Figure~\ref{fig:exa-view} shows example views of these actors after some particular communication protocol has been executed (domain $\pi$).

First consider the view $V_c=(\mathsf{O}_c,\leftrightarrow_c)$ on $M$ modelling personal information known by the client.
This information comprises the entries from her telephone book and the information about Alice that has been communicated.
Namely, the client knows Bob's telephone number ${teln}_{bob}$ as entry $\ctxm{teln}{ab}{12}\in \mathsf{O}_c$ in her telephone book; she also knows that this is Bob's telephone number, expressed by detectability $\ctxm{ds}{ab}{12}\in\mathsf{O}_c$ and associability $\ctxm{ds}{ab}{12}\leftrightarrow_c\ctxm{teln}{ab}{12}$.
About Alice, the client knows two context-layer representations of identifier ${id}_a$: as part of her telephone book entry ($\ctxm{id}{ab}{4}\in\mathsf{O}^c$), and as a piece of information sent in protocol instance $\pi$ ($\ctxm{id}{\pi}{su}\in\mathsf{O}^c$).
She again knows the data subject corresponding to the telephone book entry ($\ctxm{ds}{ab}{4}$), and she knows the age transmitted in the protocol ($\ctxm{attr}{\pi}{su}\in\mathsf{O}^c$).
Moreover, she can associate the information in the address book to the information from the protocol instance.

The view $V_s=(\mathsf{O}_s,\leftrightarrow_s)$ of the server also contains information about both Alice and Bob.
About Bob, the server knows two mutually associable pieces of information $\ctxm{col1}{db}{2}$, $\ctxm{key}{db}{2}$ from the database.
About Alice, the server also knows two associable pieces of information from the database.
In addition, it knows the two other context-layer representations $\ctxm{id}{\pi}{su}$, $\ctxm{attr}{\pi}{su}$ of that same information as transmitted in the protocol instance $\pi$.

Now consider the view $V_{\{c,s\}}$ of the client and server if they combine their knowledge.
In this view, all information about Alice from the two actors is mutually associable because both actors know the same identifier (in the figure, all context personal items about Alice are connected by arrows).
However, information about Bob is divided into two equivalence classes: the client knows entity $bob$ and his telephone number ${teln}_{bob}$ and the server knows age $age_b$ and telephone number ${teln}_b$, but they cannot associate this information to each other (indicated by the absence of arrows between the information in the figure).\qed

\end{example}

\subsection{Privacy Requirements}\label{subsec:model-reqs}

The second step of our analysis framework is to model each relevant data minimisation requirement in terms of the views of actors and coalitions.
This includes both modelling functional requirements, i.e., modelling what \emph{should} be learned by the actors in the protocol, and modelling privacy requirements, i.e., modelling what \emph{should not} be learned.
These requirements are formulated independently from any particular system, then verified for each particular system modelled.
Thus, our framework can be used to generically verify any requirement that can be phrased in terms of views, including:
\begin{itemize}
 \item \emph{Detectability requirements} --- Can a given actor/coalition of actors detect a given piece of information, or a given context-layer representation?
 \item \emph{Linkability requirements} --- Can a given actor/coalition of actors associate given contexts, or any contexts in which he knows given pieces of information?
 \item \emph{Involvement requirements} --- Is there a domain $d$ in which an actor can associate one profile to a given context $c_1$, and another to a given context $c_2$, i.e., does he know that the actors represented by $c_1,c_2$ were both involved in domain $d$?
\end{itemize}
More complex requirements can be defined as arbitrary combinations of these elementary requirements and their negations.
The next example shows different types of requirements.

\begin{example}
We formulate three requirements for the scenario given in Example~\ref{exa:pi-model}.
Recall that we have actors $\mathcal{A}=\{c,s\}$ with views
$$V_{c}=(\mathsf{O}_c,\leftrightarrow_c), V_{s}=(\mathsf{O}_s,\leftrightarrow_s),\textrm{ and }V_{\{c,s\}}=(\mathsf{O}_{\{c,s\}},\leftrightarrow_{\{c,s\}}).$$ 
First, since the goal of the protocol is to exchange information, we can check whether the client has indeed learned the age of Alice, and whether she can link it to her telephone book entry.
This corresponds to verifying that $\ctxm{attr}{\pi}{su}\in O_c$ and $\ctxm{attr}{\pi}{su}\leftrightarrow_c \ctxm{id}{ab}{4}$ hold (a detectability requirement and a linkability requirement, respectively).
Second, since the protocol does not concern Bob, we may want to make sure that the client and server together cannot inadvertently link Bob's telephone number and age due to this protocol instance.
This corresponds to verifying that $\ctxm{teln}{ab}{12}\leftrightarrow_{c,s} \ctxm{col1}{db}{2}$ does not hold (an unlinkability requirement).

Now consider the views in the particular system from Example~\ref{exa:view}.
In this case, both properties hold.
Namely, in view $V_c$, $\ctxm{attr}{\pi}{su}\in V_c$ and $\ctxm{age}{\pi}{su}\leftrightarrow_c \ctxm{id}{ab}{4}$ are true  (Figure~\ref{fig:exa-view}, left), while in view $V_{\{c,s\}}$, $\ctxm{teln}{ab}{12}\leftrightarrow_{c,s} \ctxm{col1}{db}{2}$ is not true  (Figure~\ref{fig:exa-view}, right).\qed

\end{example}

We show additional examples of the different types requirements in Section~\ref{sec:formal-case-study} when analysing identity management systems.
In Section~\ref{sec:discussion}, we discuss what kind of requirements cannot be represented in this way.

\section{Deducing Views from Communicated Messages}\label{sec:determining-views}

In this section, we determine the views of actors by modelling and analysing the messages they have exchanged.
We present the \emph{information model}, capturing messages containing personal information; and \emph{knowledge bases}, capturing which messages an actor has observed at a certain point in time.
We then propose a formal procedure to derive an actor's view from his knowledge base.
This procedure is based on the following main assumptions:
\begin{itemize}
 \item \emph{Detecting from messages} --- We model messages using a Dolev-Yao-style black-box model of cryptography. A piece of personal information is detected using a message it occurs in by: 1) reading it from that message; 2) applying a cryptographic operation on the message that uses the information; or 3) comparing the message's contents to another message whose structure is known.

 \item \emph{Associating by identifiers} --- Contexts are associated to each other by observing that the same identifier or entity occurs in both contexts.
\end{itemize}
The modelling of cryptographic primitives and operations as ``terms'' in a black-box model is standard ever since the seminal work by Dolev and Yao~\cite{Dolev1981securityofpublic}.
Determining what knowledge can be read from such messages can be done using standard deductive systems~\cite{Clarke1998Usingstatespace,Dolev1981securityofpublic,Fiore2001ComputingSymbolicModels}.
When adapting these standard techniques to our model of personal information, 
observing the application of cryptographic operations and comparing the contents of messages are needed as extensions.
These three ways of deriving personal information also occur in the popular equational approach using static equivalences~\cite{Blanchet2008AutomatedVerificationof}; see Section~\ref{sec:related-work} for a comparison.

Defining associability by identifiers is suitable for our goal, namely, comparing different protocols with respect to the knowledge that actors learn.
Namely, protocols differ in what identifiers they use and how; our definition of associability allows us to reason about the privacy consequences this has.
Associability does not take into account probabilistic links due to (combinations of) non-identifying personal information; probabilistic linking methods are orthogonal to our approach (see Section~\ref{sec:related-work}).

The formalisation of messages and knowledge bases is described in Section~\ref{subsec:kb}; the methodology for determining views from knowledges base is described in Section~\ref{subsec:deduction}.

\subsection{Messages, Information Model, Knowledge Base}\label{subsec:kb}

Communication in privacy-enhancing protocols uses messages built up from personal and other information, e.g., nonces and session keys.
At the {\objl} layer of our three-layer model, non-personal information is modelled by a set $\mathsf{G}^c$ of \emph{context non-personal items}.
Items in $\mathsf{G}^c$ belong to a domain, but not to a profile: in this case we denote the profile as $\cdot$, e.g. $\ctxmg{shakey}{\eta}$.
At the information layer, we define set $\mathcal{G}$ of \emph{non-personal items}.

\begin{table}[tb]
\small\centering\begin{tabular}[b]{lp{3.1in}}
\textbf{Messages}              			&\textbf{Meaning}\tabularnewline\hline
$M,M_i ::= \emptyset~~|$				&empty message\tabularnewline
~~$\mathsf{p}~~|$					&information (for $\mathcal{L}^c$: $\mathsf{p}\in\mathsf{I}^c\cup\mathsf{D}^c\cup\mathsf{G}^c$; for $\mathcal{L}$: $\mathsf{p}\in\mathcal{I}\cup\mathcal{D}\cup\mathcal{G}$)\tabularnewline
~~$\mathsf{pk}(M_1)~~|$                                 &public key corresponding to private key $M_1$\tabularnewline
~~$\{M_1,\ldots,M_n\}~~|$              			&concatenation of messages $M_1,\ldots,M_n$\tabularnewline
~~$\mathcal{H}(M_1)~~|$            			&hash of message $M$\tabularnewline
~~$E'_{M_1}(M_2)~~|$ 					&symmetric encryption  of message $M_2$ with key $M_1$\tabularnewline
~~$E_{M_1}(M_2)~~|$         		&asymmetric encryption  of message $M_2$ with public key $M_1$\tabularnewline
~~$S_{M_1}(M_2)~~|$   				        &digital signature of message $M_2$ with private key $M_1$\tabularnewline
~~$E_{M_1}(M_2)_{M_3}~~|$   		&labelled asymmetric encryption of message $M_2$ with public key $M_1$ and label $M_3$\tabularnewline
~~$\textrm{AKA}(M_1;M_2;M_3;M_4)~~|$                    &derived key from authenticated key agreement (AKA) with  (SK,randomness) pairs $(M_1,M_2)$
                                                         and $(M_3,M_4)$\tabularnewline
~~$\textrm{cred}^{M_1}_{M_2}(M_3;M_4)~~|$	        &anonymous credential with user identifier $M_1$, issuer private key $M_2$,
                                                         attributes $M_3$, and randomness $M_4$\tabularnewline
~~$\textrm{ZK}(M_1;M_2;M_3;M_4)~~|$			&zero-knowledge proof of knowledge of secret $M_1$ with properties $M_3$ using public information $M_2$ and randomness $M_4$\tabularnewline
~~$\textrm{ICred}^{M_1}_{M_2}(M_3;M_4)$      		&issuing protocol for anonymous credential $\textrm{cred}^{M_1}_{M_2}(M_3;M'_4)$, where $M'_4$ is derived from $M_4$\tabularnewline
\end{tabular} 
 \caption{Grammar defining sets of cryptographic messages at {\objl} layer ($\mathcal{L}^c$) and information layer\label{tbl:primitives}}
\end{table}
Messages built from personal and non-personal information using cryptographic primitives such as encryption, signatures, and hashes, are defined using a grammar.
Figure~\ref{tbl:primitives} shows the grammar for the primitives used to model the identity management architectures presented in Section~\ref{sec:case-study}.
For instance, $S_*(*)$ represents digital signatures: if $\ctxm{k^-}{\pi}{s}\in\mathsf{I}^c$ is a private key and $\ctxm{d}{\pi}{su}\in\mathsf{D}^c$ is a data item, then $S_{\ctxm{k^-}{\pi}{s}}(\ctxm{d}{\pi}{su})$ is a digital signature on the data item using the key.
(In this case, we write $S_{\ctxm{k^-}{}{s}}(\ctxm{d}{}{su})\ctxm{}{\pi}{}$ as shorthand.)
Although we model particular primitives here, our approach in general is independent from the particular primitives that are used; in Section~\ref{sec:discussion}, we offer some insight into the effort needed to model other primitives.
As usual (e.g.,~\cite{Blanchet2013ProVerif}), the public key belonging to private key $\mathsf{k}^-$ is represented as $\mathsf{pk}(\mathsf{k}^-)$.

We model the following cryptographic primitives.
Concatenation, hashing, and (a)symmetric encryption are modelled as usual~\cite{Clarke1998Usingstatespace,Fiore2001ComputingSymbolicModels}.
Digital signatures are ``with appendix'' \cite{Menezes1996HandbookofApplied}: that is, an actor needs to know the message that was signed in order to verify the signature.
Labelled asymmetric encryption~\cite{Bangerter2004CryptographicFrameworkControlled} is asymmetric encryption to which a label is unmodifiably attached at encryption time.
For instance, the label can represent a policy specifying when the recipient is allowed to decrypt the data.
Authenticated key agreement (AKA)~\cite{Law2003EfficientProtocolAuthenticated} allows two parties to derive a unique session key based on secret keys and randomness contributed by both parties.
We consider the variant presented in \cite{Law2003EfficientProtocolAuthenticated} in which both parties send each other a random value.
Both parties can determine the session key, modelled by the $\textrm{AKA}$ primitive, from one private key, the other public key, and the randomness.
The \emph{$\textrm{cred}$} primitive models anonymous credentials \cite{Bangerter2004CryptographicFrameworkControlled}.
Message $\textrm{cred}^{M_1}_{M_2}(M_3;M_4)$ represents an endorsement with private key $M_2$ that the attributes $M_3$ belong to the user with identifier $M_1$, randomised using $M_4$.

We also model two-party cryptographic protocols.
Using these protocols, anonymous credentials can be issued without the issuer obtaining the credential or learning $M_1$; also, their ownership can be proven without revealing the credential itself.
Such protocols only have meaning when looked at as a whole, i.e., the meaning lies not in individual messages, but in their combination in a particular order.
Thus, we model the complete transcript (i.e., all messages of all participants) of such a protocol as one grammar element.
We introduce two such primitives.

First, we model a family of zero-knowledge (ZK) proofs (e.g.,~\cite{Cramer1997ModularDesignof}) by means of the $\textrm{ZK}$ primitive.
In a ZK proof for a given property, a prover wants to convince a verifier that he knows some secrets satisfying that property with respect to some given public information, without revealing anything about the secrets.
Here, we consider ZK proofs proving that (1) the public information has a certain message structure with respect to the private information, and (2) some secret attributes $\mathsf{d}_i$ satisfy some properties $\psi_k(\mathsf{d}_i)$.
For instance, $\textrm{ZK}(\{\mathsf{d},\mathsf{n}\};\mathcal{H}(\{\mathsf{d},\mathsf{n}\});\psi_2(\mathsf{d});\mathsf{n}')$ denotes a ZK proof (using randomness $\mathsf{n}'$) convincing a verifier knowing the hash $\mathcal{H}(\{\mathsf{d},\mathsf{n}\})$ that the prover knows the pre-image $\{\mathsf{d},\mathsf{n}\}$ of the hash, and that $\psi_2(\mathsf{d})$ is satisfied; without the verifier learning anything else about $\mathsf{d}$ or $\mathsf{n}$.
See Appendix~\ref{subsec:app-zero-knowledge} for a detailed discussion.

Second, we model the issuing protocol for anonymous credentials \cite{Bangerter2004CryptographicFrameworkControlled} by means of the \emph{$\textrm{ICred}$} primitive.
This protocol is run between a user and an issuer.
In advance, both parties need to know the attributes to be certified, but only the user needs to know the identifier to which the attributes are issued.
As a result of the protocol, the user obtains an anonymous credential linking the attributes to the identifier.
The issuer does not learn the credential; moreover, because he does not know the identifier, he cannot issue credentials in her name without her involvement. 
Also, by using ZK proofs for proving ownership, the credential can be ``shown'' without revealing the identifier.
See Appendix~\ref{subsec:app-anoncred} for details.

Formally, we define an \emph{information model} that extends the personal information model from Definition~\ref{def:pi-model} by adding non-personal information and messages:

\begin{definition}
 An \emph{information model} is a tuple
$$\imodel$$
 so that:
\begin{itemize}
 \item The set $\mathcal{L}^c$ of \emph{context messages} consists of sets $\mathsf{I}^c$ of \emph{context identifiers}, $\mathsf{D}^c$ of \emph{context data items}, and $\mathsf{G}^c$ of \emph{context non-personal items}, and messages built from them using the grammar of Table~\ref{tbl:primitives}. Here, $\mathsf{I}^c$ and $\mathsf{D}^c$ are as in Definition~\ref{def:pi-model}; $\mathsf{G}^c$ consists of items $\ctxm{p}{\eta}{\cdot}$ with arbitrary variable $p$ and domain $\eta$; the set $\mathsf{P}^c:=\mathsf{I}^c\cup\mathsf{D}^c\cup\mathsf{G}^c$ is the set of \emph{context items};
 \item The set $\mathcal{L}$ of \emph{information messages} consists of sets $\mathcal{I}$ of \emph{identifiers} and $\mathcal{D}$ of data items as in Definition~\ref{def:pi-model}; $\mathcal{G}$ of non-personal items; and messages built from them using the grammar of Table~\ref{tbl:primitives};
 \item Sets $\mathsf{E}^c$ of \emph{context entities} and $\mathcal{E}$ of \emph{entities}, and the \emph{related} relation $\relation$ on $\mathcal{O}=\mathcal{E}\cup\mathcal{I}\cup\mathcal{D}$ are as in Definition~\ref{def:pi-model}.
 \item
$\sigma$ is a map $\mathcal{L}^c\cup\mathsf{E}^c\to\mathcal{L}\cup\mathcal{E}$; 
as a map $\sigma|_{\mathsf{O}^c}:\mathsf{O}^c\to\mathcal{O}$, $\sigma$ is as in Definition~\ref{def:pi-model} (where  $\mathsf{O}^c=\mathsf{E}^c\cup\mathsf{I}^c\cup\mathsf{D}^c$, and $\mathcal{O}=\mathcal{E}\cup\mathcal{I}\cup\mathcal{D}$); also, $\sigma(\mathsf{G}^c)\subset\mathcal{G}$, and $\sigma$ preserves the grammar structure of messages;
 \item $\tau$ is a map $\mathcal{L}\to \mathfrak{C}$; $\tau|_{\mathcal{I}\cup\mathcal{D}}$ is as in Definition~\ref{def:pi-model};
 $\tau(E'_x(y))=\tau(z)$ iff $z=E'_{x'}(y')$, $\tau(x)=\tau(x')$ and $\tau(y)=\tau(y')$; and similarly for other primitives.
 \item $\{\psi_1,\ldots,\psi_k\}$ are partial functions $\psi_i:\mathsf{I}^c\cup\mathsf{D}^c\to\mathsf{D}^c$, $\mathcal{I}\cup\mathcal{D}\to\mathcal{D}$ as in Definition~\ref{def:pi-model}.
\end{itemize}
\end{definition}
In particular, $\pimodel$ in the above definition is a PI model.
Note that (context) entities cannot occur in messages, so we mention them explicitly in the tuple defining an information model.
The maps $\sigma$ and $\tau$ preserve grammar structure; for instance, if $\sigma(\mathsf{k}^-)=sk_{al}$ and $\sigma(\mathsf{d})={age}_{bob}$, then $\sigma(S_{\mathsf{k}^-}(\mathsf{d}))=S_{sk_{al}}({age}_{bob})$.
Like pieces of personal information, we call context messages $\mathsf{m}$ and $\mathsf{n}$ in general \emph{equivalent} iff $\sigma(\mathsf{m})=\sigma(\mathsf{n})$, and \emph{content equivalent} iff $\tau(\sigma(\mathsf{m}))=\tau(\sigma(\mathsf{n}))$.

The above restrictions on the way $\tau$ acts on encryptions and other primitives (the fifth bullet of the definition) reflect two assumptions on message contents: namely, that they are \emph{deterministic} and \emph{unique}.
The ``if'' part of the statement reflects determinism, meaning that given the same contents as input, cryptographic primitives always give the same output.
Randomness, e.g., in signing or in non-deterministic encryption, can be modelled explicitly as part of the plaintext.
By assuming deterministic messages, we can distinguish the case where an actor observes two different randomised encryptions with the same input from the case where he observes the same randomised encryption twice; in the latter case, we will allow an actor to draw certain conclusions from this.
The ``only if'' part reflects uniqueness.
Concerning uniqueness, note that differently-constructed messages could a priori have the same contents; e.g., the hashes of two different values could collide; or the hash of some value could be the same as the encryption of some other value. 
We assume that this does not happen, i.e., elements of our grammar at the contents layer uniquely represent message contents (the \emph{structural equivalence} assumption~\cite{Veeningen2011FormalPrivacyAnalysis}).

The complete knowledge of an actor is modelled by a \emph{knowledge base}.
We model this knowledge at the context layer so that we can later determine what knowledge of personal information follows from it.
Formally:
\begin{definition}
Let $I=\imodel$ be an information model.
A \emph{knowledge base} on $I$ is a set $\mathcal{C}\subset\mathcal{L}^c\cup\mathsf{E}^c$.
\end{definition}
In addition to the messages an actor has sent and received, his knowledge base needs to contain the pieces of personal information from his initial view.
This includes context entities: because they cannot occur in messages, we mention them explicitly in the definition.
Also, the knowledge base should contain other relevant material such as secret keys known by the actor, and nonces he has generated during the execution of the cryptographic protocols.
Note that we do not need to specify the order of messages: because we use contexts, we can already distinguish between messages from different protocol instances.
We use the notation $\mathcal{C}_a$ to refer to the knowledge base of an actor $a$, and $\mathcal{C}_A$ to refer to the knowledge base of coalition $A\subset\mathcal{A}$ (defined to be the union of the knowledge bases of the respective actors, see Section~\ref{sec:traces}).

In the next example, we show several context messages and the knowledge base of an actor after communication.

\begin{example}\label{exa:kb}
We consider the PI model of Example~\ref{exa:pi-model}.
We model two context messages in domain $\pi$, which represents a protocol instance.
First, we model a  symmetric encryption of Alice's identifier, encrypted using a shared key.
The shared key is modelled by a non-personal item with context-layer representation $\ctxm{shkey}{\pi}{\cdot}$.
The encryption is then denoted $$\mathsf{m}_1=E'_{\ctxm{shkey}{}{\cdot}}(\ctxm{id}{}{su})\ctxm{}{\pi}{}.$$
Second, we model a message representing an encryption under $\ctxm{shkey}{\pi}{\cdot}$ of Alice's age and a randomised signature on her age using the server's secret key.
The randomness used in the signature is represented as a non-personal item with context-layer representation $\ctxm{n}{\pi}{\cdot}$.
The secret key of the server is context identifier $\ctxm{k^-}{\pi}{srv}$.
The second message is:
$$\mathsf{m}_2=E'_{\ctxm{shkey}{}{\cdot}}(\{\ctxm{age}{}{su},\ctxm{n}{}{\cdot},S_{\ctxm{k^-}{}{srv}}(\{\ctxm{age}{}{su},\ctxm{n}{}{\cdot}\})\})\ctxm{}{\pi}{}.$$

We now consider the knowledge base of the client, supposing that he has observed (i.e., sent or received) messages $\mathsf{m}_1$ and $\mathsf{m}_2$.
We model the communication addresses that the client and the server have used as context identifiers $\ctxm{ip}{\pi}{cl}$, $\ctxm{ip}{\pi}{srv}$.
The client knows these, as well as messages $\mathsf{m}_1,\mathsf{m}_2$.
In addition, his knowledge base contains the personal and other information known at the beginning of the scenario.
Apart from his address book, we assume that this initial knowledge includes the shared key, as well as his own address and the address and public key of the server, known in some arbitrary contexts $\ctxm{*}{\cdot}{\cdot}$, $\ctxm{*}{\cdot}{me}$, $\ctxm{*}{\cdot}{srv}$.
His full knowledge base after communication is then:
\begin{align*}
 \mathcal{C}_{cli}=\{&\ctxm{ds}{ab}{12}, \ctxm{teln}{ab}{12}, \ctxm{ds}{ab}{4}, \ctxm{id}{ab}{4}, \ctxm{skey}{\cdot}{\cdot}, \ctxm{ip}{\cdot}{me}, \ctxm{ip}{\cdot}{srv}, \\&\mathsf{pk}(\ctxm{k^-}{\cdot}{srv}), \ctxm{ip}{\pi}{cl}, \ctxm{ip}{\pi}{srv},\mathsf{m}_1,\mathsf{m}_2  \},
\end{align*}
with $\ctxm{ds}{ab}{12},\ctxm{ds}{ab}{4}$ context entities and the other elements of $\mathcal{C}_{cli}$ context messages.
\qed
\end{example}

\subsection{From Knowledge Base to View}\label{subsec:deduction}

In this subsection, we show how to determine the view corresponding to a knowledge base.
The first component of the view, the set of detectable items, is defined using a deductive system.
The second component, the associability relation, is defined based on linking identifiers and entities.

\subsubsection{A Deductive System for Detectability}\label{subsec:inf1}

In this section, we define what messages containing personal information can be built from knowledge base $\mathcal{C}$.
Informally, we say that message $\mathsf{m}$ is \emph{detectable} from $\mathcal{C}$, written $\mathcal{C}\vdash\mathsf{m}$, if it can be obtained from messages in $\mathcal{C}$ using the three operations on messages described at the beginning of this section: reading information from them, applying cryptographic operations on them, and comparing the contents of messages.
In particular, detectability of context identifiers and data items in a view is defined as detectability from as messages from the corresponding knowledge base.

The semantics of $\vdash$ is given by a deductive system.
Deductive systems are commonly used in protocol analysis to reason about what messages an attacker can fabricate (see, e.g.,~\cite{Clarke1998Usingstatespace,Fiore2001ComputingSymbolicModels}).
Typically, such deductive systems consist of general \emph{axioms} stating which messages are known; and particular \emph{construction} and \emph{elimination} rules stating the functionality of cryptographic primitives: construction rules describe how a cryptographic primitive is constructed from its parts (e.g., a symmetric encryption is constructed from the key and plaintext); \emph{elimination} rules describing how parts can be obtained from a cryptographic primitive by applying cryptographic operations (e.g. the plaintext is obtained from an encryption by decrypting using the key).
However, in these works, such rules operate directly on message contents, without taking into account what information they represent, or in which context this information is known (i.e., they operate at our contents layer).
Conversely, for our purposes, we need to consider the context: hence we need to re-interpret these rules at the {\objl} layer and add additional ones.
Our formal definition of $\vdash$ is as follows:
\begin{figure*}[tb!]
\hspace{-3cm}\fbox{
\parbox{18cm}{
\centering
\small
$\ianc{}{}{\txt{\textbf{General}\\\textbf{Rules}}}$
~~~
$\axvdashc{0}{}{\mathcal{C}\vdash \mathsf{m}}{(\mathsf{m}\in\mathcal{C})~~}{}$
~~~
$\axvdashc{E$\psi$}{\mathcal{C}\vdash\mathsf{d}}{\mathcal{C}\vdash\psi_i(\mathsf{d})}{(\psi_i(\mathsf{d})\textrm{ defined})~~}$
~~~
$\ianc{\mathcal{C}\vdash \mathsf{n}_1,\mathcal{C}\vdash \mathsf{m}_1,\mathcal{C}\vdash \mathsf{m_2}}{\mathcal{C}\vdash \mathsf{n}_2}{
\txt{($\proveseq{\mathsf{m}_1}{\mathsf{m}_2}{\mathsf{m}_3}{\mathsf{m}_4}$;\\$\mathsf{n}_1=_{\mathsf{m}_3\sim\mathsf{m}_4} \mathsf{n}_2$)}
~~\textbf{($\vdash$C)}}$
\\
$\ianc{}{}{\txt{\textbf{PK}\\\avdashc{*P}}}$
~~
$\axvdashc{CP}{\mathcal{C}\vdash \mathsf{m}}{\mathcal{C}\vdash \mathsf{pk}(\mathsf{m})}{}$
~~
$\ianc{}{}{\txt{\textbf{Concate-}\\\textbf{nation}~\avdashc{*C}}}$%
~~%
$\axvdashc{CC}{\mathcal{C}\vdash \mathsf{m},\mathcal{C}\vdash \mathsf{n}}{\mathcal{C}\vdash \{\mathsf{m}{,}\mathsf{n}\}}{}$
~~
$\axvdashc{EC}{\mathcal{C}\vdash \{\mathsf{m}{,}\mathsf{n}\}}{\mathcal{C}\vdash \mathsf{m}}{}$
~~
$\axvdashc{EC'}{\mathcal{C}\vdash \{\mathsf{m}{,}\mathsf{n}\}}{\mathcal{C}\vdash \mathsf{n}}{}$
~~
$\ianc{}{}{\txt{\textbf{Hash}\\\avdashc{*H}}}$
~~
$\axvdashc{CH}{\mathcal{C}\vdash \mathsf{m}}{\mathcal{C}\vdash \mathcal{H}(\mathsf{m})}{}$
\\
$\ianc{}{}{\txt{\textbf{Symmetric en-}\\\textbf{cryption}~\avdashc{*E}}}$%
~~~%
$\axvdashc{CE}{\mathcal{C}\vdash \mathsf{m},\mathcal{C}\vdash \mathsf{n}}{\mathcal{C}\vdash E'_{\mathsf{n}}(\mathsf{m})}{}$
~~~
$\axvdashc{EE}{\mathcal{C}\vdash E'_{\mathsf{n}}(\mathsf{m}),\mathcal{C}\vdash \mathsf{n}}{\mathcal{C}\vdash \mathsf{m}}{}$
~~~
$\axvdashc{TE}{\mathcal{C}\vdash	E'_{\mathsf{n}}(\mathsf{m})
      ,\mathcal{C}\vdash		\mathsf{n}'
      }{\mathcal{C}\vdash		\mathsf{n}
      }{(				\mathsf{n}'\doteq\mathsf{n}
      )~~}$
~~~
$\ianc{}{}{\txt{\textbf{Asymmetric en-}\\\textbf{cryption}~\avdashc{*A}}}$%
~~~%
$\axvdashc{CA}{\mathcal{C}\vdash \mathsf{m},\mathcal{C}\vdash \mathsf{k}^+}{\mathcal{C}\vdash E_{\mathsf{k}^+}(\mathsf{m})}{}$
~~~
$\axvdashc{EA}{\mathcal{C}\vdash E_{\mathsf{pk}(\mathsf{k}^-)}(\mathsf{m}),\mathcal{C}\vdash \mathsf{k}^-}{\mathcal{C}\vdash \mathsf{m}}{}$
~~~
$\axvdashc{TA}{\mathcal{C}\vdash	E_{\mathsf{pk}(\mathsf{k}^-)}(\mathsf{m})
      ,\mathcal{C}\vdash		{\mathsf{k}^-}{'}
      }{\mathcal{C}\vdash		\mathsf{k}^-
      }{(				{\mathsf{k}^-}{'}\doteq \mathsf{k}^-
      )~~}$
~~~
$\ianc{}{}{\txt{\textbf{Sign}\\\avdashc{*S}}}$%
~~~%
$\axvdashc{CS}{\mathcal{C}\vdash \mathsf{m},\mathcal{C}\vdash \mathsf{k}^-}{\mathcal{C}\vdash S_{\mathsf{k}^-}(\mathsf{m})}{}$
~~~
$\axvdashc{TS}{\mathcal{C}\vdash	S_{\mathsf{k}^-}(\mathsf{m})
      ,\mathcal{C}\vdash		\{\mathsf{pk}({\mathsf{k}^-}{'}){,}\mathsf{m'}\}
      }{\mathcal{C}\vdash		\{\mathsf{pk}(\mathsf{k}^-),\mathsf{m}\}
      }{(				*'\doteq*
      )~~}$
\\
-~-~-~-~-~-~-~-~-~-~-~-~-~-~-~-~-~-~-~-~-~-~-~-~-~-~-~-~-~-~-~-~-~-~-~-~-~-~-~-~-~-~-~-~-~-~-~-~-~-~-~-~-~-~-~-~-~-~-~-~-~-~-~-~-~-~-~-~-~-~-~-~-~-~-~-~-~-~-~-~-~-~-~-~-~-~-~-~-~-~-~-~-~-~-~-~-~-~-~\\
$\ianc{}{}{\txt{\textbf{Lab. asym.}\\\textbf{enc.}~\avdashc{*L}}}$%
~
$\axvdashc{CL}{\mathcal{C}\vdash \mathsf{m},\mathcal{C}\vdash \mathsf{k}^+,\mathcal{C}\vdash \mathsf{n}}{\mathcal{C}\vdash E_{\mathsf{k}^+}(\mathsf{m})_\mathsf{n}}{}$
~
$\axvdashc{EL}{\mathcal{C}\vdash E_{\mathsf{pk}(\mathsf{k}^-)}(\mathsf{m})_\mathsf{n}}{\mathcal{C}\vdash \mathsf{n}}{}$
~
$\axvdashc{EL'}{\mathcal{C}\vdash E_{\mathsf{pk}(\mathsf{k}^-)}(\mathsf{m})_\mathsf{n},\mathcal{C}\vdash \mathsf{k}^-}{\mathcal{C}\vdash \mathsf{m}}{}$
~
$\axvdashc{TL}{\mathcal{C}\vdash	E_{\mathsf{pk}(\mathsf{k}^-)}(\mathsf{m})_\mathsf{n}
      ,\mathcal{C}\vdash		{\mathsf{k}^-}{'}
      }{\mathcal{C}\vdash		\mathsf{k}^-
      }{(				{\mathsf{k}^-}{'}\doteq \mathsf{k}^-
      )~~}$
~~~
$\ianc{}{}{\txt{\textbf{Auth Key}\\\textbf{Agr}~\avdashc{*G}}}$%
~~~%
$\axvdashc{CG}{\mathcal{C}\vdash  \{\mathsf{k}_1^-{,}\mathsf{n}_1{,} \mathsf{pk}(\mathsf{k}_2^-){,} \mathsf{n}_2\} }
      {\mathcal{C}\vdash \textrm{AKA}(\mathsf{k}_1^-;\mathsf{n}_1;\mathsf{k}_2^-;\mathsf{n}_2)}{}$
~~~
$\axvdashc{CG'}{\mathcal{C}\vdash  \{\mathsf{pk}(\mathsf{k}_1^-){,}\mathsf{n}_1{,} \mathsf{k}_2^-{,} \mathsf{n}_2\} }
      {\mathcal{C}\vdash \textrm{AKA}(\mathsf{k}_1^-;\mathsf{n}_1;\mathsf{k}_2^-;\mathsf{n}_2)}{}$
~~~
$\ianc{}{}{\txt{\textbf{Anon Cred}\\\avdashc{*R}}}$%
~~~
$\axvdashc{CR}{\mathcal{C}\vdash  \{\mathsf{k}^-{,}\mathsf{m}_1{,}\mathsf{m}_2{,}\mathsf{n}\} }
      {\mathcal{C}\vdash \mbox{cred}_{\mathsf{k}^-}^{\mathsf{m}_1}(\mathsf{m}_2;\mathsf{n})}{}$~~
$\axvdashc{TR}{\mathcal{C}\vdash	\mbox{cred}_{\mathsf{k}^-}^{\mathsf{m}_1}(\mathsf{m}_2;\mathsf{n})
      ,\mathcal{C}\vdash		\{\mathsf{pk}({\mathsf{k}^-}{'}){,}\mathsf{m}_1'{,}\mathsf{m}_2'\}
      }{\mathcal{C}\vdash		\{\mathsf{pk}(\mathsf{k}^-){,}\mathsf{m}_1{,}\mathsf{m}_2\}
      }{(				*'\doteq *
      )~~}$
~~~
$\ianc{}{}{\txt{\textbf{ZK Proof}\\\avdashc{*Z}}}$%
~~~
$\axvdashc{CZ}{\mathcal{C}\vdash\{\mathsf{m}_1{,}\mathsf{m}_2{,}\mathsf{m}_3{,}\mathsf{m}_4\}}
              {\mathcal{C}\vdash\textrm{ZK}(\mathsf{m}_1;\mathsf{m}_2;\mathsf{m}_3;\mathsf{m}_4)}
              {}$
~~~
$\axvdashc{EZ$_1$}{\mathcal{C}\vdash\mbox{ZK}(\mathsf{m}_1{;}\mathsf{m}_2{;}\mathsf{m}_3{;}\{\mathsf{n}_p{,}\mathsf{n}_v\})}
              {\mathcal{C}\vdash \mathsf{m}_3}
              {}$
~~~
$\axvdashc{EZ$_2$}{\mathcal{C}\vdash\{\mbox{ZK}(\mathsf{m}_1{;}\mathsf{m}_2{;}\mathsf{m}_3{;}\{\mathsf{n}_p{,}\mathsf{n}_v\}){,}\mathsf{n}_p\}}
              {\mathcal{C}\vdash \mathsf{m}_1}
              {}$
~~~
$\axvdashc{EZ$_3$}{\mathcal{C}\vdash	\mbox{ZK}(\mathsf{m}_1{;}\mathsf{m}_2{;}\mathsf{m}_3{;}\{\mathsf{n}_p{,}\mathsf{n}_v\})
      }{\mathcal{C}\vdash		\mathsf{m}_2
      }{}$
~~~
$\axvdashc{TZ$_1$}{\mathcal{C}\vdash	\mbox{ZK}(\mathsf{m}_1{;}\mathsf{m}_2{;}\mathsf{m}_3{;}\{\mathsf{n}_p{,}\mathsf{n}_v\})
      ,\mathcal{C}\vdash		\mathsf{n}_p'
      }{\mathcal{C}\vdash		\mathsf{n}_p
      }{(				\mathsf{n}_p'\doteq\mathsf{n}_p
      )~~}$
~~~
$\ianc{}{}{\txt{\textbf{Cred Iss}\\\avdashc{*I}}}$%
~~~
$\axvdashc{CI}{\mathcal{C}\vdash \{\mathsf{k}^-{,}\mathsf{m}_1{,}\mathsf{m}_2{,}\mathsf{n}\}}
	      {\mathcal{C}\vdash \mbox{ICred}_{\mathsf{k}^-}^{\mathsf{m}_1}(\mathsf{m}_2;\mathsf{n}\})}
              {}$
~
$\axvdashc{EI$_1$}{\mathcal{C}\vdash \{\mbox{ICred}_{\mathsf{k}^-}^{\mathsf{m}_1}(\mathsf{m}_2;\{\mathsf{n}_i\}_{i=1}^7){,}\mathsf{n}_2\}}
              {\mathcal{C}\vdash \mbox{cred}_{\mathsf{k}^-}^{\mathsf{m}_1}(\mathsf{m}_2;\{\mathsf{n}_2,\mathsf{n}_5\})}
              {}$
~
$\axvdashc{EI$_2$}{\mathcal{C}\vdash \{\mbox{ICred}_{\mathsf{k}^-}^{\mathsf{m}_1}(\mathsf{m}_2;\{\mathsf{n}_i\}_{i=1}^7){,}\mathsf{n}_3\}}
              {\mathcal{C}\vdash \{\mathsf{m}_1,\mathsf{n}_1,\mathsf{n}_2\}}
              {}$
~~~
$\axvdashc{EI$_3$}{\mathcal{C}\vdash \{\mbox{ICred}_{\mathsf{k}^-}^{\mathsf{m}_1}(\mathsf{m}_2;\{\mathsf{n}_i\}_{i=1}^7){,}\mathsf{n}_6\}}
              {\mathcal{C}\vdash \mathsf{k}^-}
              {}$
~~~
$\axvdashc{EI$_4$}{\mathcal{C}\vdash \mbox{ICred}_{\mathsf{k}^-}^{\mathsf{m}_1}(\mathsf{m}_2;\{\mathsf{n}_i\}_{i=1}^7)}
              {\mathcal{C}\vdash \{\mathsf{pk}(\mathsf{k}^-),\mathsf{m}_2,\mathcal{H}(\mathsf{m}_1,\mathsf{n}_1)\}}
              {}$
~~~
$\axvdashc{TI$_1$}{\mathcal{C}\vdash	\mbox{ICred}_{\mathsf{k}^-}^{\mathsf{m}_1}(\mathsf{m}_2;\{\mathsf{n}_i\}_{i=1}^7)
      ,\mathcal{C}\vdash		\{\mathsf{m}_1'{,}\mathsf{n}_2'\}
      }{\mathcal{C}\vdash		\{\mathsf{m}_1,\mathsf{n}_2\}
      }{(				*'\doteq *
      )~~}$
~~~
$\axvdashc{TI$_2$}{\mathcal{C}\vdash	\mbox{ICred}_{\mathsf{k}^-}^{\mathsf{m}_1}(\mathsf{m}_2;\{\mathsf{n}_i\}_{i=1}^7)
      ,\mathcal{C}\vdash		\mbox{cred}_{{\mathsf{k}^-}{'}}^{\mathsf{m}_1'}(\mathsf{m}_2';\{\mathsf{n}_2'{,}\mathsf{n}_5'\})
      }{\mathcal{C}\vdash		\mbox{cred}_{\mathsf{k}^-}^{\mathsf{m}_1}(\mathsf{m}_2;\{\mathsf{n}_2,\mathsf{n}_5\})
      }{(				*'\doteq *
      )~~}$
~~~
$\axvdashc{TI$_3$}{\mathcal{C}\vdash	\mbox{ICred}_{\mathsf{k}^-}^{\mathsf{m}_1}(\mathsf{m}_2;\{\mathsf{n}_i\}_{i=1}^7)
      ,\mathcal{C}\vdash		\mathsf{n}_2'
      }{\mathcal{C}\vdash		\mathsf{n}_2
      }{(				\mathsf{n}_2'\doteq\mathsf{n}_2
      )~~}$
~~~
$\axvdashc{TI$_4$}{\mathcal{C}\vdash	\mbox{ICred}_{\mathsf{k}^-}^{\mathsf{m}_1}(\mathsf{m}_2;\{\mathsf{n}_i\}_{i=1}^7)
      ,\mathcal{C}\vdash		\mathsf{n}_3'
      }{\mathcal{C}\vdash		\mathsf{n}_3
      }{(				\mathsf{n}_3'\doteq\mathsf{n}_3
      )~~}$
~~~
$\axvdashc{TI$_5$}{\mathcal{C}\vdash	\mbox{ICred}_{\mathsf{k}^-}^{\mathsf{m}_1}(\mathsf{m}_2;\{\mathsf{n}_i\}_{i=1}^7)
      ,\mathcal{C}\vdash		\mathsf{n}_6'
      }{\mathcal{C}\vdash		\mathsf{n}_6
      }{(				\mathsf{n}_6'\doteq\mathsf{n}_6
      )~~}$
}
}
\caption{Deductive system for detectability: $\mathsf{m}$, $\mathsf{m}_i$, $\mathsf{n}$, $\mathsf{n}_i$, $\mathsf{k}^-$, $\mathsf{k}^+$, $\mathsf{k}_i^-$ and $*'$ any context message; $\mathsf{d}\in\mathsf{I}^c\cup\mathsf{D}^c$ any context identifier or data item; $\mathsf{p}_*\in\mathsf{D}^c$ any context data item.
$*'\doteq *$ means ``for any pair of dashed and non-dashed context messages''.
Rules $\avdashc{0}$ to $\avdashc{TS}$ explained in Section~\ref{subsec:inf1}; rules $\avdashc{CL}$ to $\avdashc{TI$_6$}$ in Section~\ref{subsec:inf2}.
\label{fig:deductive-system}
}
\end{figure*}
\begin{definition}\label{def:detectability}
 Let $\mathcal{C}$ be a knowledge base, and $\mathsf{m}$ a context message.
 The \emph{detectability} relation $\mathcal{C}\vdash\mathsf{m}$ is defined by the inference rules given in Figure~\ref{fig:deductive-system}.
\end{definition}

The deductive system in Figure~\ref{fig:deductive-system} consists of three general rules $\avdashc{0}$, $\avdashc{E$\psi$}$, and $\avdashc{C}$; and particular \emph{construction}, \emph{elimination}, and \emph{testing} rules for the particular cryptographic primitives modelled.
Hence, when using our framework to analyse a system, $\avdashc{0}$, $\avdashc{E$\psi$}$, and $\avdashc{C}$ are always the same; the other rules need to be adapted to model the particular primitives used in the system.

$\avdashc{0}$ is the standard axiom to detect known messages.
Construction and elimination rules, in particular those for (standard) hashes, (a)symmetric encryption, concatenation, and signatures, are as usual \cite{Clarke1998Usingstatespace}.
For instance, rule $\avdashc{CE}$ states that symmetric encryption $E'_\mathsf{n}(\mathsf{m})$ can be detected if $\mathsf{m}$ and $\mathsf{n}$ can be detected, and rule $\avdashc{EE}$ states that plaintext $\mathsf{m}$ can be obtained from encryption $E'_\mathsf{n}(\mathsf{m})$ if key $\mathsf{n}$ is known.
However, note that because our deductive system operates at the context layer, rule $\avdashc{EE}$ only applies if the key is known \emph{in the same context as the message}.
Thus, this rule fails to capture that an actor can perform decryption using keys he knows from different contexts.
To avoid this problem, we introduce testing rules.
These rules let an actor detect a new context-layer representation of a messages whose contents he already knew by applying a cryptographic operation.
For instance, rule $\avdashc{TE}$ states that if an actor can detect encryption $E'_\mathsf{n}(\mathsf{m})$ and any message content equivalent to $\mathsf{n}$, then he can detect $\mathsf{n}$.
He can then use $\mathsf{n}$ to decrypt the message.

Example~\ref{exa:key-guessing} shows a typical example of the use of testing and elimination rules.
\begin{example}\label{exa:key-guessing}
Consider knowledge base $\mathcal{C}_{cli}$ from Example~\ref{exa:kb}.
Then $\ctxm{id}{\pi}{su}$ is detectable from $\mathcal{C}_{cli}$ by the derivation shown in Figure~\ref{fig:exakeyguessing-der}.
\begin{figure}[tb]
$$\small\ianc
  {
    \ianc{}{\mathcal{C}_{cli}\vdash E'_{\ctxm{shkey}{}{\cdot}}(\ctxm{id}{}{su})\ctxm{}{\pi}{}}{\avdashc{0}},
    \ianc
        {
          \ianc{}{\mathcal{C}_{cli}\vdash E'_{\ctxm{shkey}{}{\cdot}}(\ctxm{id}{}{su})\ctxm{}{\pi}{}}{\avdashc{0}},
          \ianc{}{\mathcal{C}_{cli}\vdash\ctxm{skey}{\cdot}{\cdot}}{\avdashc{0}}
        }
       {\mathcal{C}_{cli}\vdash\ctxm{shkey}{\pi}{\cdot}}
       {\avdashc{TE}}
  }
  {\mathcal{C}_{cli}\vdash\ctxm{id}{\pi}{su}}
  {\avdashc{EE}}
$$ 
\caption{Derivation of $\ctxm{id}{\pi}{su}$ given knowledge base $\mathcal{C}_{cli}$ from Example~\ref{exa:kb} (see Example~\ref{exa:key-guessing})\label{fig:exakeyguessing-der}}
\end{figure}
The derivation models the actor testing whether $\ctxm{skey}{\cdot}{\cdot}$ is the decryption key for $E'_{\ctxm{shkey}{}{\cdot}}(\ctxm{id}{}{su})\ctxm{}{\pi}{}$ $\avdashc{TE}$.
(Rule $\avdashc{TE}$ can be applied because $\ctxm{shkey}{\pi}{\cdot}\doteq \ctxm{skey}{\cdot}{\cdot}$.)
After learning that it is, the actor can decrypt the message $\avdashc{EE}$.\qed
\end{example}
We assume that for any cryptographic operation modelled by an elimination rule, there is a corresponding testing rule.
(This is an over-estimation in case the actor cannot distinguish between a failed and successful cryptographic operation, e.g. when certain kinds of encryption schemes are used in which the plaintext resulting from decryption cannot be recognised as valid.)
On the other hand, not all testing rules testing have a corresponding elimination rule, e.g.,~rule $\avdashc{TS}$ for signature verification.

Differently from other deductive systems, we introduce two additional general rules: $\avdashc{E$\psi$}$ to reason about properties of attributes and $\avdashc{C}$ to reason about contents of messages.
Rule $\avdashc{E$\psi$}$ states that any properties that apply to an attribute can be detected from the attribute.
Note that, because the rule only applies if image $\psi_i(\mathsf{d})$ under the partial function $\psi_i$ is defined, only applicable properties can be detected.
Rule $\avdashc{C}$ covers knowledge obtained by comparing contents of different messages.

\begin{figure}
 \centering\includegraphics[scale=0.8]{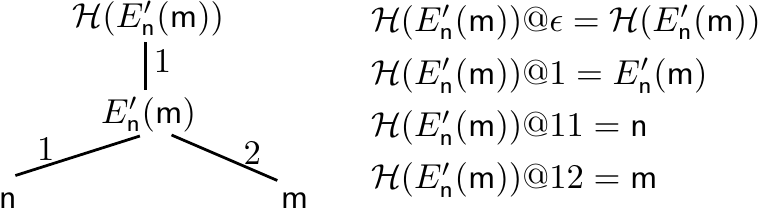}
 \caption{Parse tree of message $\mathcal{H}(E'_{\mathsf{n}}(\mathsf{m}))$ (left) and submessages (right)\label{fig:exa-parse-tree}}
\end{figure}

Before we can discuss rule $\avdashc{C}$ in detail, we first need to formalise the notion of \emph{submessages} of a message.
A message $\mathsf{m}$ has a natural syntactic structure according to the grammar in Figure~\ref{fig:deductive-system}.
This structure can be represented by a parse tree, in which the nodes are the submessages of $\mathsf{m}$; the root is the message $\mathsf{m}$ itself.
We write $\mathsf{m}@z$ for the submessage of $\mathsf{m}$ at path $z$ from $\mathsf{m}$ in its parse tree; the empty path is denoted $\epsilon$.
Figure~\ref{fig:exa-parse-tree} shows the parse tree of a message (left) and the corresponding formal representation of its submessages (right).

If two context messages $\mathsf{m}_1$ and $\mathsf{m}_2$ are content equivalent, then by the uniqueness assumption, their respective submessages are also content equivalent.
That is, if $\mathsf{m}_1\doteq\mathsf{m}_2$ and $\mathsf{m}_1@z$ and $\mathsf{m}_2@z$ are defined (i.e., there exists a submessage at path $z$), then $\mathsf{m}_1@z\doteq\mathsf{m}_2@z$.
Also, if $\mathsf{m}_1$ and $\mathsf{m}_2$ contain data items satisfying a property $\psi_k$, the content equivalence of that property is also implied.
Formally:
\begin{definition}\label{def:evidence}
 The pair $(\mathsf{m}_1,\mathsf{m}_2)$ is \emph{evidence} for $\mathsf{n}_1\doteq\mathsf{n}_2$, denoted $\proveseq{\mathsf{m_{1}}}{\mathsf{m_{2}}}{\mathsf{n}_{1}}{\mathsf{n}_{2}}$, if one of the following two conditions holds:
\begin{itemize}
 \item $\mathsf{m}_1\doteq\mathsf{m}_2$, and for some $z$, $\mathsf{m}_1@z=\mathsf{n}_1$ and $\mathsf{m}_2@z=\mathsf{n}_2$;
 \item $\proveseq{\mathsf{m_{1}}}{\mathsf{m_{2}}}{\mathsf{n}_{1}'}{\mathsf{n}_{2}'}$, and for some $i$, $\mathsf{n}_1=\psi_i(\mathsf{n}_1')$ and $\mathsf{n}_2=\psi_i(\mathsf{n}_2')$. 
\end{itemize}
\end{definition}
The ``content analysis'' inference rule $\avdashc{C}$ then states that if an actor can derive evidence $(\mathsf{m}_1,\mathsf{m}_2)$ for $\mathsf{n}_{1}\doteq\mathsf{n}_{2}$ and he can derive a message with $\mathsf{n}_1$ in it, then he can derive the same message with $\mathsf{n}_1$ replaced by $\mathsf{n}_2$, and vice versa.
The following example shows how $\avdashc{C}$ models an actor determining a piece of information by reasoning about its contents:

\begin{example}\label{exa:content-reasoning}
\begin{figure}[tb]
\centering$
\footnotesize\ianc{
  \ianc{}{\mathcal{C}_a\vdash\ctxm{id}\eta2}{\avdashc0},
  \ianc{
    \ianc{
      \ianc{}{\mathcal{C}_a\vdash\ctxm{id}\eta2}{\avdashc0},	
      \ianc{}{\mathcal{C}_a\vdash\ctxm{age}\eta3}{\avdashc0},
    }{\mathcal{C}_a\vdash\{\ctxm{id}\eta2{,}\ctxm{age}\eta3\}}{\avdashc{CC}}
  }{\mathcal{C}_a\vdash\mathcal{H}(\{\ctxm{id}\eta2{,}\ctxm{age}\eta3\})}{\avdashc{CH}},
  \ianc{}{\mathcal{C}_a\vdash\ctxm{\mathcal{H}(\{{id}{,}{age}\})}\eta1}{\avdashc0},
  }
  {\mathcal{C}_a\vdash\ctxm{id}\eta1}{\avdashc C}
$
\caption{Derivation of $\ctxm{id}\eta1$ given knowledge base $\mathcal{C}_a=\{\ctxm{\mathcal{H}(\{{id},{age}\})}\eta1,\ctxm{id}\eta2,\ctxm{age}\eta3\}$ (see Example~\ref{exa:content-reasoning}).\label{fig:deduction-ca}}
\end{figure}
Consider knowledge base $$\mathcal{C}_a=\{\ctxm{\mathcal{H}(\{{id},{age}\})}\eta1,\ctxm{id}\eta2,\ctxm{age}\eta3\},$$ where $\ctxm{id}\eta1\doteq\ctxm{id}\eta2$ and $\ctxm{age}\eta1\doteq\ctxm{age}\eta3$.
Intuitively, an actor can learn $\ctxm{id}\eta1$ from this knowledge base: he can construct the hash $\mathcal{H}(\{\ctxm{id}\eta2,\ctxm{age}\eta3\})$, note that it has the same contents as $\ctxm{\mathcal{H}(\{{id},{age}\})}\eta1$, and thus infer that $\ctxm{id}\eta1$ must have the same contents as $\ctxm{id}\eta2$, which he knows.

This series of reasoning step is captured in the derivation shown in Figure~\ref{fig:deduction-ca}.
Namely, $\mathcal{C}_a\vdash\ctxm{\mathcal{H}(\{{id},{age}\})}\eta1$ holds, and by $\avdashc{CC}$, $\avdashc{CH}$ we have $\mathcal{C}_a\vdash\mathcal{H}(\{\ctxm{id}\eta2,\ctxm{age}\eta3\})$.
By Definition~\ref{def:evidence}, the pair $(\ctxm{\mathcal{H}(\{{id},{age}\})}\eta1,\mathcal{H}(\{\ctxm{id}\eta2,\ctxm{age}\eta3\}))$ is evidence for $\ctxm{id}\eta1\doteq\ctxm{id}\eta2$ (as well as for $\ctxm{age}\eta1\doteq\ctxm{age}\eta3$).
By $\avdashc C$, he can then deduce $\ctxm{id}\eta1$. 
(In the same way, also $\mathcal{C}_a\vdash\ctxm{age}\eta1$ follows.) \qed
\end{example}

\subsubsection{Inference Rules for Non-Standard Primitives}\label{subsec:inf2}

We now discuss the inference rules for the non-standard primitives modelled in this paper.
Labelled asymmetric encryption is similar to normal asymmetric encryption; note that the label can be derived from the encryption $\avdashc{EL'}$, but to change it, the plaintext is needed, i.e., the label is unmodifiably attached.
To derive a session key using authenticated key agreement, an actor needs to know one of the private keys used, the other public key, and both parties' randomness $\avdashc{CG}$, $\avdashc{CG'}$.

Messages $\textrm{ZK}(\ldots)$ and $\textrm{ICred}^*_*(\ldots)$ represent the complete transcripts of instances of zero-knowledge proofs and credential issuing protocols, respectively.
In particular, the construction rules for these messages state which inputs are required to build the complete transcript.
(When such a protocol is run, two different parties each provide part of the input; this is captured by traces, see Section~\ref{sec:traces}.)

The inference rules for $\textrm{ZK}$ given in our deductive system model the privacy aspects of a large family of ZK proofs known as ``$\Sigma$-protocols''~\cite{Cramer1997ModularDesignof}.
$\Sigma$-protocols exist for many properties; in particular, they are used to prove properties of anonymous credentials~\cite{Bangerter2004CryptographicFrameworkControlled}.
The randomness for $\Sigma$-protocols is of the form $\{\mathsf{n}_p,\mathsf{n}_v\}$, representing contributions by the prover and verifier, respectively.
Apart from the usual construction rule, there are three elimination rules: $\avdashc{EZ$_1$}$ states that the property proven by a ZK proof can be seen from its transcript; $\avdashc{EZ$_2$}$ states that the prover's secret can be derived from the prover's randomness; and $\avdashc{EZ$_3$}$ states that the public information can be derived.
Testing rule $\avdashc{TZ$_1$}$ can be applied to obtain the prover's randomness.
We assume that parties do not reuse their randomness; also, because we are only interested in privacy aspects, we only consider deriving randomness if that randomness can be used to derive other information.
See Appendix \ref{subsec:app-zero-knowledge} for details.

The inference rules for $\mathrm{cred}$ and $\mathrm{ICred}$ model anonymous credentials and their issuing protocol based on SRSA-CL signatures \cite{Bangerter2004CryptographicFrameworkControlled}.
Anonymous credentials can be verified to correspond to a given verification key, message and secret identifier $\avdashc{TR}$.
Anonymous credentials are usually derived from the transcript of its issuing protocol $\avdashc{EI$_1$}$ by the user  (the issuer does not know $\mathsf{n}_2$ and so does not learn the credential); but they can also be constructed directly from its components $\avdashc{CR}$.
Similarly for the issuing protocol transcript itself $\avdashc{CI}$.
Before the issuing protocol takes place, the user needs to have sent a randomised commitment $\mathcal{H}(\mathsf{m}_1,\mathsf{n}_1)$ to her secret identifier to the issuer.
During the protocol, additional randomness $\mathsf{n}_2,...,\mathsf{n}_8$ is generated by the two parties; $\mathsf{n}_1,...,\mathsf{n}_8$ together form the randomness component of the $\mbox{ICred}$ primitive.
Inference rules $\avdashc{EI$_2$}$ and $\avdashc{EI$_3$}$ model the inference of secret information from the transcript using randomness.
A credential issuing protocol transcript allows for deriving and testing of various nonces and information used ($\avdashc{EI$_4$}$; $\avdashc{TI$_1$}$--$\avdashc{TI$_5$}$).
As with our model of ZK proofs, we only consider rules needed to infer personal information, and assume non-reuse of randomness.
In Appendix \ref{subsec:app-anoncred} we explain why these rules accurately capture privacy aspects.

\subsubsection{Associability and View}\label{subsec:associability}

Having discussed detectability, we consider the other part of an actor view: associability.
We determine the associability relation corresponding to a knowledge base $\mathcal{C}$ by finding out which identifiers and entities are known to be equivalent in~$\mathcal{C}$:
\begin{definition}\label{def:associability}
 Let $\mathcal{C}$ be a knowledge base.
 The \emph{associability relation $\link{}$ corresponding to $\mathcal{C}$} is the equivalence relation on $\mathsf{O}^{c}$ obtained by evaluating the following rules:
\begin{enumerate}
 \item For all $\ctxm{ds}{\eta}{k},\ctxm{ds}{\zeta}{l} \in\mathcal{C}\cap\mathsf{E}^c$: if $\sigma(\ctxm{ds}{\eta}{k})=\sigma(\ctxm{ds}{\zeta}{l})$, then $\linkable{\ctxm{ds}{\eta}{k}}{\ctxm{ds}{\zeta}{l}}{}$;
 \item For all $\ctxm{x}{\eta}{k},\ctxm{y}{\eta}{k} \in\mathsf{O}^{c}$: $\linkable{\ctxm{x}{\eta}{k}}{\ctxm{y}{\eta}{k}}{}$;
 \item If $\mathcal{C}\vdash \mathsf{m}_1$, $\mathcal{C}\vdash\mathsf{m}_2$, and $\proveseq{\mathsf{m}_1}{\mathsf{m}_2}{\mathsf{i}_1}{\mathsf{i}_2}$ for $\mathsf{i}_1,\mathsf{i}_2\in\mathsf{I}^{c}$, then $\linkable{\mathsf{i}_1}{\mathsf{i}_2}{}$.
\end{enumerate}
and taking the reflexive, symmetric, transitive closure.
\end{definition}
The first point states that any known context entities representing the same entity can be associated; the second point states that all information from the same context can be associated.
The third point captures associations by identifiers.
Actors do not need to be able to detect the identifier: instead, it is sufficient to detect evidence for content equivalence (Definition~\ref{def:evidence}).
The following example demonstrates the definition.
\begin{example}\label{exa:associability}
 We determine the associability relation $\leftrightarrow_{cli}$ corresponding to the knowledge base $\mathcal{C}_{cli}$ from Example~\ref{exa:kb}:
\begin{align*}
 \mathcal{C}_{cli}=\{&\ctxm{ds}{ab}{12}, \ctxm{teln}{ab}{12}, \ctxm{ds}{ab}{4}, \ctxm{id}{ab}{4}, \ctxm{skey}{\cdot}{\cdot}, \ctxm{ip}{\cdot}{me}, \ctxm{ip}{\cdot}{srv}, \\& \mathsf{pk}(\ctxm{k^-}{\cdot}{srv}), \ctxm{ip}{\pi}{cl}, \ctxm{ip}{\pi}{srv}, 
E'_{\ctxm{shkey}{}{\cdot}}(\ctxm{id}{}{su})\ctxm{}{\pi}{},\\&E'_{\ctxm{shkey}{}{\cdot}}(\{\ctxm{age}{}{su},\ctxm{n}{}{\cdot},S_{\ctxm{k^-}{}{srv}}(\{\ctxm{age}{}{su},\ctxm{n}{}{\cdot}\})\})\ctxm{}{\pi}{}  \}.
\end{align*}
Rule 2 from Definition~\ref{def:associability} allows association of information from the same context; thus, e.g., $\ctxm{ds}{ab}{4}\leftrightarrow_{cli}\ctxm{id}{ab}{4}$.
Rule 3 allows association of context identifiers using evidence of content equivalence.
For instance, clearly, $\mathcal{C}_{cli}\vdash \ctxm{id}{ab}{4}$, $\mathcal{C}_{cli}\vdash \ctxm{id}{\pi}{su}$ (see Example~\ref{exa:key-guessing}), and $\proveseq{\ctxm{id}{ab}{4}}{\ctxm{id}{\pi}{su}}{\ctxm{id}{ab}{4}}{\ctxm{id}{\pi}{su}}$, hence  $\ctxm{id}{ab}{4}\leftrightarrow\ctxm{id}{\pi}{su}$.
In fact, all context items about Alice that occur in $\mathcal{C}_{cli}$ turn out to be mutually associable.
However, rule~1 for associating entities does not apply since, e.g.,~$\sigma(\ctxm{ds}{ab}{12})\ne\sigma(\ctxm{ds}{ab}{4})$.
Continuing in this way, the items detectable from $\mathcal{C}_{cli}$ form the following equivalence classes under $\leftrightarrow_{cli}$:
$$
\{\ctxm{ds}{ab}{12}, \ctxm{teln}{ab}{12}\}~~
\{\ctxm{ds}{ab}{4}, \ctxm{id}{ab}{4}, \ctxm{id}{\pi}{su}, \ctxm{age}{\pi}{su}\} $$ $$
\{\ctxm{ip}{\pi}{cl}, \ctxm{ip}{\cdot}{me}, \}~~
\{\ctxm{ip}{\cdot}{srv}, \ctxm{k^-}{\cdot}{srv}, \ctxm{ip}{\pi}{srv}, \ctxm{k^-}{\pi}{srv}\},
$$
with data subjects Bob, Alice, the client, and the server, respectively.\qed
\end{example}

Note that $\link{}$ is intentionally defined on all {\objl}-layer items, and not just on detectable {\objl}-layer items.
The following example shows how this broader definition allows additional inferences to be made:

\begin{example}\label{exa:associate-undet}
Let $\mathcal{C}_a=\{\ctxm{\{E_{\ctxm{shakey}{}{\cdot}}(\ctxm{id}{}{1})$, $\ctxm{d}{}{1}\}}{\eta}{}$, $\ctxm{\{E_{\ctxm{shakey}{}{\cdot}}(\ctxm{id}{}{1})$, $\ctxm{d'}{}{1}\}}{\chi}{}\}$ be a knowledge base,
where $\ctxm{shakey}{\eta}{\cdot}\doteq\ctxm{shakey}{\chi}{\cdot}$ and $\ctxm{id}{\eta}{1}\doteq\ctxm{id}{\chi}{1}$.
Let $\link{a}$ be the associability relation corresponding to $\mathcal{C}_a$.
Intuitively, even if the key used in the encryptions in $\mathcal{C}_a$ is unknown, the fact that they have the same contents means that the two identifiers, and hence also the two data items $\ctxm{d}{\eta}{1}$, $\ctxm{d'}{\chi}{1}$, must have the same data subject.
Indeed, because
$$\proveseq{E_{\ctxm{shakey}{}{\cdot}}(\ctxm{id}{}{1})\ctxm{}{\eta}{}}{E_{\ctxm{shakey}{}{\cdot}}(\ctxm{id}{}{1})\ctxm{}{\chi}{}}{\ctxm{id}{\eta}{1}}{\ctxm{id}{\chi}{1}},$$
rule 3 from Definition~\ref{def:associability} gives $\ctxm{id}{\eta}{1}\leftrightarrow_a\ctxm{id}{\chi}{1}$; by rule 2 and transitivity, $\linkable{\ctxm{d}{\eta}{1}}{\ctxm{d'}{\chi}{1}}{a}$.
\qed
\end{example}

We can now define the view corresponding to a knowledge base:

\begin{definition}\label{def:view-of-kb}
 Let $\mathcal{C}$ a knowledge base.
 The \emph{view $V$ corresponding to $\mathcal{C}$} is the view $V=(\mathsf{O},\link{})$, where $\mathsf{O}=\{\mathsf{p}\in\mathsf{I}^c\cup\mathsf{D}^c~|~\mathcal{C}\vdash\mathsf{p}\}\cup(\mathcal{C}\cap\mathsf{E}^c)$, and $\link{}$ is as in Definition~\ref{def:associability}.
\end{definition}

\begin{example}
We determine the view $V_{cli}=(\mathsf{O}_{cli},\link{cli})$ corresponding to the knowledge base $\mathcal{C}_{cli}$ from Example~\ref{exa:kb}.
First, let us consider the view of the client on Alice and Bob.
On Alice, we have $\ctxm{id}{\pi}{su}\in \mathsf{O}_{cli}$ because $\mathcal{C}_{cli}\vdash\ctxm{id}{\pi}{su}$, as shown in Example~\ref{exa:key-guessing}.
Similarly, $\ctxm{ds}{ab}{4},\ctxm{id}{ab}{4},\ctxm{age}{\pi}{su}\in\mathsf{O}_{cli}$, and as we saw in Example~\ref{exa:associability}, they are mutually associable.
On Bob, the two items $\ctxm{ds}{ab}{12}$ and $\ctxm{teln}{ab}{12}$ are detectable and associable.
In fact, the client's view on Alice and Bob is as in Figure~\ref{fig:exa-view}.

Apart from this, the client's view also contains knowledge about the client and server.
Namely, in both cases, it contains two associable {\objl}-layer representations of the communication address:  $\ctxm{ip}{\cdot}{me},\ctxm{ip}{\pi}{cl}\in\mathsf{O}_{cli}$ on the client, and $\ctxm{ip}{\cdot}{srv},\ctxm{ip}{\pi}{srv}\in\mathsf{O}_{cli}$ on the server.
\qed
\end{example}

\subsection{Deciding Detectability and Linkability}\label{subsec:prolog}

\begin{figure}[tb]
\parbox[b]{5in}{
\small
 \begin{algorithmic}[1]
\newcommand{\LINEIF}[2]{\STATE\algorithmicif\ {#1}\ \algorithmicthen\ {#2} \algorithmicend\ \algorithmicif}
\STATE \COMMENT{Let $\vDash$ denote the deductive system without the content analysis rule}
\FORALL{context items $\mathsf{m}'$: $\mathsf{m}'\doteq\mathsf{m}$, $\mathcal{C}_a\vDash\mathsf{m}'$}
  \FORALL{context items $\mathsf{p},\mathsf{p}'$: $\mathsf{m}@z=\mathsf{p}$, $\mathsf{m}'@z=\mathsf{p}'$, $\mathsf{p}\ne\mathsf{p}'$}
    \STATE \COMMENT{Find sequence of evidence for $\mathsf{p}\doteq\mathsf{p}'$ using breadth-first search}\\
    \STATE $Q\gets\{\mathsf{p}\}$ \COMMENT{queue of items to check}; $P\gets\{\}$ \COMMENT{already checked}; $\textrm{found}\gets\FALSE$
    \WHILE{$Q\ne\{\} \wedge \neg\textrm{found}$}
      \STATE $\mathsf{q}\gets \textrm{pop}(Q)$; $P\gets P\cup\{q\}$ \COMMENT{move $\mathsf{q}$ from queue to already checked}
      \LINEIF{$\mathsf{q}=\mathsf{p}'$}{$\textrm{found}\gets\TRUE; \textbf{break}$ \COMMENT{evidence for $\mathsf{p}\doteq\mathsf{p}'$ found}}
      \FORALL{context items $\mathsf{q}'$: $\mathsf{q}'$ occurs in message in $\mathcal{C}_a$, $\mathsf{q}'\doteq\mathsf{q}$, $\mathsf{q}'\notin P\cup Q$}
        \STATE \COMMENT{Try to find evidence for $\mathsf{q}\doteq\mathsf{q}'$}
        \FORALL{context items $\mathsf{n}$: $\mathcal{C}_a\vDash\mathsf{n}$, $\mathsf{n}$ is minimal w.r.t. $\mathsf{q}$}
          \LINEIF{$\exists\mathsf{n}':\mathcal{C}_a\vDash \mathsf{n}': \proveseq{\mathsf{n}}{\mathsf{n}'}{\mathsf{q}}{\mathsf{q}'}$}{$Q\gets Q\cup\{\mathsf{q}'\}$}
        \ENDFOR
        \FORALL{context items $\mathsf{n}'$: $\mathcal{C}_a\vDash\mathsf{n}'$, $\mathsf{n}'$ is minimal w.r.t. $\mathsf{q}'$}
          \LINEIF{$\exists\mathsf{n}:\mathcal{C}_a\vDash \mathsf{n}: \proveseq{\mathsf{n}}{\mathsf{n}'}{\mathsf{q}}{\mathsf{q}'}$}{$Q\gets Q\cup\{\mathsf{q}'\}$}
        \ENDFOR
      \ENDFOR
    \ENDWHILE
    \LINEIF{$\neg\textrm{found}$}{$\textbf{break}$ \COMMENT{No such $\mathsf{p}'$ found: try next $\mathsf{m}'$}}
  \ENDFOR
  \RETURN \TRUE \COMMENT{Actor has evidence that $\mathsf{m}\doteq\mathsf{m}'$ for a $\mathsf{m}'$ such that $\mathcal{C}_a\vDash\mathsf{m}'$}
\ENDFOR
\RETURN \FALSE \COMMENT{For all $\mathsf{m}'$ such that $\mathsf{m}\doteq\mathsf{m}'$, $\mathcal{C}_a\vDash\mathsf{m}'$: actor has no evidence for $\mathsf{m}\doteq\mathsf{m}'$}
\end{algorithmic}
}
\caption{\label{fig:pseudocode-ca}Algorithm implementing the deductive system: given knowledge base $\mathcal{C}_a$ and context message $\mathsf{m}$, check whether $\mathcal{C}_a\vdash\mathsf{m}$}
\end{figure}
In this section, we present the algorithms to decide detectability and linkability used in our tool.
Our tool consists of a series of Prolog scripts\footnote{The implementation, along with its documentation, can be downloaded at \url{http://www.mobiman.me/publications/downloads/}.} for the automatic verification of privacy requirements for a set of architectures.
The most technically challenging part of this task is to compute the views of actors (i.e., the sets of detectable items and associability relations) from their knowledge bases.
Here, we describe our algorithms and their efficiency in general terms; for details, refer to the documentation of the implementation.

Our deductive system is essentially a traditional deductive system~\cite{Clarke1998Usingstatespace,Fiore2001ComputingSymbolicModels} to which testing rules and the content analysis rule have been added.
Let us first ignore content analysis, and only consider the construction, testing and elimination rules.
Construction rules generally derive messages from submessages; testing and elimination rules derive submessages from messages using some ``additional prerequisites'' (e.g., the key for the decryption rule $\avdashc{EE}$).
As testing/elimination and construction cancel each other out, there is no point in applying testing/elimination to the result of construction rule.
Thus, to check the derivability of a message $\mathsf{m}$, we try to find a message $\mathsf{n}$ in which it occurs as submessage, and try to derive $\mathsf{m}$ from it using elimination and testing.
If this does not work, we repeat the procedure for $\mathsf{m}$'s submessages: if successful, then $\mathsf{m}$ can be obtained from them with a construction rule.

While trying elimination or testing rules, we need to check the derivability of the additional prerequisites $\mathsf{n}$.
We claim that this check can be done at the contents layer (so a simple deductive system suffices).
For the testing rule this is clear; however, it also holds for elimination rules because their additional prerequisites can always be obtained from a content equivalent message using the testing rule.

Thus, in terms of evaluation, our deductive system differs from standard systems in two ways.
First, for elimination rules, the additional prerequisites are evaluated not using the deductive system itself, but using a (standard) deductive system at the contents layer.
Second, testing rules are added which are evaluated in the same way as elimination rules.
Intuitively, our deductive system is thus not much harder to evaluate than a corresponding standard deductive system.
(However, typically it will be run on a larger message set because information has multiple representations.)

We now turn our implementation of the deductive system without content analysis into an implementation of the full deductive system.
Note that any deduction in the full deductive system can be transformed into a deduction deriving the same message satisfying the following conditions:
\begin{itemize}
 \item After content analysis rules, no other rules are applied to a message
 \item In any application of $\avdashc{C}$, the message $\mathsf{n}_2$ and the message $\mathsf{n}_1$ from which it is derived only differ by one context item at one position
 \item In any application of $\avdashc{C}$, the messages $\mathsf{m}_1$ and $\mathsf{m}_2$ are derived without content analysis; also, $\mathsf{m}_1$ is minimal with respect to $\mathsf{n}_1$ in the sense that no elimination or testing rule can be applied to it to obtain a submessage containing $\mathsf{n}_1$; and/or $\mathsf{n}_2$ is minimal with respect to $\mathsf{m}_2$.
\end{itemize}
The algorithm in Figure~\ref{fig:pseudocode-ca} is an imperative translation of our Prolog implementation; by the above properties, it implements derivability in our full deductive system.
Namely, to derive $\mathsf{m}$ from a given knowledge base $\mathcal{C}_a$, it takes all messages $\mathsf{m}'\doteq\mathsf{m}$ such that $\mathcal{C}_a\vdash\mathsf{m}'$, and tries to obtain $\mathsf{m}'$ from $\mathsf{m}$ by content analysis in a context-item-by-context-item fashion.
For all positions $z$ at which $\mathsf{m}$ and $\mathsf{m}'$ differ, the algorithm performs a breadth-first search for messages obtained from $\mathsf{m}$ by content analysis at position $z$, until it finds $\mathsf{m}$ with $\mathsf{m}@z$ replaced by $\mathsf{m}'@z$.
The breadth-first search is performed by first searching for a minimal message using testing and elimination rules (lines 10 and 13); and then searching for a content equivalent message using  testing, elimination and construction rules (lines 11 and 14).
We did not optimise this algorithm in terms of complexity.
Indeed, in practice, most context items are content equivalent only to few other items, so the search space for the algorithm is very limited.

The algorithm for checking the associability of two contexts is similar to the previous algorithm.
In particular, it starts with one context $(\eta,k)$ and uses breadth-first search to find associable contexts.
This involves finding all identifiers and entities that occur in $(\eta,k)$ and all other contexts in which that identifier/entity occurs.
The algorithm then searches evidence for content equivalence of the different representations of the identifier/entity.

\section{States, Traces, and System Evolution}\label{sec:traces}

In this section, we complete our formal framework for the analysis of data minimisation by modelling communication in an information system.
In Section~\ref{sec:determining-views}, we showed how to determine what knowledge of personal information actors have given their knowledge bases.
In this section, we show how these knowledge bases, collected in a \emph{state}, can be derived from a model of exchanged messages given as \emph{traces}.
(The approach of our framework is to model messages based on protocol descriptions.
In an alternative type of analysis, the knowledge base of an actor could be derived from communication logs and then analysed using the methods presented in Section~\ref{sec:determining-views}.)

A state collects the knowledge of all actors in an information system at a certain point in time.
Each actor has his own knowledge base.
The knowledge about personal information by an actor, captured by his view, follows from his knowledge base.
The knowledge of coalitions of actors follows from the union of their respective knowledge bases:
\begin{definition}
 Let $\mathcal{A}$ be a set of actors, and $I$ an information model.
\begin{itemize}
 \item A \emph{state of $I$ involving $\mathcal{A}$} is a collection $\{\mathcal{C}_x\}_{x\in\mathcal{A}}$ of knowledge bases about $I$.
 \item The \emph{view of actor $a\in\mathcal{A}$ in state $\{\mathcal{C}_x\}_{x\in\mathcal{A}}$} is the view corresponding to knowledge base $\mathcal{C}_a$ (Definition~\ref{def:view-of-kb}).
 \item The \emph{view of coalition $\{a_1,\ldots,a_k\}\subset\mathcal{A}$ of actors in state $\{\mathcal{C}_x\}_{x\in\mathcal{A}}$} is the view corresponding to knowledge base $\mathcal{C}_{a_1}\cup\ldots\cup\mathcal{C}_{a_k}$.
\end{itemize}

\end{definition}
We assume that information model $I$ is fixed. 
That is, changes in knowledge during the system evolution are modelled by different states of the same information model $I$.

A \emph{trace} is a series of communication steps.
Each communication step is modelled by a \emph{message transmission} involving two parties that both use a particular communication address modelled as a context identifier.
We consider three types of message transmissions.
The simplest type $(1)$ captures an actor using address $\mathsf{a}$ to send a message $\mathsf{m}$ to another actor using address $\mathsf{b}$.
Two other types model the execution of cryptographic protocols:
type (2) denotes a zero-knowledge proof with prover using address $\mathsf{a}$ and verifier using address $\mathsf{b}$; type (3) denotes a credential issuing protocol with user $\mathsf{a}$ and issuer $\mathsf{b}$.
\begin{definition}
 A \emph{message transmission} is of one of the following three types:
$$
(1)~\mathsf{a}\to\mathsf{b}:\mathsf{m};~~
(2)~\transmission{\mathsf{a}}{\mathsf{b}}{\textrm{ZK}(\mathsf{m}_1;\mathsf{m}_2;\mathsf{m}_3;\mathsf{m}_4)};$$ $$
(3)~\transmission{\mathsf{a}}{\mathsf{b}}{\textrm{ICred}^{\mathsf{m}_1}_{\mathsf{m}_2}(\mathsf{m}_3;\mathsf{m}_4)},$$
  with $\mathsf{a},\mathsf{b}$ context identifiers, and $\mathsf{m}_i$ context messages.
\end{definition}
\begin{definition}
A \emph{trace} $\mathfrak{T}$ is a sequence $\mathfrak{t}_1;\cdots;\mathfrak{t}_k$ of message transmissions.
\end{definition}
States \emph{evolve} by traces so that the actors involved learn the messages exchanged:
\begin{definition}
An \emph{evolution} from state $\{\mathcal{C}^0_x\}_{x\in\mathcal{A}}$ into state $\{\mathcal{C}^k_x\}_{x\in\mathcal{A}}$ by trace $\mathfrak{t}_1;\cdots;\mathfrak{t}_k$ is a series of steps (let $\mathfrak{t}_i=\mathsf{a}_i\to\mathsf{b}_i:\mathsf{m}_i$ or $\mathfrak{t}_i=\transmission{\mathsf{a}_i}{\mathsf{b}_i}{\mathsf{m}_i}$):
$$\{\mathcal{C}^0_x\}_{x\in\mathcal{A}} \stackrel{\mathfrak{t}_1}{\rightarrow} \{\mathcal{C}^1_x\}_{x\in\mathcal{A}} \stackrel{\mathfrak{t}_2}{\rightarrow} \cdots \stackrel{\mathfrak{t}_n}{\rightarrow} \{\mathcal{C}^n_x\}_{x\in\mathcal{A}},$$
 where for every actor $z\in\mathcal{A}$, $\mathcal{C}^i_z=\mathcal{C}^{i-1}_z\cup\{\mathsf{a}_i,\mathsf{b}_i,\mathsf{m}_i\}$ if $z\leftrightarrow \sigma(\mathsf{a}_i)$ or $z\leftrightarrow \sigma(\mathsf{b}_i)$, and $\mathcal{C}^i_z=\mathcal{C}^{i-1}_z$ otherwise.
\end{definition}

The following example demonstrates traces, states, and message transmissions.
\begin{example}\label{exa:trace}
Consider again the PI model from Example~\ref{exa:pi-model}, extended into an information model in Example~\ref{exa:kb}.
We model a complete system evolution as a trace executed from an initial state.

We are interested in the knowledge of two actors $\mathcal{A}=\{cli,srv\}$: the client and server.
The initial state $\{\mathcal{C}^0_x\}_{x\in\mathcal{A}}$ consists of initial knowledge of the client and server.
As discussed before, this needs to include all used communication addresses and keys.
As in Example~\ref{exa:kb}, for the client we take:
\begin{align*}
 \mathcal{C}^0_{cli}=\{&\ctxm{ds}{ab}{12}, \ctxm{teln}{ab}{12}, \ctxm{ds}{ab}{4}, \ctxm{id}{ab}{4}, \ctxm{skey}{\cdot}{\cdot}, \ctxm{ip}{\cdot}{me},\\& \ctxm{ip}{\cdot}{srv}, \mathsf{pk}(\ctxm{k^-}{\cdot}{srv})\}.
\end{align*}
Similarly, for the server we define:
\begin{align*}
 \mathcal{C}^0_{srv}=\{\ctxm{key}{db}{1},\ctxm{col1}{db}{1},\ctxm{col1}{db}{2},\ctxm{key}{db}{2}, \ctxm{n}{\cdot}{\cdot}, \ctxm{skey}{\cdot}{\cdot}, \ctxm{ip}{\cdot}{srv}, \ctxm{k^-}{\cdot}{srv}\}.
\end{align*}
($\ctxm{n}{\cdot}{\cdot}$ is the nonce from the server's reply.) The communication described in Example~\ref{exa:kb} is now formalised by trace $\mathfrak{t}$ consisting of the following message transmissions:
\begin{align*}
&\ctxm{ip}{\cdot}{cli}\to \ctxm{ip}{\cdot}{srv}:~E'_{\ctxm{shkey}{}{\cdot}}(\ctxm{id}{}{su})\ctxm{}{\pi}{}; \\
&\ctxm{ip}{\cdot}{srv}\to \ctxm{ip}{\cdot}{cli}:~E'_{\ctxm{shkey}{}{\cdot}}(\{\ctxm{age}{}{su},\ctxm{n}{}{\cdot},S_{\ctxm{k^-}{}{srv}}(\{\ctxm{age}{}{su},\ctxm{n}{}{\cdot}\})\})\ctxm{}{\pi}{}. 
\end{align*}
Then, state $\{\mathcal{C}^0_x\}_{x\in\mathcal{A}}$ evolves by $\mathfrak{t}$ into state $\{\mathcal{C}_x\}_{x\in\mathcal{A}}$, where:
\begin{align*}
 \mathcal{C}_{srv}=&\mathcal{C}^0_{srv}\cup \{\ctxm{ip}{\pi}{cli}, \ctxm{ip}{\pi}{srv}, E'_{\ctxm{shkey}{}{\cdot}}(\ctxm{id}{}{su})\ctxm{}{\pi}{},\\& ~~~~~E'_{\ctxm{shkey}{}{\cdot}}(\{\ctxm{age}{}{su},\ctxm{n}{}{\cdot},S_{\ctxm{k^-}{}{srv}}(\{\ctxm{age}{}{su},\ctxm{n}{}{\cdot}\})\})\ctxm{}{\pi}{}\},\\
 \mathcal{C}_{cli}=&\mathcal{C}^0_{cli}\cup \{\ctxm{ip}{\pi}{cli}, \ctxm{ip}{\pi}{srv}, E'_{\ctxm{shkey}{}{\cdot}}(\ctxm{id}{}{su})\ctxm{}{\pi}{},\\& ~~~~~E'_{\ctxm{shkey}{}{\cdot}}(\{\ctxm{age}{}{su},\ctxm{n}{}{\cdot},S_{\ctxm{k^-}{}{srv}}(\{\ctxm{age}{}{su},\ctxm{n}{}{\cdot}\})\})\ctxm{}{\pi}{}\}.
\end{align*}
Note that $\mathcal{C}_{cli}$ is as in Example~\ref{exa:kb}.
The views of $cli$, $srv$ and the coalition $\{cli, srv\}$ about Alice and Bob in this state are as shown in Figure~\ref{fig:exa-view}.
\qed
\end{example}

\section{Case Study: Privacy in Identity Management Systems}\label{sec:case-study}

Having described our privacy comparison framework, we now introduce a case study to demonstrate its operation.
In Sections~\ref{sec:pi-model}--\ref{sec:traces}, we have presented the various formalisms needed to perform the four steps of our privacy comparison framework (Figure~\ref{fig:steps}).
In the case study, we will demonstrate these four steps by comparing the data minimisation characteristics of several identity management (IdM) systems.
In this section, we introduce the case study.
First, we provide an overview of IdM systems ($\S$\ref{subsec:idm-intro}).
Then, we discuss the requirements related to privacy by data minimisation that are relevant for IdM systems ($\S$\ref{subsec:casestudy-reqs}); and present the four IdM systems we analyse ($\S$\ref{subsec:casestudy-systems}).

\subsection{Identity Management Systems}\label{subsec:idm-intro}

As providers of on-line services are offering more and more customisation to their users, they need to collect more and more of their personal information.
Traditionally, each service provider would manage the accounts of users separately.
However, this identity management model, called the \emph{isolated user identity management model}~\cite{JoesangUser-CentricIdentityManagement}, has disadvantages for both users and service providers: the user has to manually provide and update her information and keep authentication tokens for each service provider, whereas it is hard for the service provider to obtain guarantees that the information given by the user is correct.

This problem is commonly addressed using an \emph{Identity Management (IdM) System}.
Intuitively, the task of managing and endorsing identity information is delegated to \emph{identity providers}.
Identity management is split up in two phases: \emph{registration} and \emph{service provision}.
At registration, users establish accounts at (possibly multiple) identity providers.
(This includes \emph{identification}: i.e., the user transfers her attributes to the identity provider, and the identity provider possibly checks them. However, both the transfer and checking of attributes performed by the identity provider are out of scope of this work.)
Service provision is the phase when a user requests a service from a service provider: at this point, user attributes required for the service provision need to be collected and sent to the service provider.

\begin{figure}[tb]
 \centering\includegraphics[scale=0.6]{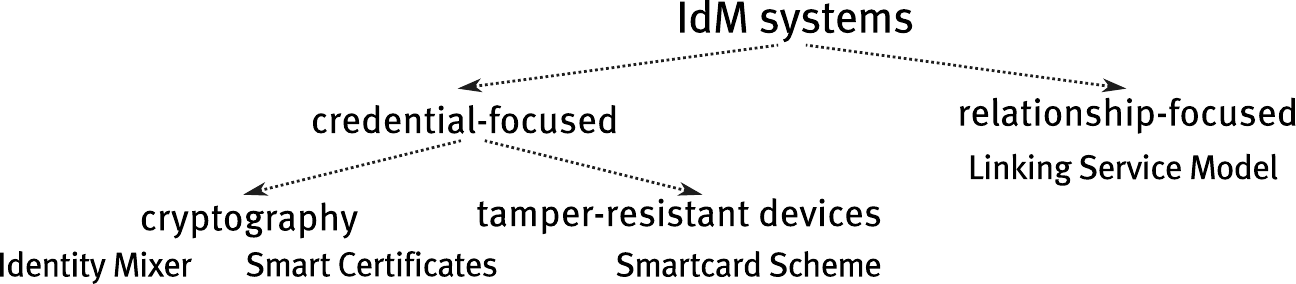}
 \caption{\label{fig:architecture-types}Taxonomy of IdM systems}
\end{figure}
IdM systems can be divided into two main categories~\cite{Bhargav-Spantzel2007Usercentricity:taxonomy} depending on whether or not the identity providers are involved in the service provision phase: \emph{credential-focused} and \emph{relationship-focused} systems (also know as network-based and claim-based systems \cite{Alpar2011IdentityCrisis}).
Figure~\ref{fig:architecture-types} shows a taxonomy of IdM systems.

In credential-focused IdM systems, the user gets long-term credentials from the identity provider in the registration phase that she can directly present to the service providers in the service provision phase.
These credentials contain her identity attributes.
We can distinguish between two mechanisms employed to prevent the user from tampering with them, namely \emph{cryptography} and \emph{tamper-resistant devices}.
Credential-focused systems relying on cryptography include CardSpace \cite{Nanda2007TechnicalReferenceInformation}, U-Prove \cite{Paquin2010U-ProveCTPWhite} and Identity Mixer \cite{Bangerter2004CryptographicFrameworkControlled}.
The system presented in \cite{Vossaert2010User-centricidentitymanagement} relies on the use of a smartcard as a tamper-resistant device.

In relationship-focused IdM systems, in contrast, identity providers present the attributes to service providers.
During the registration phase, identity providers establish shared identifiers to refer to each other's identity of the user.
During the service provision phase, the user authenticates to an identity provider.
The identity provider then sends attributes to the service provider (possibly indirectly via the user).
If needed, the shared identifiers established during registration are used to collect (or \emph{aggregate} \cite{Chadwick2009AttributeAggregationin}) attributes held by other identity providers without the user having to authenticate to them as well.
The combination of reliance on authentication performed by another party and exchange of identity information is sometimes referred to as \emph{federated identity management} \cite{JoesangUser-CentricIdentityManagement,Smedinghoff2009FederatedIdentityManagement}.
(Note that this term is also used to describe the general concept of sharing information between different domains \cite{Alpar2011IdentityCrisis} or the mere use of multiple identity providers \cite{EU2003IdentityManagementSystems}. To avoid confusion, we will not use it further.)
Relationship-focused systems include Liberty Alliance \cite{LibertyIDWSF}, Shibboleth \cite{shibboleth}, and the linking service model \cite{Chadwick2009AttributeAggregationin}.

Because in IdM systems, large amounts of personal information are processed by many different parties, privacy has become a major concern~\cite{Hansen200435,Spiekermann2009EngineeringPrivacy}.
In such systems, privacy threats posed by authorised insiders are nowadays considered to be a critical problem besides outsider attacks on cryptographic protocols~\cite{Fyffe2008Addressinginsiderthreat}.
Insiders may compile comprehensive user profiles to sell or use for secondary purposes such as marketing.
These profiles can include sensitive information that is explicitly transferred by the user, but also information that is transferred \emph{implicitly}~\cite{Spiekermann2009EngineeringPrivacy}.
For instance, the mere fact that a user performed a transaction at a certain service provider may be privacy-sensitive.
In addition, profiles held by different parties may be combined~\cite{Spiekermann2009EngineeringPrivacy} to compile even more comprehensive profiles.
\emph{Privacy-enhancing IdM systems} (e.g.,~\cite{Bangerter2004CryptographicFrameworkControlled,Chadwick2009AttributeAggregationin,Vossaert2010User-centricidentitymanagement}) aim to minimise the amount of information disclosed as well as prevent that different pieces of information can be linked together~\cite{Hansen200435}.

\subsection{Requirements}\label{subsec:casestudy-reqs}

We now present a set of privacy requirements for IdM systems.
We have elicited these requirements by
analysing the information that actors can learn;
considering which of this knowledge should be avoided; and 
systematically grouping this knowledge into requirements according to what kind of knowledge it is, and who should or should not learn it.
We validate our set of requirements in two different ways.
First, we check if they cover relevant privacy requirements discussed in the literature.
For this, we have studied taxonomies of privacy in identity management~\cite{Bhargav-Spantzel2007Usercentricity:taxonomy,Hansen200435} and the proposals for the identity management systems analysed in this paper~\cite{Bangerter2004CryptographicFrameworkControlled,Chadwick2009AttributeAggregationin,Vossaert2010User-centricidentitymanagement}, and verified if all requirements discussed in these works are covered by our requirements.
Second, we check if they cover all possible situations expressible in our model that can lead to privacy risks.
For this, we have systematically considered all elementary detectability, linkability and involvement requirements expressible in our model, checked which of these can lead to privacy risks, and verified that the relevant ones are covered by our requirements.

Table~\ref{tbl:privacy-properties-informal} lists our privacy requirements, also showing in which existing works they are discussed.
We first present our requirements, then discuss if they cover all relevant requirements from the literature mentioned above.
The analysis of coverage of situations expressible in our model is presented in Section~\ref{subsec:formal-requirements}.

\newcolumntype{P}[1]{>{\raggedright}p{#1}}

\begin{table}[tb]
\small
\hspace{-2.5cm}
\begin{tabular}[b]{P{2.2in}p{3.2in}p{0.9in}}
\textbf{Functional requirements}          &\textbf{Description}&\textbf{References}\tabularnewline\hline
Attribute exchange (AX)                    &The service provider learns the value of the required attributes/properties of the user requesting the service.
					   &\cite{Bangerter2004CryptographicFrameworkControlled,Bhargav-Spantzel2007PrivacyRequirementsin,Chadwick2009AttributeAggregationin,Park1999SmartCertificates:Extending,Vossaert2010User-centricidentitymanagement}
					   \tabularnewline\tabularnewline
\textbf{Privacy requirements}              &\textbf{Description}\tabularnewline\hline
Irrelevant attribute undetectability (SID) &The service provider does not learn anything about attribute values irrelevant to the transaction.
					   &\cite{Bangerter2004CryptographicFrameworkControlled,Bhargav-Spantzel2007PrivacyRequirementsin,Park1999SmartCertificates:Extending,Vossaert2010User-centricidentitymanagement}
					   \tabularnewline
Property-attribute undetectability (SPD)   &The service provider does not learn anything about attributes apart from the properties he needs to know.
					   &\cite{Bangerter2004CryptographicFrameworkControlled,Bhargav-Spantzel2007PrivacyRequirementsin,Park1999SmartCertificates:Extending,Vossaert2010User-centricidentitymanagement}
					   \tabularnewline
IdP attribute undetectability (ID)         &Identity providers do not learn anything about the user's attributes from other identity providers.
					   &-
					   \tabularnewline\hline
Mutual IdP involvement undetectability (IM)&One identity provider does not learn whether a given user also has an account at another identity provider.
					   &\cite{Chadwick2009AttributeAggregationin}
					   \tabularnewline
IdP-SP involvement undetectability (ISM)   &Identity providers do not learn which service providers a user uses.
					   &-
					   \tabularnewline\hline
Session unlinkability (SL)                 &A service provider cannot link different sessions of the same user.
					   &\cite{Bangerter2004CryptographicFrameworkControlled,Bhargav-Spantzel2007PrivacyRequirementsin,Hansen200435,Chadwick2009AttributeAggregationin,Vossaert2010User-centricidentitymanagement}
					   \tabularnewline
IdP service access unlinkability   (IL)    &Identity providers cannot link service access to the user profile they manage.
					   &\cite{Hansen200435}
					   \tabularnewline
IdP profile unlinkability (IIL)            &Collaborating identity providers cannot link user profiles.
					   &\cite{Hansen200435,Vossaert2010User-centricidentitymanagement}
					   \tabularnewline
IdP-SP unlinkability (ISL)                 &Identity providers and service provider cannot link service accesses to user profile at identity provider.
					   &\cite{Bangerter2004CryptographicFrameworkControlled,Hansen200435,Vossaert2010User-centricidentitymanagement}
					   \tabularnewline\tabularnewline
\textbf{Accountability requirements}       &\textbf{Description}\tabularnewline\hline
Anonymity revocation (AR)                  &Service provider and identity providers (possibly with help from trusted third party) can reconstruct link between service access and user profile. 
					   &\cite{Bangerter2004CryptographicFrameworkControlled,Bhargav-Spantzel2007PrivacyRequirementsin,Hansen200435,Vossaert2010User-centricidentitymanagement}
					   \tabularnewline
\end{tabular}
\caption{Requirements for IdM systems\label{tbl:privacy-properties-informal}}
\end{table}

The basic \emph{functional requirement} for IdM systems is that the service provider learns the attributes it needs \cite{Bhargav-Spantzel2007PrivacyRequirementsin}: \emph{attribute exchange} (AX).
Note that in one service provision, a service provider may need attributes from several identity providers.

\emph{Privacy requirements} cover that certain personal information should not be learned by certain actors.
Privacy by data minimisation attempts to minimise the amount of information learned, and the extent to which it can be linked together \cite{Hansen200435}.
The first aspect, information learned, can be further divided into explicitly and implicitly transferred information \cite{Spiekermann2009EngineeringPrivacy}.
\emph{Detectability} requirements capture explicitly transferred information: information about the user's attributes.
\emph{Involvement} requirements capture information about whether actors know about each other's involvement with the user: a kind of implicitly transferred personal information.
The second aspect is captured in \emph{linkability} requirements: namely, requirements that (combinations of) parties should be able to link personal information from different sessions, databases, etc. as little as possible.

We define three detectability requirements.
The first are about the service provider learning no more than strictly necessary: no attribute that he does not need to know (\emph{irrelevant attribute undetectability}, SID), and no complete attribute value if all he needs to know is whether or not an attribute satisfies a certain property \cite{Bangerter2004CryptographicFrameworkControlled} (\emph{property-attribute undetectability}, SPD).
These properties limit the user profile a service provider can construct.
In addition, IdM systems should guarantee that identity providers do not learn any value or property of attributes stored at other identity identity providers: we call this requirement \emph{IdP attribute undetectability} (ID).

Involvement requirements address the fact that the mere interaction of a user with certain identity or service providers implies a business relation which can be privacy-sensitive.
For instance, ownership of credentials can be sensitive~\cite{Seamons2003ProtectingPrivacyduring} in domains such as healthcare, insurance, or finance.
In addition, even if individual credentials are not sensitive, the precise combination of credentials held by a user may help identify her.
It is natural in identity management that the service provider learns which identity providers certify the user's attributes: this allows him to judge their correctness.
However, one can aim to achieve that identity providers do not know the identity of other identity providers the user has an account at~\cite{Chadwick2009AttributeAggregationin}: we define this as \emph{mutual IdP involvement undetectability} (IM).
In the same way, a user might want to keep hidden from her identity providers the fact that she interacts with a certain service provider: we call this requirement \emph{IdP-SP involvement undetectability} (ISM).

Linkability is another fundamental privacy concern because it determines what user profiles can be constructed from the data that is collected \cite{Pfitzmann2009terminologytalkingabout}.
To prevent a service provider from  accumulating (behavioural) information, an IdM system should ensure it cannot link different service provisions to the same user: \emph{session unlinkability} (SL).
Indeed, in many cases the service provider does not need to know the identity of the user: for instance, if a user wishes to read an on-line article, the only information that is required is that she has a valid subscription.

Another concern is that parties can build more comprehensive user profiles by sharing their personal information.
To prevent this, they should not know which profiles are about the same user~\cite{Hansen200435}.
A very strong privacy guarantee in this vein is that identity providers and service providers cannot link service provisions to the user: \emph{IdP-SP unlinkability} (ISL).
\emph{IdP profile unlinkability} (IIL) is a weaker privacy guarantee requiring that two collaborating identity providers (without help from the service provider) cannot link their profiles.
\emph{IdP service access unlinkability} (IL) is about the link between a service provision and the user profile at an identity provider, thus measuring whether identity providers are aware of individual service provisions.

An \emph{accountability requirement} counterbalances the privacy guaranteed by the ISL requirement.
Namely, if the user misbehaves, it should be possible to identify her~\cite{Bangerter2004CryptographicFrameworkControlled}.
Several IdM systems \cite{Bangerter2004CryptographicFrameworkControlled,Vossaert2010User-centricidentitymanagement} introduce a trusted third party that, in such cases, can help with the identification.
The \emph{anonymity revocation} (AR) requirement states that, possibly with the help of this trusted third party, the service provider and identity providers are able to revoke the anonymity of a transaction.
(Note that in particular, AR also holds if the service provider and identity providers can revoke anonymity without needing the trusted third party.)

When comparing our requirements to those found in existing taxonomies~\cite{Bhargav-Spantzel2007PrivacyRequirementsin,Hansen200435}, we find that our requirements are generally more detailed.
In \cite{Bhargav-Spantzel2007PrivacyRequirementsin}, three requirement on data minimisation are presented: conditional release, selective disclosure, and unlinkability.
These three requirements correspond to anonymity revocation and IdP-SP unlinkability; irrelevant attribute and property-attribute undetectability; and session unlinkability, respectively (for selective disclosure, the authors do not distinguish between attributes and properties).
The authors also mention policy support, which we do not cover.
On the other hand, our other requirements are not addressed.
In \cite{Hansen200435}, ``user-controlled linkage of personal data'' is mentioned as the underlying principle of privacy-enhancing identity management.
This includes our unlinkability properties (although \cite{Hansen200435} does not identify them separately), but also a ``control'' aspect of privacy which we do not cover.
The authors of~\cite{Hansen200435} also stress that the desired degree of linkability depends on the application, mentioning revocation in particular.

As shown in the table, many of our requirements are discussed by designers of IdM systems~\cite{Bangerter2004CryptographicFrameworkControlled,Chadwick2009AttributeAggregationin,Vossaert2010User-centricidentitymanagement}.
We compare our requirements to those claimed by designers (including the ones we do not cover) when discussing these systems in Section~\ref{sec:architectures}.

\subsection{Four Systems}\label{sec:architectures}\label{subsec:casestudy-systems}

We now present the four IdM systems we formally analyse.
We consider one traditional system, \emph{smart certificates}~\cite{Park1999SmartCertificates:Extending}, for whose development privacy was not a primary concern; it can be classified as credential-focused and relying on cryptography.
We then consider three systems designed with privacy in mind: the \emph{linking service model}~\cite{Chadwick2009AttributeAggregationin}, a relationship-focused IdM system; \emph{Identity Mixer}~\cite{Bangerter2004CryptographicFrameworkControlled}, a credential-focused system relying on cryptographic protocols; and a credential-focused IdM system based on smartcards ~\cite{Vossaert2010User-centricidentitymanagement} we will refer to as the \emph{Smartcard scheme}.

For our analysis, we aimed to cover differ kinds of IdM systems that exist in the literature.
In particular, this means selecting credential-focused and relationship-focused systems~\cite{Alpar2011IdentityCrisis,Bhargav-Spantzel2007PrivacyRequirementsin}.
For the former type, Identity Mixer has received a lot of attention in the research community.
For the latter type, the linking service is one of few proposals supporting multiple identity providers that takes privacy into account~\cite{Chadwick2009AttributeAggregationin}.
We then also included the smartcard scheme because it is a recent proposal in a completely different direction than the previous two.
Of course, our formal results are about these particular systems; however, when analysing the results, we will also informally discuss to what extent they generalise to similar systems.

We now briefly discuss these systems and the privacy guarantees that they have been designed to provide.

\subsubsection{Smart Certificates}\label{subsub:architectures-smartcert}

Park et al.~\cite{Park1999SmartCertificates:Extending} proposed an IdM system built on top of a Public Key Infrastructure (PKI).
In a PKI, a certificate authority (CA) issues certificates stating that a certain public key belongs to a certain user.
A user authenticates by proving knowledge of the secret key corresponding to this public key.
Identity providers issue certificates that link attributes to the public key certificate.
In our analysis, we consider one particular variant described in \cite{Park1999SmartCertificates:Extending}: the user-pull model with long-lived certificates obtained during registration.

\begin{figure}[tb]
\centering
\subfigure[Registration phase\label{fig:sc-reg}]{\includegraphics[scale=0.55]{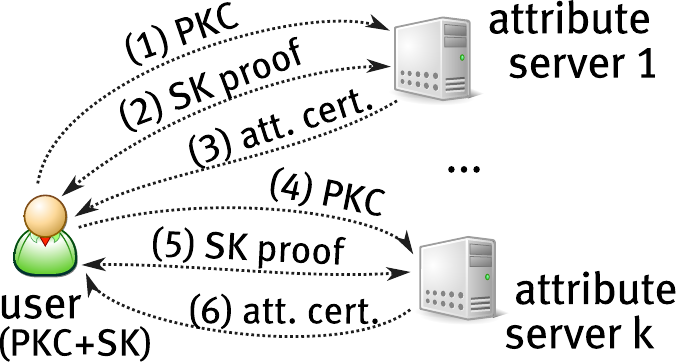}}
\subfigure[Service provision phase\label{fig:sc-sp}]{\raisebox{0.2\height}{~~~~~~~~~~~\includegraphics[scale=0.55]{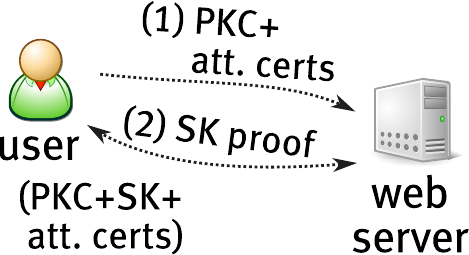}~~~~~~~~~~}}
 \caption{\label{fig:architecture-attcert}Smart certificates}
\end{figure}

The flow of information is summarised in Figure~\ref{fig:architecture-attcert}.
In the registration phase (Figure~\ref{fig:sc-reg}), the user gets an attribute certificate from an identity provider (the ``attribute server'' in \cite{Park1999SmartCertificates:Extending}), which enables her to present her attributes to others.
This involves three steps: (1) the user presents her public key certificate; (2) she proves that she also knows the corresponding secret key (this is an interactive protocol shown as a two-sided arrow in the figure); and (3) the attribute server issues an attribute certificate.
The process is then repeated with other identity providers (steps (4) to (6)).
The attributes in the certificate are signed using the attribute server's secret key and hence cannot be tampered with by the user.
During service provision (Figure~\ref{fig:sc-sp}), the user exchanges attributes with the service provider (``web server'') in two steps: (1) she presents her public key certificate and the attribute certificates containing the attributes needed; and (2) she proves knowledge of the corresponding secret key.

The system presented in \cite{Park1999SmartCertificates:Extending} is mainly designed to satisfy the attribute exchange requirement (AX) in a secure way (``the attributes of individual users are provided securely'').
Privacy concerns are addressed in an extension of the system in which some attributes in a credential are encrypted in such a way that they can only be read by an ``appropriate'' server, corresponding to our SID/SPD properties.
However, we will consider the original scheme in which SID/SPD are not claimed to hold.

\subsubsection{Linking Service Model}\label{subsec:architectures-ls}

\begin{figure}[tb]
\centering
\subfigure[Registration phase\label{fig:tas3-reg}]{\includegraphics[scale=0.55]{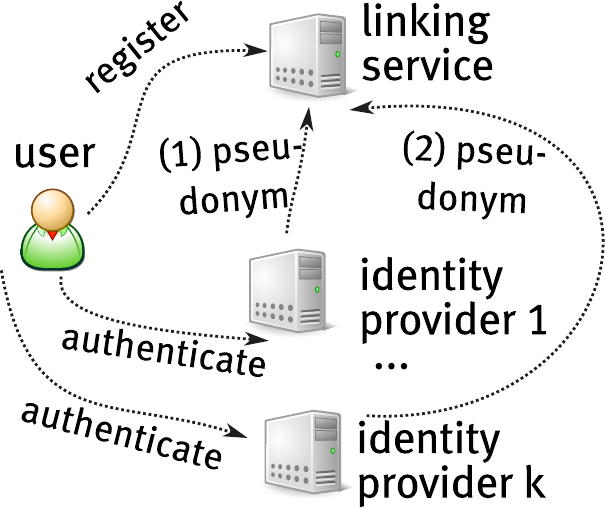}}~~~
\subfigure[Service provision phase\label{fig:tas3-sp}]{\includegraphics[scale=0.55]{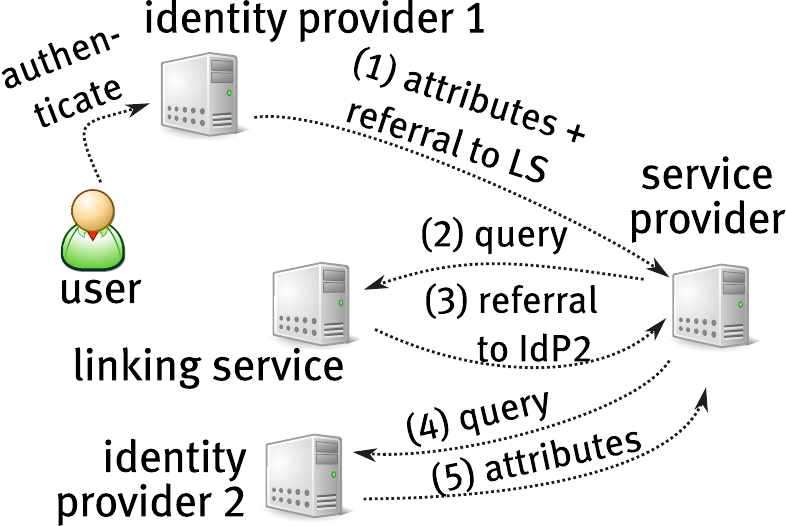}}
 \caption{\label{fig:architecture-tas3}Linking service model}
\end{figure}

The linking service model~\cite{Chadwick2009AttributeAggregationin} is a relationship-focused IdM system.
Its main goal is to facilitate the collection of user attributes from different identity providers in a privacy-friendly way without the user having to authenticate to each identity provider separately.
To this end, this model includes a \emph{linking service} which is responsible for holding the links between profiles of the user at the different identity providers without knowing any personal information about the user.

The flow of information is summarised in Figure~\ref{fig:architecture-tas3}.
During registration (Figure~\ref{fig:tas3-reg}), the user first creates an anonymous account at the linking service LS.
LS requests the identity providers to authenticate the user; each identity provider generates a pseudonym for the user and sends it to LS (steps (1) and (2)).
(The specific method of authentication between the user and the identity providers and linking service is out of our scope.)
In the service provision phase (Figure~\ref{fig:tas3-sp}), the user authenticates to one particular identity provider IdP$_1$.
IdP$_1$ provides the service provider SP with an ``authentication assertion'' containing the attributes requested from it, and a referral to LS (1).
The referral is an encryption of the pseudonym shared between IdP$_1$ and LS that only LS can decrypt.
SP sends this referral to LS (2), which responds by sending a similar referral to other identity providers (3).
Finally, SP requests (4) and obtains (5) the required attributes from the other identity providers (for simplicity, we just show one other identity provider in the figure).

The linking service model aims to satisfy the attribute exchange requirement (AX) as well as a number of privacy requirements~\cite{Chadwick2009AttributeAggregationin}.
In particular, the main goal of the linking service model is to guarantee that identity providers do not know the involvement of other identity providers (IM).
Moreover, the model aims to achieve session unlinkability (SL) through the use of random user identifiers.
Finally, the linking service should not learn the partial identities of the user for the service providers; that is, it does not learn any personal information about the user.
We call this requirement \emph{LS attribute undetectability} (LD); it is not listed in Table~\ref{tbl:privacy-properties-informal} because it is only relevant for this system; however, our analysis will include the verification of this requirement.

\subsubsection{Identity Mixer}\label{subsec:architectures-idemix}

\begin{figure}[tb]
\centering
\subfigure[Registration phase\label{fig:idemix-reg}]{\includegraphics[scale=0.55]{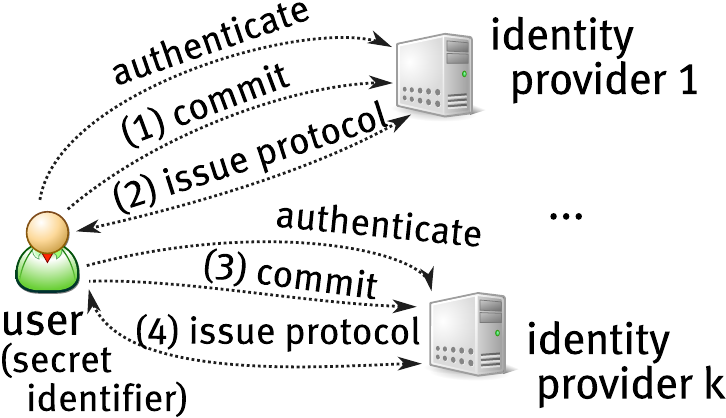}}~~~
\subfigure[Service provision phase\label{fig:idemix-sp}]{\raisebox{0.2\height}{\includegraphics[scale=0.55]{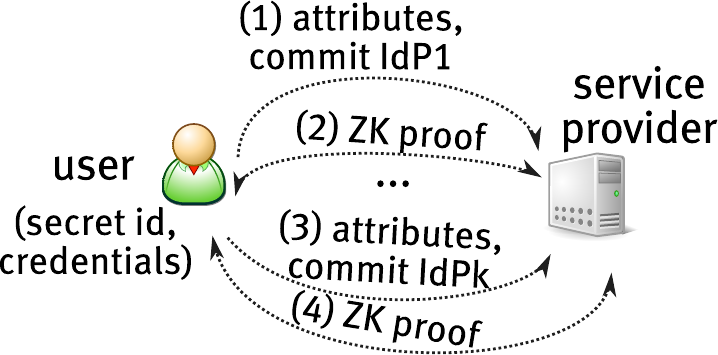}}}
 \caption{\label{fig:architecture-idemix}Identity Mixer}
\end{figure}
Identity Mixer~\cite{Bangerter2004CryptographicFrameworkControlled} is a credential-focused IdM system using a cryptographic primitive called anonymous credentials.
These credentials link attributes to a user identifier, but are issued by identity providers and shown to service providers using protocols ensuring that neither party learns that identifier.
Thus, nobody but the user knows whether different issuing or showing protocols were performed by the same user, while integrity of the attributes is still assured.

Figure~\ref{fig:architecture-idemix} shows the information flows in Identity Mixer.
During registration (Figure~\ref{fig:idemix-reg}), the user first sends a commitment to her (secret) identifier to an identity provider IdP$_1$ (1), after which the user and IdP$_1$ together run the credential issuing protocol (2).
From this, the user obtains a credential with her attributes linked to her secret identifier, without IdP$_1$ learning the identifier.
Communication with other identity providers is analogous (steps (3) and (4)).
In the service provision phase (Figure~\ref{fig:idemix-sp}), the user shows information from several credentials to the service provider SP.
She first shows her credential from one identity provider.
To this end, she sends a message containing the attributes she wants to reveal, and ``commitments'' to the secret identifier and all other attributes (1).
Next, she performs a zero-knowledge proof (2) which proves to SP that the attributes and commitments come from a valid credential issued by the identity provider, while revealing nothing else about the credential.
Credentials issued by other identity providers are shown in the same way (steps (3) and (4)).

Identity Mixer is designed to satisfy a number of privacy requirements~\cite{Bangerter2004CryptographicFrameworkControlled}.
In particular, it aims to satisfy both session unlinkability and IdP/SP unlinkability (together called ``multi-show unlinkability'' in \cite{Bangerter2004CryptographicFrameworkControlled}) and irrelevant attribute and property-attribute undetectability (together called  ``selective show of data items'' in \cite{Bangerter2004CryptographicFrameworkControlled}).
The system allows for providing the service provider with an encryption of some attributes for a trusted third party (``conditional showing of data items'' in \cite{Bangerter2004CryptographicFrameworkControlled}) that can be used for anonymity revocation.
Apart from the data minimisation requirements we defined, the system additionally allows credential issuing where an identity provider copies attributes from another certificate without knowing their values (``blind certification'' in \cite{Bangerter2004CryptographicFrameworkControlled}).
The main motivation for this functionality comes from the use of these certificates for e-cash~\cite{Bangerter2004CryptographicFrameworkControlled}.
In traditional identity management scenarios, such as ours, identity providers should know the attributes they endorse, so we do not consider this requirement in this work.

\subsubsection{Smartcard Scheme}\label{subsec:architectures-smartcard}

\begin{figure}[tb!]
 \centering
\subfigure[Registration phase\label{fig:smartcard-reg}]{\includegraphics[scale=0.55]{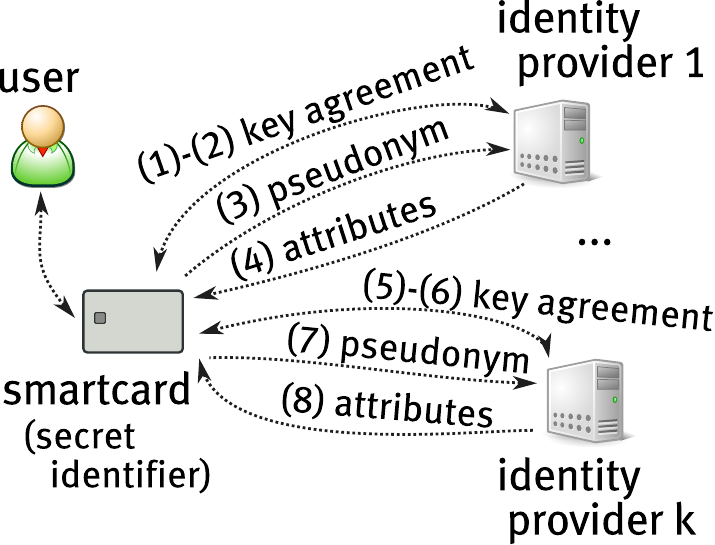}}~~~
\subfigure[Service provision phase\label{fig:smartcard-sp}]{\raisebox{0.05\height}{\includegraphics[scale=0.55]{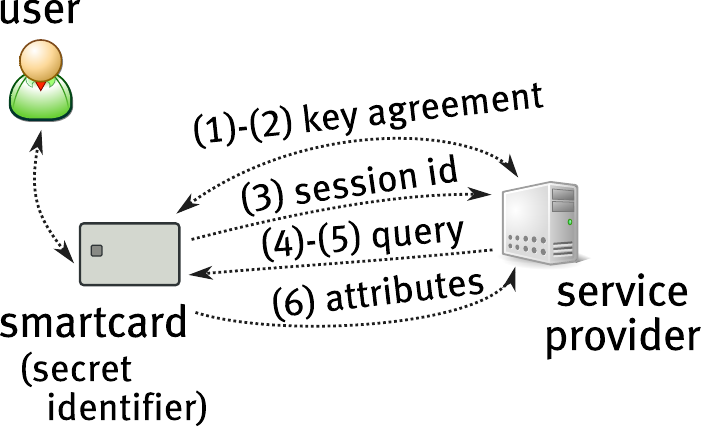}}}
 \caption{\label{fig:architecture-smartcard}Smartcard scheme}
\end{figure}
Vossaert et al~\cite{Vossaert2010User-centricidentitymanagement} proposed a credential-focused IdM system which relies on PKI for authentication and on smartcards (or other tamper-resistant devices) to ensure that attributes are not modified and observed during their transmission from the identity provider to the service provider.
Identity providers and service providers only communicate via the smartcard, and each has a different pseudonym of the user based on a secret user identifier stored on the smartcard.

The information flow defined in the scheme is shown in Figure~\ref{fig:architecture-smartcard}.
In the registration phase (Figure~\ref{fig:smartcard-reg}), the smartcard SC and the first identity provider IdP$_1$ establish a secure, authenticated channel using a key agreement protocol (steps (1) and (2)).
Over this secure channel, SC sends a pseudonym based on its secret identifier specific for IdP$_1$ (3); IdP$_1$ sends its attributes (4).
Registration at other identity providers is similar (steps (5) to (8)).
In a service provision (Figure~\ref{fig:smartcard-sp}), SC and service provider SP establish a secure, authenticated channel as in the registration phase (steps (1) and (2)).
SC generates a random session identifier (3); SP then specifies what attributes he wants, and how long they may have been cached (steps (4) and (5)).
SC responds by giving the requested attributes.
For anonymity revocation purposes, this response also includes Alice's identifier encrypted for the trusted third party (6).

The system is designed to meet several requirements related to the knowledge of personal information~\cite{Vossaert2010User-centricidentitymanagement}.
The requirements specified correspond to our notions of attribute exchange, session unlinkability, and anonymity revocation.
Irrelevant property and property-attribute undetectability follow from their more general notion of ``restricting released personal data''.
The Smartcard scheme also aims to fulfil IdP profile unlinkability and IdP/SP unlinkability by preventing collusion of identity and service providers.

\subsubsection{Privacy Requirements Claimed by Systems}

Table~\ref{tbl:claims} summarises the privacy claims for the systems.
One goal of our formal analysis will be to verify whether these claims actually hold.
In addition, we will analyse the systems against the complete range of identified requirements in order to achieve a comprehensive comparison of their privacy features.

\newcolumntype{C}{>{\centering\arraybackslash}m{0.35cm}<{}}
\begin{table}[tb]
\small
 \centering\begin{tabular}[b]{l|CC|CCC|CC|CCCC}
  \textbf{Scheme}        &$\textbf{AX}$&$\textbf{AR}$&$\textbf{SID}$&$\textbf{SPD}$&$\textbf{ID}$&$\textbf{IM}$&$\textbf{ISM}$&$\textbf{SL}$&$\textbf{IL}$&$\textbf{IIL}$&$\textbf{ISL}$\\\hline
  Smart certificates     &   \tick     &             &              &              &             &             &              &             &             &              &              \\
  Linking service model~~&   \tick     &             &              &              &             &    \tick    &              &  \tick      &             &              &              \\
  Identity Mixer         &   \tick     &     \tick   &    \tick     &    \tick     &             &             &              &  \tick      &             &              &    \tick     \\
  Smartcard scheme       &   \tick     &   \tick     &    \tick     &    \tick     &             &             &              &  \tick      &             &   \tick      &    \tick     \\
\end{tabular}
\caption{Comparison of privacy requirements claimed by the various systems\label{tbl:claims}}
\end{table}

\section{Formal Analysis of the Case Study}\label{sec:formal-case-study}

In this section, we formally analyse and compare the IdM systems presented in the previous section.
To make this comparison, we perform the four steps of our privacy analysis framework (Figure~\ref{fig:steps}).
In Section~\ref{subsec:formal-scenario}, we model the personal information in a scenario (step 1).
In Section~\ref{subsec:formal-requirements}, we model the privacy requirements (step 2; we also discuss whether the requirements identified in Section~\ref{subsec:casestudy-reqs} cover all privacy risks expressible in our model).
In Section~\ref{subsec:formalizing-architectures}, we model the communication in each IdM system (step 3).
In Section~\ref{subsec:analysis-results}, we verify which requirements hold in which system, and analyse the results (step 4).

\subsection{Step 1: Model Personal Information in Scenario}\label{subsec:formal-scenario}

\begin{figure}[tb]
\footnotesize
\begin{center}
\subfigure[Actors/entities\label{fig:casestudy-entities}]{
\begin{tabular}[b]{ll}
$e\in\mathcal{E}$&\textbf{Actor/entity} \\\hline
$al$&Alice \\
$ii$&Address provider \\
$is$&Subscription provider \\
$bs$&E-book store \\
$ttp$&Trusted third party\\
\\
\end{tabular}
}
~
\subfigure[Domains\label{fig:casestudy-domains}]{\begin{tabular}[b]{ll}
\textbf{Dom.}&\textbf{Description}\\\hline
$\cdot$&identifiers/keys\\
$\iota$&Alice's knowledge\\
$\kappa$&$ii$'s user database\\
$\mu$&$is$'s user database\\\hline
$\pi$&registration at $ii$\\
$\eta$&registration at $is$\\
$\zeta$,$\xi$&service provisions
\end{tabular}
}
~
\subfigure[Profiles\label{fig:casestudy-profiles}]{
\begin{tabular}[b]{l|cc}
 &\multicolumn{2}{c}{\textbf{Domains}}\\
 \textbf{Entity}&$\cdot$,$\iota$,$\kappa$,$\mu$&$\pi$,$\eta$,$\kappa$,$\mu$ \\\hline
 $al$ & $al$ & $u$ \\
 $ii$ & $ii$ & $idp1$ \\
 $is$ & $is$ & $idp2$ \\
 $bs$ & $bs$ & $sp$ \\
 $ttp$ & $ttp$ & $ttp$\\
\multicolumn{2}{c}{}\\
\end{tabular}
}
\\
\subfigure[Information about $ii$, $is$, $bs$, $ttp$: anybody knows identifiers and public keys $\mathsf{pk}(\ctxm{k^-}{\cdot}{*})$; actor knows own private key\label{fig:casestudy-nonalice}]{
 ~~~~~~~~~~~~~~~~~~~~~~~~~~~~~~~~~~~~~~~~~~~~~~~~$\{\ctxm{ip}{\cdot}{ii}$, $\ctxm{k^-}{\cdot}{ii}$, $\ctxm{ip}{\cdot}{is}$,
 $\ctxm{k^-}{\cdot}{is}$, $\ctxm{ip}{\cdot}{bs}$, $\ctxm{k^-}{\cdot}{bs}$, $\ctxm{k^-}{\cdot}{ttp}\}$~~~~~~~~~~~~~~~~~~~~~~~~~~~~~~~~~~~~~~~~~~~~~~
}
~~~
\subfigure[Information about Alice: each row is a piece of information (for $\mathbf{d_7}$ and $\mathbf{ip}$: different pieces of information in each domain); columns $al$, $ii$, $is$, and $bs$ show the initial knowledge of actors about the information\label{fig:casestudy-alice}]{
\begin{tabular}{l|llll|l}
\textbf{Info}  & $al$ & $ii$ & $is$  & $bs$ & \textbf{Description}\\\hline
$al$      & $\{\ctxm{al}{\iota}{al}, \ctxm{al}{\zeta}{u}, \ctxm{al}{\xi}{u}\}$ & - & - & - & Entity \\
$i_{ii}$  & $\{\ctxm{i_{ii}}{\iota}{al}, \ctxm{i_{ii}}{\pi}{u}\}$ & $\{\ctxm{i_{ii}}{\kappa}{al}, \ctxm{i_{ii}}{\pi}{u}\}$ & - & - & Identifier at $ii$ \\
$d_1$     & $\ctxm{d_{1}}{\iota}{al}$ & $\ctxm{d_{1}}{\kappa}{al}$ & - & - & City \\
$d_2$     & $\ctxm{d_{2}}{\iota}{al}$ & $\ctxm{d_{2}}{\kappa}{al}$ & - & - & Age \\
$d_2{>}60$& $\ctxm{d_{2}{>}60}{\iota}{al}$ & $\ctxm{d_{2}{>}60}{\kappa}{al}$ & - & - & age ``$>60$'' property \\
$d_3$     & $\ctxm{d_{3}}{\iota}{al}$ & $\ctxm{d_{3}}{\kappa}{al}$ & - & - & Address \\
$i_{is}$  & $\{\ctxm{i_{is}}{\iota}{al}, \ctxm{i_{is}}{\eta}{u}\}$&-&$\{\ctxm{i_{is}}{\mu}{al},\ctxm{i_{is}}{\eta}{u}\}$ & - & Identifier at $is$ \\
$d_5$     & $\ctxm{d_{5}}{\iota}{al}$ & - & $\ctxm{d_{5}}{\mu}{al}$ & - & Subscription date \\
$d_6$     & $\ctxm{d_{6}}{\iota}{al}$ & - & $\ctxm{d_{6}}{\mu}{al}$ & - & Subscription type \\
$\mathbf{d_7}$   & - & - & - & $\{\mathbf{\ctxm{d_{7}}{\zeta}{u}},\mathbf{\ctxm{d_{7}}{\xi}{u}}\}$ & Transaction details \\
$\mathbf{ip}$    & $\{\mathbf{\ctxm{ip}{\pi}{u}},\mathbf{\ctxm{ip}{\eta}{u}},\mathbf{\ctxm{ip}{\zeta}{u}},\mathbf{\ctxm{ip}{\xi}{u}}\}$ & - & - & - & IP address 
\end{tabular}
}

\end{center}
\caption{Schematic representation of PI model and initial actor views\label{fig:casestudy}}
\end{figure}
Step 1 of our analysis method is to model the personal information in a scenario.
The scenario needs to be designed in such a way that all privacy properties to be verified (i.e.,~in this case, the ones in Table~\ref{tbl:privacy-properties-informal}) can be phrased in terms of personal information occurring in the scenario.
Thus, we include attributes that should be disclosed (for AX), should not be disclosed (for SPD), and of which only a property should be disclosed (for SID); and we consider multiple identity providers (for IM, IL, and IIL) and sessions (for SL).
Given these constraints, we design the scenario to look as realistic as possible.

In particular, we consider a scenario with four main actors: a user: Alice, a 65 year-old woman; a service provider: an e-book store; and two identity providers: one for Alice's address (the address provider) and one for Alice's subscription at some society (the subscription provider).
In the \emph{registration phase} of this scenario, Alice creates an account at both identity providers.
The address provider stores three identity attributes of the user: the street, city, and age.
The subscription provider stores two user attributes: date of subscription and subscription type.

In the \emph{service provision phase}, Alice purchases books from the e-book store on two separate occasions.
To this end, she needs to provide her personal information, endorsed by the identity providers, to the e-book store.
The service provider, for statistical purposes, demands to know the city that Alice comes from.
Moreover, the e-store offers a discount to customers that are over 60 years old.
As Alice is 65 years old, she is eligible for the discount.
The e-book store, however, does not necessarily need to learn her exact birth date or age; Alice can just prove that she is over 60 years old.
Moreover, the e-book store does not need to know that the purchases are both made by the same user.
On the other hand, in case of abuse, the service provider does want to be able to link the purchase to Alice's profile at the address provider with the help of a trusted third party.
(Note that the scenario does not cover the separate issue of anonymous payment of the e-book.)

Our formalisation of this scenario as (views on) a PI model is shown schematically in Figure~\ref{fig:casestudy}.
Figure~\ref{fig:casestudy-entities} lists the actors/entities in the system (in this case study, actors and entities coincide).
The trusted third party $ttp$ is included because of the anonymity revocation requirement; however, note that it only occurs in the Identity Mixer and Smartcard schemes.

Figures~\ref{fig:casestudy-domains}~and~\ref{fig:casestudy-profiles} summarise the contexts we use to model different representations of personal information.
Figure~\ref{fig:casestudy-domains} lists all domains.
The ``$\cdot$'' domain contains publicly known identifiers for the identity and service providers, and their private keys.
The $\iota$, $\kappa$, and $\lambda$ domains represent databases of user information held by the respective parties.
The $\pi$, $\eta$, $\zeta$, and $\xi$ domains represent the communication protocols that are executed during the scenario.
For simplicity, all communication related to one service provision is modelled in a single domain.
This expresses that parties involved in service provision without communicating directly (e.g.,~the linking service and IdP$_2$ in the linking service model) are able to link their views of the protocol.
Alternatively, each pair of communication partners could have a separate domain.
Figure~\ref{fig:casestudy-profiles} shows the profiles representing the actors in the different domains.
For instance, in the $\cdot$, $\iota$, $\kappa$ and $\lambda$ domains, Alice represented by the $al$ profile;
in the $\pi$, $\eta$, $\zeta$, and $\xi$ domains, she is represented by $u$.
By naming these profiles differently, we emphasise that actors learn the information not as information about Alice, but as information about ``the purchaser in transaction $x$'', etc.

Figures~\ref{fig:casestudy-nonalice}~and~\ref{fig:casestudy-alice} define the pieces of personal information in the scenario, and the knowledge about them that actors hold in the initial state.
For simplicity, we give an explicit {\objl}-layer representation, and use some notational conventions to implicitly describe the information and contents layers.
Namely, when context items about the same entity using the same variable are denoted in the normal font (e.g. $\ctxm{i_{ii}}{\pi}{u}$ and $\ctxm{i_{ii}}{\kappa}{al}$), they are equivalent; when denoted in boldface (e.g. $\mathbf{\ctxm{ip}{\pi}{u}}$, $\mathbf{\ctxm{ip}{\eta}{u}}$), they are all pairwise non-equivalent.
Items of the form $\ctxm{i}{*}{*}$, $\ctxm{i_*}{*}{*}$, $\ctxm{k^-}{*}{*}$, and $\ctxm{ip}{*}{*}$ (for any $*$) are identifiers; items $\ctxm{d_*}{*}{*}$ are data items; other items are non-personal information.
All representations of a single piece of information use the same variable.
Because this scenario includes only one data subject, all pieces of information have unique contents, i.e., the information and contents layers coincide.
We have one attribute property $\psi_1$ representing if an age is over 60.
At the information layer, $\psi_1(d_2)=d_2{>}60$; at the context layer, $\psi_1(\ctxm{d_2}{\pi}{u})=\ctxm{d_2{>}60}{\pi}{u}$, and similarly for other contexts.

Figure~\ref{fig:casestudy-nonalice} defines the information available about $ii$, $is$, and $bs$.
This information consists of a private key for each of the actors, and an identifier for $ii$, $is$, and $bs$.
All actors know each other's identifiers and the public keys $\mathsf{pk}(\ctxm{k^-}{\cdot}{*})$ corresponding to each private key; each actor also knows his own private key.

Figure~\ref{fig:casestudy-alice} defines the personal information known initially about Alice.
Each row except the last two shows different context-layer representations of one piece of information, indicating which actor initially knows which representation.
For instance, $d_1$ represents a city; Alice knows her city as $\ctxm{d_{1}}{\iota}{al}$ and $ii$ knows it as $\ctxm{d_{1}}{\kappa}{al}$.
We assume that the actual attribute exchange between user and identity provider during registration has already taken place, as shown in the $\kappa$ and $\mu$ domains.
In the last two rows, each context item represents a different piece of information; e.g., the transaction details $\mathbf{\ctxm{d_{7}}{\zeta}{u}},\mathbf{\ctxm{d_{7}}{\xi}{u}}$ of the two service provisions are different.
We assume some initial knowledge about Alice in the $\pi$, $\eta$, $\zeta$ and $\xi$ domains representing protocols.
Knowledge of $\ctxm{i_{ii}}{\pi}{u}$, $\ctxm{i_{is}}{\eta}{u}$ held by Alice and the respective identity providers represents the fact that Alice has authenticated to them.
In the context of the two service provisions, Alice knows that she is the data subject ($\ctxm{al}{\zeta}{u}$, $\ctxm{al}{\xi}{u}$); the service provider knows transaction details $\mathbf{\ctxm{d_{7}}{\zeta}{u}}$, $\mathbf{\ctxm{d_{7}}{\xi}{u}}$.
Alice knows her own IP address $\mathbf{\ctxm{ip}{*}{u}}$, where $*\in\{\pi,\eta,\zeta,\xi\}$; note that it is assumed to change dynamically between sessions.

\subsection{Step 2: Model Privacy Requirements}\label{subsec:formal-requirements}

\begin{table}[tb]
 \small\centering\begin{tabular}[b]{ll}
  \textbf{Requirement}&\textbf{Formalisation}\\\hline
  Attribute exchange (AX)                    & $\ctxm{d_1}{\zeta}{u},\ctxm{d_2{>}60}{\zeta}{u},\ctxm{d_6}{\zeta}{u},\ctxm{d_1}{\xi}{u},\ctxm{d_2{>}60}{\xi}{u},\ctxm{d_6}{\xi}{u}\in \mathsf{O}_{bs}$\\
  Anonymity revocation (AR)                  & $\linkable{\ctxm{*}{\kappa}{al}}{\ctxm{*}{\zeta}{u}}{\{bs,ii,is,ttp\}}\linkable{}{\ctxm{*}{\xi}{u}}{\{bs,ii,is,ttp\}}$ \\\hline
  Irrelevant attribute undetectability (SID) & $\ctxm{d_3}{*}{*}\notin\mathsf{O}_{bs} \wedge \ctxm{d_5}{*}{*}\notin\mathsf{O}_{bs}$ \\
  Property-attribute undetectability (SPD)   & $\ctxm{d_2}{*}{*}\notin\mathsf{O}_{bs}$ \\
  IdP attribute undetectability (ID)         & $\ctxm{d_1}{*}{*}\notin\mathsf{O}_{is} \wedge \ctxm{d_2}{*}{*}\notin\mathsf{O}_{is} \wedge \ctxm{d_3}{*}{*}\notin\mathsf{O}_{is} \wedge$ \\
                                             & ~~~~~$\ctxm{d_2{>}60}{*}{*}\notin\mathsf{O}_{is} \wedge \ctxm{d_5}{*}{*}\notin\mathsf{O}_{ii} \wedge \ctxm{d_6}{*}{*}\notin\mathsf{O}_{ii}$ \\\hline
  Mutual IdP involvement undetectability (IM)& $\neg (\exists p:\linkable{\ctxm{*}{\cdot}{is}}{\ctxm{*}{p}{idp2}}{ii} \wedge \linkable{\ctxm{*}{p}{u}}{\ctxm{*}{\kappa}{al}}{ii}) \wedge$\\
                                             & $  \neg (\exists p:\linkable{\ctxm{*}{\cdot}{ii}}{\ctxm{*}{p}{idp1}}{is} \wedge \linkable{\ctxm{*}{p}{u}}{\ctxm{*}{\mu}{al}}{is})$ \\
  IdP-SP involvement undetectability (ISM)   & $\neg (\exists p:\linkable{\ctxm{*}{\cdot}{bs}}{\ctxm{*}{p}{sp}}{ii} \wedge \linkable{\ctxm{*}{p}{u}}{\ctxm{*}{\kappa}{al}}{ii}) \wedge$ \\
                                             & $\neg (\exists p:\linkable{\ctxm{*}{\cdot}{bs}}{\ctxm{*}{p}{sp}}{is} \wedge \linkable{\ctxm{*}{p}{u}}{\ctxm{*}{\mu}{al}}{is})$ \\\hline
  Session unlinkability (SL)                 & $\ctxm{*}{\zeta}{u}\nleftrightarrow_{bs}\ctxm{*}{\xi}{u}$ \\
  IdP service access undetectability (IL)    & $\ctxm{*}{\kappa}{al}\nleftrightarrow_{ii}\ctxm{*}{\zeta}{u}) \wedge \ctxm{*}{\kappa}{al}\nleftrightarrow_{ii}\ctxm{*}{\xi}{u} \wedge$ \\
                                             & $\ctxm{*}{\mu}{al}\nleftrightarrow_{is}\ctxm{*}{\zeta}{u} \wedge \ctxm{*}{\mu}{al}\nleftrightarrow_{is}\ctxm{*}{\xi}{u}$ \\
  IdP profile unlinkability (IIL)            & $\ctxm{*}{\kappa}{al} \nleftrightarrow_{\{ii,is\}} \ctxm{*}{\mu}{al}$ \\
  IdP/SP unlinkability (ISL)                 & $\ctxm{*}{\kappa}{al} \nleftrightarrow_{\{ii,is,bs\}} \ctxm{*}{\zeta}{u} \wedge \ctxm{*}{\mu}{al} \nleftrightarrow_{\{ii,is,bs\}} \ctxm{*}{\zeta}{u} \wedge $\\
                                             & $\ctxm{*}{\kappa}{al} \nleftrightarrow_{\{ii,is,bs\}} \ctxm{*}{\xi}{u} \wedge \ctxm{*}{\mu}{al} \nleftrightarrow_{\{ii,is,bs\}} \ctxm{*}{\xi}{u}$
 \end{tabular}
 \caption{Formalisation of requirements in our scenario ($\mathsf{m}\nleftrightarrow_a\mathsf{n}$ means $\neg(\linkable{\mathsf{m}}{\mathsf{n}}{a})$; $*$~means for all possible values)\label{tbl:formalization}}
\end{table}
Step~2 of our framework is to formalise the requirements from Table~\ref{tbl:privacy-properties-informal} in terms of actor views.
As above, the view of an actor $a\in\mathcal{A}$ and a coalition $A\subset\mathcal{A}$ are denoted $V_a=(\mathsf{O}_a,\leftrightarrow_a)$ and $V_A=(\mathsf{O}_A,\leftrightarrow_A)$, respectively.
The formalisation of our requirements in terms of these views is shown in Table~\ref{tbl:formalization}.
AX and AR are detectability and linkability requirements (see Section~\ref{subsec:model-reqs}), respectively.
(For AX, note that $bs$ can always associate the personal information of the user to the purchase because of the common context $(\zeta,u)$ or $(\xi,u)$, so we do not check this.)
SID, SPD and SID are undetectability requirements; SL, IL, IIL, and ISL are unlinkability requirements.
\mbox{(Un-)}detectability requirements are straightforward to formalise; e.g., property-attribute undetectability means undetectability by $bs$ of the context item $\ctxm{d_2}{\delta}{p}$ in any context $(\delta,p)$.
\mbox{(Un-)}linkability requirements translate to contexts not being associable by an actor or coalition.
IM and ISM are non-involvement requirements: formally, they translate to two associations that should not hold simultaneously; for instance, for IM, there should be no domain $p$ in which $ii$ can link the $idp2$ profile to $\ctxm{}{\cdot}{idp2}$ and the $u$ profile to $\ctxm{}{\kappa}{al}$.

We now analyse whether the above privacy requirements cover all privacy risks expressible in our model.
To this end, we consider all coalitions and all possible knowledge (in terms of elementary detectability, involvement, and linkability aspects; see Section~\ref{subsec:model-reqs}); and verify if they represent a privacy risk, and if so, by which privacy requirement they are captured.
The result is shown in Table~\ref{tbl:coverage} (at the end of this paper). 
The first group of columns indicates the coalition with respect to which a requirement is defined; the next groups list the detectability, involvement, and linkability aspects that it entails.

First consider detectability requirements.
With respect to $bs$, all personal information is required to be either detectable by AX, or undetectable by SID and SPD (except for $d_7$, which $bs$ can always detect by definition of the scenario).
Similarly, identity providers can detect attributes they endorse by definition of the scenario, but no others by ID.
(Undetectability of endorsed attributes would be a requirement for the blind certification \cite{Bangerter2004CryptographicFrameworkControlled} feature of the Identity Mixer scheme as discussed in Section~\ref{subsec:architectures-idemix}.)
There are no detectability requirements with respect to $ttp$, or about the transaction details $d_7$.
In fact, these aspects would not produce relevant results because $ttp$ never learns any attributes, and $bs$ never communicates any transaction details.

Involvement requirements do not cover $ttp$ or $al$: the involvement of $ttp$ is publicly known, and Alice's involvement is covered by linkability.
For identity providers, there are involvement requirements about all remaining parties, i.e., the other identity provider and the service provider.
Usually, service providers assess trustworthiness of user attributes by considering which identity provider endorsed them; hence we do not regard involvement requirements with respect to the service provider as important.
(Among the analysed systems, only the Smartcard scheme would satisfy them.)

Linkability requirements capture associations by coalitions of actors.
Clearly, at least $ii$ and $is$ are needed to associate $\kappa$ and $\mu$; IIL states that without help of others, they cannot.
There is no requirement about when $bs$ helps them with this; as it turns out, this help never makes a difference.
Linkability between user databases and service provisions is defined with respect to the respective identity providers, and with respect to a coalition of all identity and service providers.
Considering other coalitions would not reveal interesting differences in the systems we analyse.
Similarly, no requirement involves $ii$ or $is$ in linking the service provisions to each other; in practice, an identity provider would link service provisions to each other by first linking them to its own user profile, which is covered by IL.
Finally, AR requires linking the service provisions to $\kappa$ and not to $\mu$; this is an arbitrary choice made in the definition of the scenario.

\subsection{Step 3: Model Communication in IdM systems}\label{subsec:formalizing-architectures}

Step 3 of our framework is to model the communication in the systems we want to analyse ($\S$\ref{sec:architectures}).
For each system, this formalisation consists of two parts.
First, we define an initial state  $\{\mathcal{C}_a^0\}_{a\in\mathcal{A}}$ capturing the initial knowledge of all actors, extending the knowledge from  Figure~\ref{fig:casestudy} with respect to the specific system.
Second, we define a trace $\mathsf{Scenario}$ that models the communication that takes place in the system in the system from the initial state when registration at $ii$, registration at $is$, and two service provisions at $bs$, are consecutively performed.

We introduce the abbreviation $\textrm{MS}_{\mathsf{k}^-}(\mathsf{m}):=\{\mathsf{m},S_{\mathsf{k}^-}(\mathsf{m})\}$ to denote a message along with its signature, capturing both X.509 certificates \cite{Housley2002InternetX.509Public} and signed SAML assertions \cite{SAML20}.

\subsubsection{Smart Certificates}

\begin{figure}[bt]
\small
\begin{align*}
\mathcal{C}^0_{al} &= \textrm{(Fig.\ref{fig:casestudy})} \cup \{ \textrm{MS}_{\ctxm{k^-}{\cdot}{ca}}(\ctxm{i}{\cdot}{al}, \mathsf{pk}(\ctxm{k^-}{\cdot}{u}),\ctxm{n_{c}}{\cdot}{\cdot}), \ctxm{k^-}{\cdot}{al},    
\\&~~~~~~~~~~~~~~~~~~~~~~~~~~~~~~ \mathbf{\ctxm{n_{z,a}}{\pi}{\cdot}}, \mathbf{\ctxm{n_{z,a}}{\eta}{\cdot}}, \mathbf{\ctxm{n_{z,a}}{\zeta}{\cdot}}, \mathbf{\ctxm{n_{z,a}}{\xi}{\cdot}} \}; \\
\mathcal{C}^0_{ii} &= \textrm{(Fig.\ref{fig:casestudy})} \cup \{ \mathbf{\ctxm{n_{z,b}}{\pi}{\cdot}}, \ctxm{n_{a}}{\pi}{\cdot}   \};\\
\mathcal{C}^0_{is} &= \textrm{(Fig.\ref{fig:casestudy})} \cup \{ \mathbf{\ctxm{n_{z,b}}{\eta}{\cdot}}, \ctxm{n_{b}}{\eta}{\cdot}  \};\\
\mathcal{C}^0_{bs} &= \textrm{(Fig.\ref{fig:casestudy})} \cup \{ \mathbf{\ctxm{n_{z,b}}{\zeta}{\cdot}}, \mathbf{\ctxm{n_{z,b}}{\xi}{\cdot}} \}
\end{align*}
\begin{align*}
&\mathsf{Scenario}:= \ctxm{\mathsf{Reg}_1}{\pi}{};~~\ctxm{\mathsf{Reg}_2}{\eta}{};~~\ctxm{\mathsf{ServProv}}{\zeta}{};~~\ctxm{\mathsf{ServProv}}{\xi}{}~~~~~~~~~~~~~\\
&\mathsf{Reg}_1:=\\
&~~~~\mathbf{\ctxm{ip}{}{u}}\to \ctxm{ip}{}{idp1}:\textrm{MS}_{\ctxm{k^-}{}{ca}}(\ctxm{i}{}{u}, \mathsf{pk}(\ctxm{k^-}{}{u}), \ctxm{n_{c}}{}{\cdot});&(1)\\
&~~~~\transmission{\mathbf{\ctxm{ip}{}{u}}}{\ctxm{ip}{}{idp1}}{}\textrm{ZK}(\ctxm{k^-}{}{u};\mathsf{pk}(\ctxm{k^-}{}{u});{\emptyset};\{\mathbf{\ctxm{n_{z,a}}{}{\cdot}}, \mathbf{\ctxm{n_{z,b}}{}{\cdot}}\});&(2)\\
&~~~~\ctxm{ip}{}{idp1}\to \mathbf{\ctxm{ip}{}{u}}:\textrm{MS}_{\ctxm{k^-}{}{idp1}}(\ctxm{i}{}{u}, \ctxm{d_{1}}{}{u}, \ctxm{d_{2}}{}{u}, \ctxm{d_{3}}{}{u}, \ctxm{n_{a}}{}{\cdot})&(3)\\
&\mathsf{Reg}_2:=\\
&~~~~\mathbf{\ctxm{ip}{}{u}}\to \ctxm{ip}{}{idp2}:\textrm{MS}_{\ctxm{k^-}{}{ca}}(\ctxm{i}{}{u}, \mathsf{pk}(\ctxm{k^-}{}{u}), \ctxm{n_{c}}{}{\cdot});&(4)\\
&~~~~\transmission{\mathbf{\ctxm{ip}{}{u}}}{\ctxm{ip}{}{idp2}}{}\textrm{ZK}(\ctxm{k^-}{}{u};\mathsf{pk}(\ctxm{k^-}{}{u});{\emptyset};\{\mathbf{\ctxm{n_{z,a}}{}{\cdot}}, \mathbf{\ctxm{n_{z,b}}{}{\cdot}}\});&(5)\\
&~~~~\ctxm{ip}{\eta}{idp2}\to \mathbf{\ctxm{ip}{\eta}{u}}:\textrm{MS}_{\ctxm{k^-}{\eta}{idp2}}(\ctxm{i}{\eta}{u}, \ctxm{d_{5}}{\eta}{u}, \ctxm{d_{6}}{\eta}{u}, \ctxm{n_{b}}{\eta}{\cdot})&(6)\\
&\mathsf{ServProv}:=\\
&~~~~\mathbf{\ctxm{ip}{}{u}}\to \ctxm{ip}{}{sp}:\textrm{MS}_{\ctxm{k^-}{}{ca}}(\ctxm{i}{}{u}, \mathsf{pk}(\ctxm{k^-}{}{u}), \ctxm{n_{c}}{}{\cdot}),\\ &\mathignore{~~~~\mathbf{\ctxm{ip}{}{u}}\to \ctxm{ip}{}{sp}:}\textrm{MS}_{\ctxm{k^-}{}{idp1}}(\ctxm{i}{}{u}, \ctxm{d_{1}}{}{u}, \ctxm{d_{2}}{}{u}, \ctxm{d_{3}}{}{u}, \ctxm{n_{a}}{}{\cdot}), &(1)\\
&\mathignore{~~~~\mathbf{\ctxm{ip}{}{u}}\to \ctxm{ip}{}{sp}:}\textrm{MS}_{\ctxm{k^-}{}{idp2}}(\ctxm{i}{}{u}, \ctxm{d_{5}}{}{u}, \ctxm{d_{6}}{}{u}, \ctxm{n_{b}}{}{\cdot});\\
&~~~~\transmission{\mathbf{\ctxm{ip}{}{u}}}{\ctxm{ip}{}{sp}}{}\textrm{ZK}(\ctxm{k^-}{}{u};\mathsf{pk}(\ctxm{k^-}{}{u});{\emptyset};\{\mathbf{\ctxm{n_{z,a}}{}{\cdot}}, \mathbf{\ctxm{n_{z,b}}{}{\cdot}}\})&(2)
\end{align*}
\caption{Formalisation of smart certificates: initial knowledge and trace\label{fig:formalization-smart-certificates}}
\end{figure}
Figure~\ref{fig:formalization-smart-certificates} displays our formalisation of smart certificates ($\S$\ref{subsub:architectures-smartcert}).
In addition to the knowledge from Figure~\ref{fig:casestudy}, Alice initially knows her public key certificate $$\textrm{MS}_{\ctxm{k^-}{\cdot}{ca}}(\ctxm{i}{\cdot}{al}, \mathsf{pk}(\ctxm{k^-}{\cdot}{u}), \ctxm{n_{c}}{\cdot}{\cdot})$$ ($\ctxm{n_{c}}{\cdot}{\cdot}$ represents additional information in the certificate such as the validity date), and the corresponding private key $\ctxm{k^-}{\cdot}{al}$.
The other items of initial knowledge are the contributions $\mathbf{\ctxm{n_{z,*}}{*}{\cdot}}$ to Alice's proof of knowledge of $\ctxm{k^-}{\cdot}{al}$, and additional information $\ctxm{n_{a}}{\pi}{\cdot}$, $\ctxm{n_{b}}{\eta}{\cdot}$ put in the attribute certificates issued by $ii$ and $is$.

The messages in the traces $\mathsf{Reg}_1$ and $\mathsf{Reg}_2$ correspond to those in Figure~\ref{fig:sc-reg};
the messages in the trace $\mathsf{ServProv}$ correspond to those in Figure~\ref{fig:sc-sp}.
We model the proof that Alice knows the secret key corresponding to her public key as a ZK proof with secret information $\ctxm{k^-}{\pi}{u}$ and public information $\mathsf{pk}(\ctxm{k^-}{\pi}{u})$.

\subsubsection{Linking Service Model}

\begin{figure}[tb]
\small
\begin{align*}
\mathcal{C}^0_{ii} &= \textrm{(Fig.\ref{fig:casestudy})} \cup \{ \ctxm{ip}{\cdot}{ls}, \mathsf{pk}(\ctxm{k^-}{\cdot}{ls}), \ctxm{i}{\cdot}{ls}, \\&~~~~~~~~~~~~~~~~~~~~\ctxm{i}{\cdot}{is}, \ctxm{i_{i1,ls}}{\pi}{u}, \mathbf{\ctxm{n}{\pi}{\cdot}}, 
                    \ctxm{i_{ii}}{\zeta}{u}, \mathbf{\ctxm{i_{sess}}{\zeta}{u}}, \mathbf{\ctxm{n}{\zeta}{\cdot}}, \ctxm{i_{ii}}{\xi}{u}, \mathbf{\ctxm{i_{sess}}{\xi}{u}}, \mathbf{\ctxm{n}{\xi}{\cdot}} \};\\
\mathcal{C}^0_{is} &= \textrm{(Fig.\ref{fig:casestudy})} \cup \{ \ctxm{ip}{\cdot}{ls}, \mathsf{pk}(\ctxm{k^-}{\cdot}{ls}), \ctxm{i}{\cdot}{ls}, \ctxm{i}{\cdot}{is}, \ctxm{i_{i2,ls}}{\eta}{u}, \mathbf{\ctxm{n}{\eta}{\cdot}} \};\\
\mathcal{C}^0_{bs} &= \textrm{(Fig.\ref{fig:casestudy})} \cup \{  \ctxm{ip}{\cdot}{ls}, \mathsf{pk}(\ctxm{k^-}{\cdot}{ls}), \ctxm{i}{\cdot}{ls}, \ctxm{i}{\cdot}{is} \};\\
\mathcal{C}^0_{ls} &= \textrm{(Fig.\ref{fig:casestudy})} \cup \{ \ctxm{ip}{\cdot}{ls}, \mathsf{pk}(\ctxm{k^-}{\cdot}{ls}), \ctxm{k^-}{\cdot}{ls}, \ctxm{i}{\cdot}{ls}, \ctxm{i}{\cdot}{is}, \ctxm{i_{l}}{\nu}{al}, \ctxm{i_{l}}{\pi}{u}, \ctxm{i_{l}}{\eta}{u}, \mathbf{\ctxm{n'}{\zeta}{\cdot}}, \mathbf{\ctxm{n'}{\xi}{\cdot}} \}
\end{align*}
\begin{align*}
&\mathsf{Scenario}:= \ctxm{\mathsf{Reg}_1}{\pi}{};~~\ctxm{\mathsf{Reg}_2}{\eta}{};~~\ctxm{\mathsf{ServProv}}{\zeta}{};~~\ctxm{\mathsf{ServProv}}{\xi}{}~~~~~~~~~~~~~\\
&\mathsf{Reg}_1:=\\
  &~~~~\ctxm{ip}{}{idp1}\to \ctxm{ip}{}{ls}:\textrm{MS}_{\ctxm{k^-}{}{idp}}(\ctxm{i_{i1,ls}}{}{u}, \mathbf{\ctxm{n}{}{\cdot}})&(1)\\
&\mathsf{Reg}_2:=\\
  &~~~~\ctxm{ip}{}{idp2}\to \ctxm{ip}{}{ls}:\textrm{MS}_{\ctxm{k^-}{}{idp}}(\ctxm{i_{i2,ls}}{}{u}, \mathbf{\ctxm{n}{}{\cdot}})&(2)\\
&\mathsf{ServProv}:=\\
  &~~~~\ctxm{ip}{}{idp1}\to \ctxm{ip}{}{sp}:\textrm{MS}_{\ctxm{k^-}{}{idp1}}(\mathbf{\ctxm{i_{sess}}{}{u}}, \ctxm{d_{1}}{}{u}, \ctxm{d_{2}}{}{u}, \ctxm{i}{}{ls}, \\&~~~~~~~~~~~~~~~~~~~~~~~~~~~~~~~~~~~~~~~~~E_{\mathsf{pk}(\ctxm{k^-}{}{ls})}(\ctxm{i_{i1,ls}}{}{u},\mathbf{\ctxm{n}{}{\cdot}})) &(1)\\
  &~~~~\ctxm{ip}{}{sp}\to \ctxm{ip}{}{ls}:E_{\mathsf{pk}(\ctxm{k^-}{}{ls})}(\ctxm{i_{i1,ls}}{}{u}, \mathbf{\ctxm{n}{}{\cdot}}), 
\textrm{MS}_{\ctxm{k^-}{}{idp1}}(\mathbf{\ctxm{i_{sess}}{}{u}},   \\&
~~~~~~~~~~~~~~~~~~~~~~~~~~~~~~~~~~~~~~~~~\ctxm{d_{1}}{}{u}, \ctxm{d_{2}}{}{u},\ctxm{i}{}{ls},E_{\mathsf{pk}(\ctxm{k^-}{}{ls})}(\ctxm{i_{i1,ls}}{}{u}, \mathbf{\ctxm{n}{}{\cdot}})); &(2)\\
  &~~~~\ctxm{ip}{}{ls}\to \ctxm{ip}{}{sp}:\ctxm{i}{}{idp2}, E_{\mathsf{pk}(\ctxm{k^-}{}{idp2})}(\ctxm{i_{i2,ls}}{}{u}, \mathbf{\ctxm{n'}{}{\cdot}});&(3)\\
  &~~~~\ctxm{ip}{}{sp}\to \ctxm{ip}{}{idp2}:E_{\mathsf{pk}(\ctxm{k^-}{}{idp2})}(\ctxm{i_{i2,ls}}{}{u}, \mathbf{\ctxm{n'}{}{\cdot}}),\textrm{MS}_{\ctxm{k^-}{}{idp1}}(\mathbf{\ctxm{i_{sess}}{}{u}}, \\&
 ~~~~~~~~~~~~~~~~~~~~~~~~~~~~~~~~~~~~~~~~~\ctxm{d_{1}}{}{u}, \ctxm{d_{2}}{}{u},\ctxm{i}{}{ls}, E_{\mathsf{pk}(\ctxm{k^-}{}{ls})}(\ctxm{i_{i1,ls}}{}{u}, \mathbf{\ctxm{n}{}{\cdot}}));&(4)\\
  &~~~~\ctxm{ip}{}{idp2}\to \ctxm{ip}{}{sp}:\textrm{MS}_{\ctxm{k^-}{}{idp2}}(\mathbf{\ctxm{i_{sess}}{}{u}}, \ctxm{d_{6}}{}{u})&(5)
\end{align*}
\caption{Formalisation of linking service model: initial knowledge and trace\label{fig:formalization-ls}}
\end{figure}
Figure~\ref{fig:formalization-ls} displays the formalisation of the linking service model ($\S$\ref{subsec:architectures-ls}).
This system introduces the linking service $ls$ as an additional actor: it has an address and a private/public key pair.
$ls$ and $is$ have publicly known identifiers $\ctxm{i}{\cdot}{ls}$, $\ctxm{i}{\cdot}{is}$ used in the referrals.
The user database of $ls$, modelled by domain $\nu$, contains an entry for the user containing only the identifier $\ctxm{i_{l}}{\nu}{al}$.
User authentication to $ls$ during registration is modelled by $ls$'s knowledge of $\ctxm{i_{l}}{\pi}{u}$; the pseudonyms generated by the identity providers are modelled as $\ctxm{i_{i1,ls}}{\pi}{u}$ and $\ctxm{i_{i2,ls}}{\pi}{u}$.
Alice's authentication at $ii$ during service provision is modelled by the fact that $ii$ knows the identifiers $\ctxm{i_{ii}}{*}{u}$, $*\in\{\zeta,\xi\}$.

The registration and service provision phases in the trace correspond to Figures~\ref{fig:tas3-reg} and \ref{fig:tas3-sp}, respectively.
To prove authenticity, the identity providers sign information for $bs$ using their private key.
$bs$ forwards the authentication assertion from $ii$ to $ls$ and $is$ to prove that the user has authenticated.
The referrals by $ii$ and $is$ include random nonces $\mathbf{\ctxm{n}{}{\cdot}}$, $\mathbf{\ctxm{n'}{}{\cdot}}$ to ensure that $bs$ cannot link different sessions by comparing them.

The linking service model aims to satisfy a privacy requirement specifically about the linking service, which we call \emph{LS attribute undetectability} (LD).
We can express this requirement formally in a similar way to the SID, SPD, and ID requirements: $\ctxm{d_1}{*}{*}\notin\mathsf{O}_{ls}\wedge...\wedge\ctxm{d_6}{*}{*}\notin\mathsf{O}_{ls}$.

The linking service model in general is independent from message formats.
However, the authors also present an instantiation using the SAML 2.0~\cite{SAML20} and Liberty ID-WSF 2.0~\cite{LibertyIDWSF} standards.
Our model captures that instantiation.

\subsubsection{Identity Mixer}

\begin{figure}[tbh!]
\small
\begin{align*}
\mathcal{C}^0_{al} &= \textrm{(Fig.\ref{fig:casestudy})} \cup\{ \ctxm{i}{\cdot}{al}, \\&~~~~~~~~ \ctxm{n_{c1,1}}{\pi}{\cdot}, \ctxm{n_{c1,2}}{\pi}{\cdot}, \ctxm{n_{c1,3}}{\pi}{\cdot}, \ctxm{n_{c1,7}}{\pi}{\cdot}, \ctxm{n_{c2,1}}{\eta}{\cdot}, \ctxm{n_{c2,2}}{\eta}{\cdot}, \ctxm{n_{c2,3}}{\eta}{\cdot}, \ctxm{n_{c2,7}}{\eta}{\cdot},\\
                   &~~~~~~~~ \mathbf{\ctxm{n_{v}}{\zeta}{\cdot}}, \ctxm{cnd}{\zeta}{\cdot}, \mathbf{\ctxm{n}{\zeta}{\cdot}}, \mathbf{\ctxm{n_{1,1}}{\zeta}{\cdot}}, \mathbf{\ctxm{n_{1,2}}{\zeta}{\cdot}}, \mathbf{\ctxm{n_{1,3}}{\zeta}{\cdot}}, \mathbf{\ctxm{n_{1,a}}{\zeta}{\cdot}}, \mathbf{\ctxm{n_{2,1}}{\zeta}{\cdot}},\mathbf{\ctxm{n_{2,a}}{\zeta}{\cdot}}, \\
                   &~~~~~~~~ \mathbf{\ctxm{n_{v}}{\xi}{\cdot}}, \ctxm{cnd}{\xi}{\cdot}, \mathbf{\ctxm{n}{\xi}{\cdot}}, \mathbf{\ctxm{n_{1,1}}{\xi}{\cdot}}, \mathbf{\ctxm{n_{1,2}}{\xi}{\cdot}}, \mathbf{\ctxm{n_{1,3}}{\xi}{\cdot}}, \mathbf{\ctxm{n_{1,a}}{\xi}{\cdot}}, \mathbf{\ctxm{n_{2,1}}{\xi}{\cdot}}, \mathbf{\ctxm{n_{2,a}}{\xi}{\cdot}} \} \\
\mathcal{C}^0_{ii} &= \textrm{(Fig.\ref{fig:casestudy})} \cup \{ \ctxm{n_{c1,4}}{\pi}{\cdot}, \ctxm{n_{c1,5}}{\pi}{\cdot}, \ctxm{n_{c1,6}}{\pi}{\cdot}  \}; \\
\mathcal{C}^0_{is} &= \textrm{(Fig.\ref{fig:casestudy})} \cup \{ \ctxm{n_{c2,4}}{\eta}{\cdot}, \ctxm{n_{c2,5}}{\eta}{\cdot}, \ctxm{n_{c2,6}}{\eta}{\cdot} \}; \\
\mathcal{C}^0_{bs} &= \textrm{(Fig.\ref{fig:casestudy})} \cup \{ \mathbf{\ctxm{n_{1,b}}{\zeta}{\cdot}}, \mathbf{\ctxm{n_{2,b}}{\zeta}{\cdot}}, \mathbf{\ctxm{n_{1,b}}{\xi}{\cdot}}, \mathbf{\ctxm{n_{2,b}}{\xi}{\cdot}}  \}\end{align*}
\begin{align*}
&\mathsf{Scenario}:= \ctxm{\mathsf{Reg}_1}{\pi}{};~~\ctxm{\mathsf{Reg}_2}{\eta}{};~~\ctxm{\mathsf{ServProv}}{\zeta}{};~~\ctxm{\mathsf{ServProv}}{\xi}{}~~~~~~~~~~~~~\\
&\mathsf{Reg}_1:=\\
  &~~~~\mathbf{\ctxm{ip}{}{u}}\to \ctxm{ip}{}{idp1}:\mathcal{H}(\ctxm{i}{}{u}, \ctxm{n_{c1,1}}{}{\cdot});&(1)\\
  &~~~~\transmission{\mathbf{\ctxm{ip}{}{u}}}{\ctxm{ip}{}{idp1}}{}\textrm{ICred}^{\ctxm{i}{}{u}}_{\ctxm{k^-}{}{idp1}}(\ctxm{i_{ii}}{}{u}, \ctxm{d_{1}}{}{u}, \ctxm{d_{2}}{}{u}, \ctxm{d_{3}}{}{u};\{\ctxm{n_{c1,i}}{}{\cdot}\}_{i=1}^7) &(2)\\
&\mathsf{Reg}_2:=\\
  &~~~~\mathbf{\ctxm{ip}{}{u}}\to \ctxm{ip}{}{idp2}:\mathcal{H}(\ctxm{i}{}{u}, \ctxm{n_{c2,1}}{}{\cdot});&(3)\\
  &~~~~\transmission{\mathbf{\ctxm{ip}{}{u}}}{\ctxm{ip}{}{idp2}}{}\textrm{ICred}^{\ctxm{i}{}{u}}_{\ctxm{k^-}{}{idp2}}(\ctxm{d_{5}}{}{u}, \ctxm{d_{6}}{}{u};\{\ctxm{n_{c2,i}}{}{\cdot}\}_{i=1}^7)&(4)\\
&\mathsf{ServProv}:=\\
  &~~~~\mathbf{\ctxm{ip}{}{u}}\to \ctxm{ip}{}{sp}:\mathcal{H}(\ctxm{i}{}{u}, \mathbf{\ctxm{n}{}{\cdot}}), \mathcal{H}(\ctxm{i_{ii}}{}{u}, \mathbf{\ctxm{n_{1,2}}{}{\cdot}}), \mathcal{H}(\ctxm{d_{2}}{}{u}, \mathbf{\ctxm{n_{1,1}}{}{\cdot}}), &(1) \\
  &\mathignore{~~~~\mathbf{\ctxm{ip}{}{u}}\to \ctxm{ip}{}{sp}:} \mathcal{H}(\ctxm{d_{3}}{}{u}, \mathbf{\ctxm{n_{1,3}}{}{\cdot}}),\ctxm{d_{1}}{}{u}, \ctxm{d_2{>}60}{}{u}, \ctxm{cnd}{}{\cdot},\\
  &\mathignore{~~~~\mathbf{\ctxm{ip}{}{u}}\to \ctxm{ip}{}{sp}:} \mathsf{pk}(\ctxm{k^-}{}{ttp}), E_{\mathsf{pk}(\ctxm{k^-}{}{ttp})}(\ctxm{i_{ii}}{}{u}, \mathbf{\ctxm{n_{v}}{}{\cdot}})_{\ctxm{cnd}{}{\cdot}};\\
  &~~~~\transmission{\mathbf{\ctxm{ip}{}{u}}}{\ctxm{ip}{}{sp}}{}\textrm{ZK}(\textrm{cred}^{\ctxm{i}{}{u}}_{\ctxm{k^-}{}{idp1}}(\ctxm{i_{ii}}{}{u}, \ctxm{d_{1}}{}{u}, \ctxm{d_{2}}{}{u}, \ctxm{d_{3}}{}{u};\ctxm{n_{c1,2}}{}{\cdot}, &(2) \\
  &\mathignore{~~~~\transmission{\mathbf{\ctxm{ip}{}{u}}}{\ctxm{ip}{}{sp}}{}\textrm{ZK}(}\ctxm{n_{c1,5}}{}{\cdot}),\ctxm{i}{}{u}, \ctxm{i_{ii}}{}{u}, \ctxm{d_{1}}{}{u}, \ctxm{d_{2}}{}{u}, \ctxm{d_{3}}{}{u},\mathbf{\ctxm{n}{}{\cdot}}, \mathbf{\ctxm{n_{1,2}}{}{\cdot}}, \\
  &\mathignore{~~~~\transmission{\mathbf{\ctxm{ip}{}{u}}}{\ctxm{ip}{}{sp}}{}\textrm{ZK}(}\mathbf{\ctxm{n_{1,1}}{}{\cdot}}, \mathbf{\ctxm{n_{1,3}}{}{\cdot}};\mathcal{H}(\ctxm{i}{}{u},\mathbf{\ctxm{n}{}{\cdot}}), \mathcal{H}(\ctxm{i_{ii}}{}{u}, \mathbf{\ctxm{n_{1,2}}{}{\cdot}}),  \\ 
  &\mathignore{~~~~\transmission{\mathbf{\ctxm{ip}{}{u}}}{\ctxm{ip}{}{sp}}{}\textrm{ZK}(}\mathcal{H}(\ctxm{d_{2}}{}{u}, \mathbf{\ctxm{n_{1,1}}{}{\cdot}}),\mathcal{H}(\ctxm{d_{3}}{}{u}, \mathbf{\ctxm{n_{1,3}}{}{\cdot}}), \ctxm{d_{1}}{}{u}, \\
  &\mathignore{~~~~\transmission{\mathbf{\ctxm{ip}{}{u}}}{\ctxm{ip}{}{sp}}{}\textrm{ZK}(}\mathsf{pk}(\ctxm{k^-}{}{idp1}), \mathsf{pk}(\ctxm{k^-}{}{ttp}),E_{\mathsf{pk}(\ctxm{k^-}{}{ttp})}(\ctxm{i_{ii}}{}{u}, \\
  &\mathignore{~~~~\transmission{\mathbf{\ctxm{ip}{}{u}}}{\ctxm{ip}{}{sp}}{}\textrm{ZK}(}\mathbf{\ctxm{n_{v}}{}{\cdot}})_{\ctxm{cnd}{}{\cdot}};\ctxm{d_2{>}60}{}{u};\{\mathbf{\ctxm{n_{1,a}}{}{\cdot}}, \mathbf{\ctxm{n_{1,b}}{}{\cdot}}\});\\
  &~~~~\mathbf{\ctxm{ip}{}{u}}\to \ctxm{ip}{}{sp}:\mathcal{H}(\ctxm{i}{}{u}, \mathbf{\ctxm{n}{}{\cdot}}), \mathcal{H}(\ctxm{d_{5}}{}{u}, \mathbf{\ctxm{n_{2,1}}{}{\cdot}}), \ctxm{d_{6}}{}{u}, \ctxm{cnd}{}{\cdot};&(3)\\
  &~~~~\transmission{\mathbf{\ctxm{ip}{}{u}}}{\ctxm{ip}{}{sp}}{}\textrm{ZK}(\textrm{cred}^{\ctxm{i}{}{u}}_{\ctxm{k^-}{}{idp2}}(\ctxm{d_{5}}{}{u}, \ctxm{d_{6}}{}{u};\ctxm{n_{c2,2}}{}{\cdot},\ctxm{n_{c2,5}}{}{\cdot}),\ctxm{i}{}{u}, &(4) \\
  &\mathignore{~~~~\transmission{\mathbf{\ctxm{ip}{}{u}}}{\ctxm{ip}{}{sp}}{}\textrm{ZK}(}\ctxm{d_{5}}{}{u}, \ctxm{d_{6}}{}{u}, \mathbf{\ctxm{n}{}{\cdot}}, \mathbf{\ctxm{n_{2,1}}{}{\cdot}};\mathcal{H}(\ctxm{i}{}{u}, \mathbf{\ctxm{n}{}{\cdot}}),\\
  &\mathignore{~~~~\transmission{\mathbf{\ctxm{ip}{}{u}}}{\ctxm{ip}{}{sp}}{}\textrm{ZK}(}\mathcal{H}(\ctxm{d_{5}}{}{u}, \mathbf{\ctxm{n_{2,1}}{}{\cdot}}),\ctxm{d_{6}}{}{u}, \mathsf{pk}(\ctxm{k^-}{}{idp2});\\
  &\mathignore{~~~~\transmission{\mathbf{\ctxm{ip}{}{u}}}{\ctxm{ip}{}{sp}}{}\textrm{ZK}(} \emptyset;\{\mathbf{\ctxm{n_{2,a}}{}{\cdot}}, \mathbf{\ctxm{n_{2,b}}{}{\cdot}}\})
\end{align*}

\caption{Formalisation of Identity Mixer: initial knowledge and trace\label{fig:formalization-idemix}}
\end{figure}
The formalisation of the scenario when using Identity Mixer ($\S$\ref{subsec:architectures-idemix}) is shown in Figure~\ref{fig:formalization-idemix}.
The most notable piece of initial knowledge is Alice's secret identifier $\ctxm{i}{\cdot}{al}$.
In the trace, registration follows the steps of Figure~\ref{fig:idemix-reg}; service provision is as in Figure~\ref{fig:idemix-sp}.
For our purposes, we can represent the commitment to Alice's secret identifier in the first message by a hash $\mathcal{H}(\ctxm{i}{\pi}{u}, \ctxm{n_{c1,1}}{\pi}{\cdot})$.
By inference rule $\avdashc{EI$_2$}$, Alice learns a credential from the issuing protocol linking her attributes to her secret identifier.
For instance, from message (2) she can derive
$$\ctxm{\textrm{cred}^{\ctxm{i}{}{u}}_{\ctxm{k^-}{}{idp1}}(\ctxm{i_{ii}}{}{u}, \ctxm{d_{1}}{}{u}, \ctxm{d_{2}}{}{u}, \ctxm{d_{3}}{}{u};\ctxm{n_{c1,2}}{}{\cdot},\ctxm{n_{c1,5}}{}{\cdot},\ctxm{n_{c1,6}}{}{\cdot})}{\pi}{}.$$
Note that this credential contains Alice's identifier $\ctxm{i_{ii}}{\pi}{u}$ as an additional attribute: it is used later for anonymity revocation.

In the first message of service provision, again we represent the commitments to Alice's secret identifier and attributes by hashes.
For anonymity revocation purposes, the first message additionally includes an encryption of the identifier $\ctxm{i_{ii}}{\pi}{u}$ for the trusted third party, with a condition $\ctxm{cnd}{}{\cdot}$ attached describing when the anonymity of the transaction may be revoked.
The ZK proof in message (2) convinces $bs$ that:
\begin{itemize}
 \item Alice owns a credential, signed with $ii$'s private key;
 \item the secret identifier and attributes in the credential correspond to the values or commitments sent previously;
 \item the property $\ctxm{d_2{>}60}{}{u}$ is satisfied;
 \item the encrypted message sent previously is encrypted using $\mathsf{pk}(\ctxm{k^-}{}{ttp})$ and contains the identifier in the credential.
\end{itemize}
The second ZK proof is similar.
Note that the commitment $\mathcal{H}(\ctxm{i}{}{u}, \mathbf{\ctxm{n}{}{\cdot}})$ in messages (1) and (3) is the same, guaranteeing $bs$ that the two certificates are indeed of the same user.

\subsubsection{Smartcard Scheme}

The Smartcard scheme ($\S$\ref{subsec:architectures-smartcard}) is formalised in Figure~\ref{fig:formalization-smartcard}.
In this system, the user's personal information is exchanged on her behalf by a tamper-resistant smartcard.
The smartcard is modelled as actor $al$.
The smartcard has a certified private key; however, this private key is shared between different smartcards so it does not identify the user.
Instead, the smartcard has a secret user identifier $\ctxm{i}{\cdot}{al}$, generated on the card, which is used to generate pseudonyms.
The actors $ii$, $is$, and $bs$ each have a private key and a corresponding public key certificate signed by the certification authority.

The messages from the registration part of the trace correspond to Figure~\ref{fig:smartcard-reg}; the messages from the service provision part correspond to Figure~\ref{fig:smartcard-sp}.
Parties derive a shared session key using authenticated key agreement based on public key certificates and exchanged randomness.
The smartcard generates pseudonyms of Alice with respect to the two identity providers using hashes.
In the service provision phase, $\mathbf{\ctxm{q}{}{\cdot}}$ and $\mathbf{\ctxm{dm}{}{\cdot}}$ represent $bs$'s query: what information it needs, and how recent it should be.

Note that in~\cite{Vossaert2010User-centricidentitymanagement}, the exact format of the encrypted message to the trusted third party for anonymity revocation is not specified.
We chose an encryption of the user's identifier at the address provider because this is most appropriate for our scenario.
Also, it is not specified how attributes are sent to the smartcard for caching; we chose to add one additional message to the registration phase containing all attributes.

\begin{figure}[tbh!]
\small
\begin{align*}
\mathcal{C}^0_{al} &= \mathcal{C}^s_{al} \cup \{ \mathsf{pk}(\ctxm{k^-}{\cdot}{ca}), \textrm{MS}_{\ctxm{k^-}{\cdot}{ca}}(\ctxm{i_{c}}{\cdot}{\cdot}, \mathsf{pk}(\ctxm{k^-_{c}}{\cdot}{\cdot}), \ctxm{n_{c}}{\cdot}{\cdot}), \ctxm{k^-_{c}}{\cdot}{\cdot}, \ctxm{i}{\cdot}{al},  \\
                   &~~~~~~~~~~~~~~~~~\mathbf{\ctxm{n_{a}}{\pi}{\cdot}}, \mathbf{\ctxm{n_{a}}{\eta}{\cdot}},\mathbf{\ctxm{n_{v}}{\zeta}{\cdot}}, \mathbf{\ctxm{i_{sess}}{\zeta}{u}}, \mathbf{\ctxm{n_{a}}{\zeta}{\cdot}}, \mathbf{\ctxm{n_{v}}{\xi}{\cdot}}, \mathbf{\ctxm{i_{sess}}{\xi}{u}}, \mathbf{\ctxm{n_{a}}{\xi}{\cdot}} \}; \\
\mathcal{C}^0_{ii} &= \mathcal{C}^s_{ii} \cup \{ \mathsf{pk}(\ctxm{k^-}{\cdot}{ca}), \textrm{MS}_{\ctxm{k^-}{\cdot}{ca}}(\ctxm{i}{\cdot}{ii}, \mathsf{pk}(\ctxm{k^-}{\cdot}{ii}), \ctxm{n_{ii}}{\cdot}{\cdot}), \mathbf{\ctxm{n_{b}}{\pi}{\cdot}}  \};\\
\mathcal{C}^0_{is} &= \mathcal{C}^s_{is} \cup \{ \mathsf{pk}(\ctxm{k^-}{\cdot}{ca}), \textrm{MS}_{\ctxm{k^-}{\cdot}{ca}}(\ctxm{i}{\cdot}{is}, \mathsf{pk}(\ctxm{k^-}{\cdot}{is}), \ctxm{n_{is}}{\cdot}{\cdot}), \mathbf{\ctxm{n_{b}}{\eta}{\cdot}} \};~~\\
\mathcal{C}^0_{bs} &= \mathcal{C}^s_{bs} \cup \{ \mathsf{pk}(\ctxm{k^-}{\cdot}{ca}), \textrm{MS}_{\ctxm{k^-}{\cdot}{ca}}(\ctxm{i}{\cdot}{bs}, \mathsf{pk}(\ctxm{k^-}{\cdot}{bs}), \ctxm{n_{bs}}{\cdot}{\cdot}), \\&~~~~~~~~~~~~~~~~~\mathbf{\ctxm{n_{b}}{\zeta}{\cdot}}, \mathbf{\ctxm{dm}{\zeta}{\cdot}}, \mathbf{\ctxm{q}{\zeta}{\cdot}},\mathbf{\ctxm{n_{b}}{\xi}{\cdot}}, \mathbf{\ctxm{dm}{\xi}{\cdot}}, \mathbf{\ctxm{q}{\xi}{\cdot}} \}\end{align*}
\begin{align*}
&\mathsf{Scenario}:= \ctxm{\mathsf{Reg}_1}{\pi}{};~~\ctxm{\mathsf{Reg}_2}{\eta}{};~~\ctxm{\mathsf{ServProv}}{\zeta}{};~~\ctxm{\mathsf{ServProv}}{\xi}{}~~~~~~~~~~~~~\\
&\mathsf{Reg}_1:=\\
  &~~~~\mathbf{\ctxm{ip}{}{u}}\to \ctxm{ip}{}{idp1}:\textrm{MS}_{\ctxm{k^-}{}{ca}}(\ctxm{i_{c}}{}{\cdot}, \mathsf{pk}(\ctxm{k^-_{c}}{}{\cdot}), \ctxm{n_{c}}{}{\cdot}), \mathbf{\ctxm{n_{a}}{}{\cdot}};&(1)\\
  &~~~~\ctxm{ip}{}{idp1}\to \mathbf{\ctxm{ip}{}{u}}:\textrm{MS}_{\ctxm{k^-}{}{ca}}(\ctxm{i}{}{idp1}, \mathsf{pk}(\ctxm{k^-}{}{idp1}), \ctxm{n_{idp1}}{}{\cdot}), \mathbf{\ctxm{n_{b}}{}{\cdot}};&(2)\\
  &~~~~\mathbf{\ctxm{ip}{}{u}}\to \ctxm{ip}{}{idp1}:E'_{\textrm{AKA}(\ctxm{k^-_{c}}{}{\cdot};\mathbf{\ctxm{n_{a}}{}{\cdot}};\ctxm{k^-}{}{idp1};\mathbf{\ctxm{n_{b}}{}{\cdot}})}(\mathcal{H}(\ctxm{i}{}{u}, \ctxm{i}{}{idp1}));&(3)\\
  &~~~~\ctxm{ip}{}{idp1}\to \mathbf{\ctxm{ip}{}{u}}:E'_{\textrm{AKA}(\ctxm{k^-_{c}}{}{\cdot};\mathbf{\ctxm{n_{a}}{}{\cdot}};\ctxm{k^-}{}{idp1};\mathbf{\ctxm{n_{b}}{}{\cdot}})}(\ctxm{d_{1}}{}{u}, \ctxm{d_{2}}{}{u}, \ctxm{d_{3}}{}{u})&(4)\\
&\mathsf{Reg}_2:=\\
  &~~~~\mathbf{\ctxm{ip}{}{u}}\to \ctxm{ip}{}{idp2}:\textrm{MS}_{\ctxm{k^-}{}{ca}}(\ctxm{i_{c}}{}{\cdot}, \mathsf{pk}(\ctxm{k^-_{c}}{}{\cdot}), \ctxm{n_{c}}{}{\cdot}), \mathbf{\ctxm{n_{a}}{}{\cdot}};&(5)\\
  &~~~~\ctxm{ip}{}{idp2}\to \mathbf{\ctxm{ip}{}{u}}:\textrm{MS}_{\ctxm{k^-}{}{ca}}(\ctxm{i}{}{idp2}, \mathsf{pk}(\ctxm{k^-}{}{idp2}), \ctxm{n_{idp2}}{}{\cdot}), \mathbf{\ctxm{n_{b}}{}{\cdot}};&(6)\\
  &~~~~\mathbf{\ctxm{ip}{}{u}}\to \ctxm{ip}{}{idp2}:E'_{\textrm{AKA}(\ctxm{k^-_{c}}{}{\cdot};\mathbf{\ctxm{n_{a}}{}{\cdot}};\ctxm{k^-}{}{idp2};\mathbf{\ctxm{n_{b}}{}{\cdot}})}(\mathcal{H}(\ctxm{i}{}{u}, \ctxm{i}{}{idp2}));&(7)\\
  &~~~~\ctxm{ip}{}{idp2}\to \mathbf{\ctxm{ip}{}{u}}:E'_{\textrm{AKA}(\ctxm{k^-_{c}}{}{\cdot};\mathbf{\ctxm{n_{a}}{}{\cdot}};\ctxm{k^-}{}{idp2};\mathbf{\ctxm{n_{b}}{}{\cdot}})}(\ctxm{d_{5}}{}{u}, \ctxm{d_{6}}{}{u})&(8)\\
  &\mathsf{ServProv}:=\\
  &~~~~\mathbf{\ctxm{ip}{}{u}}\to \ctxm{ip}{}{sp}:\textrm{MS}_{\ctxm{k^-}{}{ca}}(\ctxm{i_{c}}{}{\cdot}, \mathsf{pk}(\ctxm{k^-_{c}}{}{\cdot}), \ctxm{n_{c}}{}{\cdot}), \mathbf{\ctxm{n_{a}}{}{\cdot}};&(1)\\
  &~~~~\ctxm{ip}{}{sp}\to \mathbf{\ctxm{ip}{}{u}}:\textrm{MS}_{\ctxm{k^-}{}{ca}}(\ctxm{i}{}{sp}, \mathsf{pk}(\ctxm{k^-}{}{sp}), \ctxm{n_{sp}}{}{\cdot}), \mathbf{\ctxm{n_{b}}{}{\cdot}};&(2)\\
  &~~~~\mathbf{\ctxm{ip}{}{u}}\to \ctxm{ip}{}{sp}:E'_{\textrm{AKA}(\ctxm{k^-_{c}}{}{\cdot};\mathbf{\ctxm{n_{a}}{}{\cdot}};\ctxm{k^-}{}{sp};\mathbf{\ctxm{n_{b}}{}{\cdot}})}(\mathbf{\ctxm{i_{sess}}{}{u}});&(3)\\
  &~~~~\ctxm{ip}{}{sp}\to \mathbf{\ctxm{ip}{}{u}}:\mathbf{\ctxm{i_{sess}}{}{u}}, E'_{\textrm{AKA}(\ctxm{k^-_{c}}{}{\cdot};\mathbf{\ctxm{n_{a}}{}{\cdot}};\ctxm{k^-}{}{sp};\mathbf{\ctxm{n_{b}}{}{\cdot}})}(\mathbf{\ctxm{dm}{}{\cdot}});&(4)\\
  &~~~~\ctxm{ip}{}{sp}\to \mathbf{\ctxm{ip}{}{u}}:\mathbf{\ctxm{i_{sess}}{}{u}}, E'_{\textrm{AKA}(\ctxm{k^-_{c}}{}{\cdot};\mathbf{\ctxm{n_{a}}{}{\cdot}};\ctxm{k^-}{}{sp};\mathbf{\ctxm{n_{b}}{}{\cdot}})}(\mathbf{\ctxm{q}{}{\cdot}});&(5)\\
  &~~~~\mathbf{\ctxm{ip}{}{u}}\to \ctxm{ip}{}{sp}:E'_{\textrm{AKA}(\ctxm{k^-_{c}}{}{\cdot};\mathbf{\ctxm{n_{a}}{}{\cdot}};\ctxm{k^-}{}{sp};\mathbf{\ctxm{n_{b}}{}{\cdot}})}(\ctxm{d_{1}}{}{u}, \ctxm{d_2{>}60}{}{u}, \ctxm{d_{6}}{}{u}, \\&
  \mathignore{~~~~\mathbf{\ctxm{ip}{}{u}}\to \ctxm{ip}{}{sp}:}E_{\mathsf{pk}(\ctxm{k^-}{}{ttp})}(\mathcal{H}(\ctxm{i}{}{u}, \ctxm{i}{}{idp1}), \mathbf{\ctxm{n_{v}}{}{\cdot}})) &(6)
\end{align*}

\caption{Formalisation of Smartcard scheme: initial knowledge and trace\label{fig:formalization-smartcard}}
\end{figure}

\subsection{Step 4: Verify Privacy Properties \& Analysis of Results}\label{subsec:analysis-results}

\newcolumntype{C}{>{\centering\arraybackslash}m{0.35cm}<{}}
\begin{table}[tb]
 \small\centering\begin{tabular}{l|CC|CCC|CC|CCCC}
  \textbf{Scheme}        &$\textbf{AX}$&$\textbf{AR}$&$\textbf{SID}$&$\textbf{SPD}$&$\textbf{ID}$&$\textbf{IM}$&$\textbf{ISM}$&$\textbf{SL}$&$\textbf{IL}$&$\textbf{IIL}$&$\textbf{ISL}$\\\hline
  Smart certificates     &   \tick     &\gtick&\gcross     & \gcross     & \gtick     & \gtick     &  \gtick       &  \gcross     &    \gtick    &  \gcross      &   \gcross     \\
  Linking service model~~&   \tick     &   \gtick     &    \gtick     &   \gcross     &  \gcross    &   \cross    &   \gcross     &   \tick     &    \gcross   &  \gcross      &   \gcross     \\
  Identity Mixer         &   \tick     &   \tick     &    \tick     &   \tick      &   \gtick     &   \gtick     &   \gtick      &   \tick     &    \gtick    &  \gtick       &\tick$^\dagger$\\
  Smartcard scheme       &   \tick     &   \tick     &    \tick     &   \tick      &   \gtick     &   \gtick     &   \gtick      &   \tick     &    \gtick    &  \tick       &   \tick      \\
 \end{tabular}
 \caption{Comparison of privacy requirements claimed and satisfied by the various systems. Filled check-mark: satisfied and claimed; empty check-mark: satisfied and not claimed; filled cross: not satisfied and claimed; empty cross: not satisfied and not claimed (see Table~\ref{tbl:claims})
\label{tbl:results}. $\dagger$: may not be satisfiable efficiently depending on non-privacy-related requirements.
}
 \end{table}
Step 4 of our framework is to verify which requirements are satisfied by the analysed systems.
This step is performed automatically using our Prolog tool ($\S$\ref{subsec:prolog}): given the formalised requirements ($\S$\ref{subsec:formal-requirements}) and communication in the systems ($\S$\ref{subsec:formalizing-architectures}), the tool automatically determines which requirements hold in which systems.
(More precisely, it computes the state that the given initial state evolves into by the given trace, also checking trace validity  (see Appendix~\ref{sec:trace-validity}); then computes the views of actors and coalitions in this state; and finally, verifies which of the given requirements hold in these views.)
The results are shown in Table~\ref{tbl:results}: we now analyse them. 

\subsubsection{Non-privacy requirements}
The two non-privacy requirements \emph{attribute exchange (AX)} and \emph{anonymity revocation (AR)} are satisfied in all systems.
Indeed, attribute exchange is the basic requirement of an IdM system.
It is worth noting the relationship between AR and ISL.
In smart certificates and the linking service model, ISL does not hold.
In this case, AR holds automatically because the service provider and identity providers can link service accesses to user profiles (even without the help of the trusted third party).
In the two systems satisfying ISL (the Identity Mixer and Smartcard systems), the transmission of an identifier encrypted for the trusted third party is necessary to fulfil this requirement.

\subsubsection{Detectability requirements} 
The detectability requirements with respect to the service provider, \emph{property-attribute undetectability (SPD)} and \emph{irrelevant attribute undetectability (SID)},  verify the possibility to reveal properties of attributes without revealing the exact value; and to reveal some but not all attributes.
In smart certificates, the complete certificate is transmitted, so it satisfies neither requirement.
To address SID, the identity provider could issue a separate credential for each user attribute.
To partially address SPD, the identity provider could issue several credentials proving common properties of attributes, e.g. an ``age $>60$'' credential.
These latter credentials could be obtained during the service provision phase, in effect transforming smart certificates into a relationship-focused system.
Indeed, this variant is discussed in \cite{Park1999SmartCertificates:Extending}.
Another possibility is to use certificates that allow efficient proofs of knowledge, as in the Identity Mixer system.

In the linking service model, SPD does not hold.
Actually, the linking service model focuses primarily on involvement and linkability issues, leaving the details of the actual attribute exchange to underlying standards.
However, in these standards (in particular, SAML) it is not possible to exchange properties of an attribute instead of its value.
Recently, an extension to SAML to achieve this has been proposed~\cite{NevenAttributePredicateProfile}.
With this extension (or other instantiations), the requirement may hold.

IdP attribute undetectability (ID) and LS attribute undetectability (LD) also do not hold in the linking service model.
This is because the linking service and the subscription provider both receive the signed authentication assertion from the address provider as guarantee that the user has logged in.
However, in the SAML standard, the attributes are part of this signed message, so they also need to be forwarded.
Technically, this could be easily solved by signing the attributes separately from the authentication information.
Again, this problem is due to the instantiation of the model with SAML.
Note that although ID is not explicitly claimed by the other IdM systems, they do satisfy it.

\subsubsection{Involvement requirements}

The involvement requirements state that an identity provider should not know about the user's involvement with other identity providers (\emph{mutual IdP involvement undetectability}, IM) or service providers (\emph{IdP-SP involvement undetectability}, ISM).
In credential-focused systems, this is natural: the identity provider issues a credential to the user without involving others, and it is not involved in service provisions.
Indeed, smart certificates, Identity Mixer and the Smartcard scheme all satisfy IM and ISM.

In the linking service model, ISM does not hold because there is direct communication between the identity providers and the service provider.
In a variant of the model~\cite{Chadwick2009AttributeAggregationin}, the identity providers and service provider communicate indirectly via the linking service.
However, here the identity providers encrypt the attributes for the service provider (to preserve privacy with respect to the linking service), and so still need to know its identity.
To prevent this, some kind of trusted intermediary (like the smartcard in the Smartcard scheme) seems to be necessary.

Moreover, the linking service model does not satisfy IM.
The subscription provider learns from the authentication assertion that the user has an account at the address provider (but not the other way round).
This problem is also mentioned in \cite{Chadwick2009AttributeAggregationin}: while ``multiple [identity providers] must give [a service provider] the aggregated set of attributes without knowing about one another's involvement'', the authors concede that ``linked [identity providers] may become aware of just one other [identity provider] -- the authenticating [identity provider] -- during service provision''.
IM can be satisfied (within the standards used) if the subscription provider trusts the linking service to verify the address provider's signature.
Another possibility to satisfy the requirement may be to use group signatures~\cite{Chaum1991GroupSignatures} for the authentication assertion from the address provider.
This solution prevents the subscription provider from learning at which identity provider the user authenticated, but at the cost of reduced accountability. 

\subsubsection{Linkability requirements}
Finally, we discuss the results for the linkability requirements.
\emph{Session unlinkability (SL)} is a natural requirement for relationship-focused systems, because the identity provider generates a new signature over the attributes at every service provision.
Indeed, it holds for the linking service model.
It also holds for the credential-focused Identity Mixer system because rather than showing the credential (which would allow linking), the user just proves the validity of properties using ZK proofs.
In the Smartcard scheme, the smartcard is trusted to correctly send attributes from the credentials it knows.
In the smart certificates scheme, however, the complete credential is shown so the requirement is not satisfied.
\emph{IdP service access unlinkability (IL)}, in contrast, is natural if the identity provider is not involved in service provision, i.e.,~for the credential-focused smart certificates, Identity Mixer, and Smartcard schemes.
It is less natural for relationship-focused systems such as the linking service model.
In this case, private information retrieval~\cite{Chor1995PrivateInformationRetrieval} can be used so that at least the non-authenticating identity provider does not learn which user he is providing attributes of.

To achieve \emph{IdP profile unlinkability (IIL)}, global identifiers should be avoided in credential-focused as well as relationship-focused systems.
Smart certificates, being based on the user's public key certificate, do not satisfy this requirement.
In Identity Mixer, IIL holds because the identity providers do not learn the identifiers of the credentials they issue.
In the Smartcard scheme, it holds because each identity provider learns a different identifier based on a secret known only by the smartcard.
In the linking service model, the authenticating identity provider generates a session identifier and includes it in the authentication assertion sent to the other identity provider.
This forwarding of the assertion can be avoided if identity providers trust the linking service to verify the authentication assertion:
identity providers can then issue attributes under different session identifiers, and the linking service can assert the link between them.
However, this only partially solves the problem: identity providers are still both involved in service provision, so they may link using timing information.
Indeed, just eliminating global identifiers does not fix IIL in our model.

\emph{IdP-SP unlinkability (ISL)} does not hold for the same two systems that also do not satisfy IIL, and for similar reasons.
In smart certificates, all parties learn the user's public key certificate; in the linking service model, the service provider learns the session identifier from the authenticating identity provider.
The other systems satisfy it: in Identity Mixer, not even the issuer of the credential can recognise a ZK proof about it; in the Smartcard scheme, the smartcard ensures that the information flow between identity providers and service providers is restricted to just the attributes.

However, as a consequence of ISL holding, extra work is needed to achieve accountability in two respects.
First, a message encrypted to a trusted third party is provided to the service provider to achieve anonymity revocation.
Second, although service providers do not learn a credential identifier, they do need assurance that the credential has not been revoked.
In the Smartcard scheme, the suggested solution is to let the smartcard perform a regular revocation check.
Similarly, in the Identity Mixer system, credentials can be given a short lifetime and be checked for revocation at re-issuing \cite{Camenisch2009AccumulatorBasedBilinear}.
In both cases, revocation is not immediate.

For Identity Mixer, two proposals for immediate revocation have been done \cite{Camenisch2006GeneralCertificationFramework}.
The first proposal is to include a serial number in the credential.
The credential can be issued so that either the identity provider learns this serial number or not.
The former case makes ISL not satisfied.
In the latter case, ISL holds but the credential cannot be revoked if the user loses her serial number or does not wish to participate.
Depending on the situation at hand, this latter behaviour may not be acceptable.
The second proposal is to use a ZK proof that the credential is on a public list of valid credentials~\cite{Camenisch2009AccumulatorBasedBilinear}.
This allows revocation without the user's help while not breaking ISL; however, the user needs to keep track of all revoked credentials in the system, and despite recent advances \cite{Camenisch2009AccumulatorBasedBilinear} this may still not be efficient enough.
Note that the Smartcard scheme does not support immediate revocation at all.

\section{Discussion}\label{sec:discussion}

In this section, we discuss several applicability aspects of our analysis framework: what privacy requirements can be verified, how the scenario should be defined, and what systems can be modelled.
We also discuss possible generalisations, and effort needed to analyse a new system.

\paragraph{Privacy Requirements} Our framework can be used to verify any data minimisation requirement expressible in terms of the elementary detectability, linkability, and involvement requirements described in Section~\ref{subsec:model-reqs}.
Although the case study demonstrates that this includes many relevant requirements proposed in the literature, there are also privacy aspects that our model does not capture.
Most significantly, we allow only limited reasoning (via attribute properties) about the meaning of pieces of personal information other than identifiers.
For instance, we do not allow a piece of information to be inferred from several others, e.g. ``address'' follows from ``street name'' and ``house number''.
Also, we do not consider (probabilistic) links due to combinations of non-identifying attributes, e.g., matching name and post code from two profiles imply a link with high probability.
This choice reflects the goal of our approach, namely to compare the relative privacy of different systems (that differ in what identifiers are used and how).
On the other hand, to obtain a full understanding of the privacy of users that does take such inferences into account, our approach can be complemented with orthogonal (e.g.,~probabilistic) methods (see Section~\ref{sec:related-work}).

Apart from explicitly transferred information, i.e., the user's attributes, we analyse one particular kind of implicitly transferred information; namely, involvement requirements.
However, other kinds may be of interest as well.
For instance, the number of transactions performed by a user may be privacy-sensitive, as may be the mere date and time of certain activities (see, e.g.,~privacy issues in smart metering systems \cite{Rial2011Privacy-preservingsmartmetering}).
Knowledge about numbers of transactions can be expressed in our model; date and time may be appended as ``tags'' to communication.

\paragraph{Scenario-Dependence}
Our analysis framework requires the specification of a scenario.
In particular, this scenario needs to be designed in such a way that all privacy properties to be verified can be phrased in terms of personal information occuring in the scenario.
It is straightforward to analyse variants of the scenario by modifying it, but this does involve some work in practice.
Our analysis framework and its implementation are designed to verify properties of particular elements in a particular trace; both need to be modified for other scenarios.
This task can be lightened by exploiting Prolog's programming features.
For instance, the scenario in our case study involves two traces of service provisions, which are almost the same; in our implementation of the model of the systems, both are generated by one Prolog predicate which takes the variable elements as input.
This approach can also be used to generate traces with more actors or protocols, and to generate lists of checks that need to be performed for a given privacy requirement.
Since the conclusions of an analysis depend on the scenario, it should be chosen carefully to capture all relevant privacy aspects.
We refer to~\cite{Veeningen2013SymbolicPrivacyAnalysis} for a symbolic extension of our framework which is independent from a particular scenario.

\paragraph{Adaptation and Generalisation} 
Our framework is designed to be general enough for the analysis of any system in which actors use communication protocols to exchange personal information. 
If the message format of the communication protocols in the system is available, then the main difficulty in modelling the system is to make sure the cryptographic primitives used in the protocols are accurately modelled.
Although the present work models several frequently-used primitives, the model may need to be adapted to reflect characteristics of the particular implementations used (e.g., digital signatures may be with message recovery instead of with appendix, meaning that the message can be derived from the signature~\cite{Menezes1996HandbookofApplied}); or new cryptographic primitives may need to be added.
Once this is done, modelling the actual protocols is usually a matter of industrious bookkeeping.

To give the reader a flavour of the effort needed to model new primitives, we draw upon our experiences in extending the basic formal model of \cite{Veeningen2011FormalPrivacyAnalysis} to perform the case study in this paper.
Some operations are easily expressible in terms of standard primitives.
For instance, for our purposes, commitments can be modelled as hashes because they satisfy the same inference rules.
When modelling primitives, it is helpful to look at existing formalisations, e.g. using deductive systems~\cite{Clarke1998Usingstatespace,Fiore2001ComputingSymbolicModels} or equational theories \cite{Abadi2001Mobilevaluesnew,Blanchet2008AutomatedVerificationof}: they can usually be translated to the three-layer model.
For instance, the formalisation of labelled encryption used in this work is based on \cite{Camenisch2010formalmodelof}.
Special attention should be paid to testing rules.
Deductive systems do not usually consider testing; 
equational theories can include rules, e.g., for signature verification (e.g.,~\cite{Delaune2009Verifyingprivacy-typeproperties}), which translate to testing rules in the three-layer model, but may include only those rules that were relevant to the analysis at hand.
Thus, to obtain a complete set of testing rules, one needs to take a lower-level look at the operation of the primitive.
In addition, note that existing formalisations (e.g.,~\cite{Camenisch2010formalmodelof}) may not explicitly model randomness in non-deterministic primitives;
however, in our model this is needed because we assume messages to be deterministic.

In some cases, no suitable existing formalisation of a cryptographic primitive may be available.
In such a case, the general (security) definition of the primitive (e.g.,~\cite{Cramer1997ModularDesignof} for ZK proofs) generally suffices for obtaining a description for the language $\mathcal{L}^c$.
However, different implementations of a primitive may give rise to different inference rules.
Thus, to obtain inference rules, one needs to consider the particular implementation used in the protocol under analysis.
In our experience, this is feasible.
Note that because we are only interested in privacy aspects of the primitives, usually some simplifications can be made.
See Appendix~\ref{appendix} for two examples: ZK proofs and anonymous credentials.

As mentioned in Section~\ref{sec:determining-views}, our model imposes several assumptions on the cryptographic primitives and operations modelled.
In particular, because we assume that differently-constructed messages cannot have coinciding contents, we cannot model some operations such as ``exclusive or'' (which satisfies that $x\oplus (x\oplus y)=y$).
Also, our visible failure assumption may cause an over-approximation the knowledge of actors: in our model, actors can draw conclusions from the fact that a cryptographic operation was applied successfully, in practice, this may not be possible.
These limitations may be overcome by generalising our model through its connection with static equivalence (see Section~\ref{sec:related-work}); we leave this as future work.

\section{Related Work}\label{sec:related-work}

We discuss related work on our privacy analysis framework ($\S$\ref{subsec:related-framework}), and on the identity management case study ($\S$\ref{subsec:related-idm}).

\subsection{Privacy Analysis Framework}\label{subsec:related-framework}

The analysis of privacy entails two orthogonal concerns: what information is leaked by how identifiers and other pieces of information are exchanged in communication protocols; and what inferences can be made from the information learned in this way.
The present work addresses the former concern, which we discuss first; afterwards, we briefly discuss the latter concern.

Formal analysis techniques have been applied to communication protocols for many years, mainly to verify security properties~
\cite{Abadi2001Mobilevaluesnew,Burrows1990logicofauthentication,Meadows2003Formalmethodscryptographic,Paulson1998inductiveapproachto}.
Most formal methods rely on two basic ideas: the Dolev-Yao attacker model and state exploration techniques.
In the Dolev-Yao attacker model, one considers communication messages using idealised cryptographic primitives, and an attacker who controls some or all communication channels between legitimate parties (meaning that he can insert and suppress messages at will, and fabricate messages based on his observations).
The reasoning that the attacker performs to fabricate messages can be described by deductive systems (e.g.,~\cite{Clarke1998Usingstatespace,Fiore2001ComputingSymbolicModels}) or equational theories (e.g.,~\cite{Abadi2001Mobilevaluesnew,Blanchet2008AutomatedVerificationof}).
State space exploration techniques assess the system security by analysing all possible evolutions of a given system in the presence of a Dolev-Yao attacker.
The requirements of a system are then verified by checking whether any of the states that can be reached by the system correspond to an attack (e.g., the attacker knows a secret, or has succeeded in impersonating a legitimate user).
Several process algebras \cite{Abadi2001Mobilevaluesnew,Boreale2001SymbolicTraceAnalysis,Milner1999Communicatingandmobile} provide machinery to perform state space exploration.
Other approaches have also been proposed, e.g., using induction \cite{Paulson1998inductiveapproachto}.

Recently, more and more work has focused on the use of these techniques for privacy properties, in application domains such as electronic toll collection~\cite{Dahl2011FormalAnalysisof}, e-voting~\cite{Delaune2009Verifyingprivacy-typeproperties,JannikDreier2011FormalTaxonomyof}, RFID systems~\cite{Bruso2010FormalVerificationof}, and Direct Anonymous Attestation~\cite{Smyth2011Formalanalysisof}.
These proposals express privacy in terms of ``experiments'': slightly different settings for the execution of the same protocol that should be indistinguishable to an attacker.
For instance, in electronic toll collection, an attacker should not be able to distinguish a setting in which a first car takes a left road and a second car takes a right road from a situation in which the first car takes the right road and the second car takes the left road.
Similarly, in Direct Anonymous Attestation, an attacker should not be able to distinguish a signature produced by one trusted platform module from a signature produced by another one.

Conceptually, our work differs from these existing formal methods in several ways.
We provide general definitions for detectability and associability that take into account different data subjects that may occur in a single protocol instance; conversely, existing works either provide specific definitions tailored to a particular setting or protocol~\cite{Dahl2011FormalAnalysisof,Delaune2009Verifyingprivacy-typeproperties,JannikDreier2011FormalTaxonomyof}, or only consider links between messages and their senders~\cite{Arapinis2010AnalysingUnlinkabilityand}.
Moreover, we explicitly model the knowledge of (coalitions of) legitimate actors in the system as needed for analysis of data minimisation, whereas existing methods focus on (malicious) outsiders.
Also, we consider knowledge in a particular scenario, whereas existing methods focus on a family of scenarios.
Although particular queries in our analysis framework could be translated to queries using these existing formal methods (e.g., using frameworks like~\cite{Chevalier2010Compilingandsecuring} to convert a trace to a set of actions by protocol roles), we expect that it is infeasible to design a completely automatic translation to queries that the tools available today are able to handle.
Conversely, our privacy analysis framework achieves practical privacy analysis and comes with an implementation.

At a technical level, however, there are similarities between our model for reasoning about knowledge and existing models.
Existing models reason about knowledge of an attacker about message contents.
Three popular definitions cover whether an attacker knows the contents of a given piece of information: weak secrecy~\cite{Blanchet2004AutomaticProofof}, resistance against guessing attacks~\cite{Corin2005AnalysingPasswordProtocol}, and strong secrecy~\cite{Blanchet2004AutomaticProofof}.
The weakest definition, weak secrecy, defines secrecy as non-derivability using a contents-layer deductive system; as shown in~\cite{Veeningen2011FormalPrivacyAnalysis}, this property holds exactly if no context-layer representation of the contents can be derived using our three-layer model.

The other two existing definitions strengthen the concept of weak secrecy by employing the notion of static equivalence~\cite{Abadi2001Mobilevaluesnew} of frames.
A frame captures the knowledge of an actor at a certain point in time.
Intuitively, two frames are statically equivalent if an actor cannot distinguish between the situations modelled by the two frames.
Resistance against guessing attacks~\cite{Corin2005AnalysingPasswordProtocol} of a frame models that an actor should not have any way to verify if a guess he makes about the contents of a particular piece of information is correct.
This is formalised by adding the actor's guess to the frame, and verifying that the situation in which the guess is correct is statically equivalent to the situation in which the guess is incorrect.
Intuitively, in our model, the contents $c$ of a piece of information is resistant to guessing attacks if and only if there is no context-layer item with contents $c$ that is known to be content equivalent to a guess with contents $c$.
This link can be made more precise, and can be used to generalise the approach presented in this paper (see~\cite{Veeningen2014} for details).
One strong point of static equivalence is that it can be formally linked to computational models of cryptography~\cite{Baudet2010Guessingattacksand}; compared to the equational theory of~\cite{Baudet2010Guessingattacksand}, our visible failure assumption on deterministic symmetric encryption is an over-approximation of knowledge.

Strong secrecy \cite{Blanchet2004AutomaticProofof} additionally takes into account that the secret may have the same contents as any arbitrary other message, as well as that the value of the secret may influence the behaviour of actors.
Our model (as well as the definition of resistance against guessing attacks) does not take these aspects into account, so strong secrecy is, formally speaking, stronger.
Strong and weak secrecy are known to coincide~\cite{Cortier2006RelatingTwoStandard} under certain conditions in a certain equational theory; an interesting direction for future work is to verify if similar results hold for the equational theories corresponding to our model.
We remark that in practice, tools verify an over-approximation of strong secrecy~\cite{Blanchet2008AutomatedVerificationof} and hence may give false positives.

Similarly, existing notions of linkability~\cite{Arapinis2010AnalysingUnlinkabilityand,Dahl2011FormalAnalysisof,Delaune2009Verifyingprivacy-typeproperties,JannikDreier2011FormalTaxonomyof} are formally based on static equivalences.
For instance, in the electronic toll collection example given above, consider any frame corresponding to a system evolution in which a first car with identifier $A$ goes left and a second car with identifier $B$ goes right.
Unlinkability means that this frame should be statically equivalent to a frame corresponding to a system evolution when the first car has identifier $B$ and the second car has identifier $A$.
In many cases, corresponding frames differ only by the use of identifiers, in which case static equivalence corresponds to the non-knowledge of content equivalence of these identifiers, like our definition of associability.
However, linkability also allows other correspondences and takes into account that the value of the identifier may influence the behaviour of actors, and is thus, formally speaking, more powerful.
Also in this case, existing tools over-approximate linkability~\cite{Blanchet2008AutomatedVerificationof}; in practice, it is difficult to avoid false positives.

There are also technical similarities between our model of particular cryptographic primitives and other models from the literature.
Labelled encryption is a straightforward extension of normal encryption; our model is similar to the one in \cite{Camenisch2010formalmodelof}.
The internals of (incorrect) protocols for authenticated key agreement have over the years proven a popular target for analysis using formal methods \cite{Burrows1990logicofauthentication,Lowe1996BreakingandFixing,Paulson1998inductiveapproachto}; however, we have not found prior works that formally model the external behaviour of (correct) authenticated key exchange protocols in a larger system.

For ZK proofs, both high-level and low-level formalisations exist.
In \cite{Li2009VerifyingAnonymousCredential}, a low-level model of the operation of ZK proofs is given; however, it cannot be used for knowledge derivation; also, questions have been raised about its technical correctness \cite{Camenisch2010formalmodelof}.
Two high-level formalisations of ZK proofs have been proposed~\cite{Backes2008Zero-KnowledgeinApplied,Camenisch2010formalmodelof} that, as ours, allow proofs of a restricted set of properties.
The equational theory in \cite{Backes2008Zero-KnowledgeinApplied} models the verification of ZK proofs (as our testing rules); the model of \cite{Camenisch2010formalmodelof} only allows correct ZK proofs to take place and does not express their verification.
The latter simplification is not suitable for our method, because verification expresses that an actor learns information in new contexts.
Note that both model ``signature proofs of knowledge'' rather than $\Sigma$-proofs; however, our methods can also capture that variant.

Three recent proposals~\cite{Camenisch2010formalmodelof,Li2009VerifyingAnonymousCredential,Smyth2011Formalanalysisof} are relevant for our formal model of anonymous credentials.
\cite{Li2009VerifyingAnonymousCredential} only considers operational aspects of anonymous credentials. 
\cite{Camenisch2010formalmodelof} models credentials and their showing protocol.
The model of credentials is similar to ours, and it includes a rule to obtain a credential from a committed message as in our low-level formalisation (Appendix~\ref{subsec:app-anoncred}).
The showing protocol is formalised in terms of ZK proofs.
However, credential issuing is not considered in \cite{Camenisch2010formalmodelof}.
Finally, Smyth et al.~\cite{Smyth2011Formalanalysisof} model joining and signing protocols for ECC-based Direct Anonymous Attestation, which are very similar to issuing and showing protocols for BM-CL-based anonymous credentials \cite{Camenisch2004SignatureSchemesand}.
Although our model is based on a different signature scheme \cite{Camenisch2003signatureschemewith} and specified at a higher level, their model of signatures generally corresponds to our model of signatures from committed messages in Appendix~\ref{subsec:app-anoncred}.

Apart from the concern of learning information leaked by communication protocols, the orthogonal concern of inferences made on learned information has also received substantial attention.
In particular, the inference of links based on non-identifiers has been approached from two directions: experimentally linking given data, or theoretically guaranteeing that such linking is impossible.
Methods to link data from two databases using non-identifiers have been investigated since the seminal paper of Fellegi and Sunter~\cite{fellegi}; see \cite{kopcke} for a recent comparative study of available implementations.
Data from more than two sources can be grouped together based on pairwise decisions using domain-dependent \cite{bhattacharya,domingos,meray,sapena} or domain-independent \cite{Bilenko2005AdaptiveProductNormalization:,Chaudhuri2005RobustIdentificationof} algorithms, or statistical techniques \cite{sadinle}.
On the other hands, statistical frameworks to guarantee that linking personal information in a disclosure to other data is impossible (i.e.,~anonymity) include k-anonymity \cite{CirianiVFS07a}, $\ell$-diversity \cite{Machanavajjhala:2007}, t-closeness \cite{LiLV07} and differential privacy \cite{Dwork06}.
Koot \cite{koot12} reports on experiments in which the actual degree of anonymity of particular disclosures is computed.
Inferences of attribute values based on other attributes is covered in~\cite{Pontes2012BewareofWhat}.
Our approach can be complemented with these techniques to obtain a full understanding of privacy leakage due to communication.

\subsection{Privacy in Identity Management}\label{subsec:related-idm}

The relevance of privacy by data minimisation in the identity management setting is well-established in the literature.
It has been recognised as a basic ``law of identity'' for the design of IdM systems \cite{Cameron2006LawsofIdentity}.
Hansen et al. \cite{Hansen200435} argue that privacy-enhancing IdM systems should satisfy a high level of data minimisation with user-controlled linkage of personal data, and by default unlinkability of different user actions.
Pfitzmann and Hansen \cite{Pfitzmann2009terminologytalkingabout} define privacy-enhancing identity management as preserving the unlinkability between user profiles.
Finally, in a general survey, Alp\'ar et al.~\cite{Alpar2011IdentityCrisis} identify three main privacy issues in identity management: linkability across domains, identity providers knowing user transactions, and violation of proportionality and subsidiarity (i.e., the exchange of minimal information needed for a certain goal).
These three issues correspond to our three kinds of privacy requirements: linkability, involvement and detectability, respectively.
In contrast to the vision of minimising actor knowledge, Landau and Moore argue that preventing service providers from collecting transaction data may not be desirable because it prevents the adoption of IdM systems in practice \cite{Landau2011EconomicTusslesin}.
This falls into a broader discussion on incentives of participants in IdM systems \cite{Anderson2011CanWeFix,Camp2010IdentityManagement'sMisaligned} that is out of scope for this work.

This work aims to improve the way privacy by data minimisation is assessed compared to existing comparisons~\cite{EU2003IdentityManagementSystems,Hoepman2008ComparingIdentityManagement}.
Both comparisons of IdM systems that we are aware of consider data minimisation as one aspect of a much more general comparison of IdM systems.
Data minimisation requirements are specified in a high-level way, and verified manually by inspecting the user interface and documentation of the systems.
For instance, \cite{EU2003IdentityManagementSystems} considers three different criteria: ``usage of pseudonyms/anonymity''; ``usage of different pseudonyms'' and ``user [is] only asked for needed data'' (judged on a yes/no scale).
\cite{Hoepman2008ComparingIdentityManagement} considers two: ``directed identity''/``pseudonymous/anonymous use'' and ``minimal disclosure'' (judged on a ++ to {-}{-} scale).
To improve the objectivity and accuracy of such assessments, scores for such criteria may instead be obtained by aggregating formal analysis results like ours.
To obtain a better understanding of privacy differences, these formal results can then be analysed as in Section~\ref{subsec:analysis-results}.
However, note that our method can only be used to assess data minimisation requirements \emph{given} what information should be exchanged; to verify if this exchange of information is really needed, or consented to by the user, other methods (e.g.,~\cite{Compagna2009Howtointegrate}) should be used.
Some other aspects of the privacy assessment in \cite{EU2003IdentityManagementSystems,Hoepman2008ComparingIdentityManagement} seem less suitable for formal verification, e.g. the user-friendliness and the use of standards in the systems.

Some formal works on privacy in identity management are available.
In \cite{Pfitzmann2009terminologytalkingabout}, privacy-enhancing identity management is defined as preserving unlinkability between different user profiles, and the meaning of linkability and its relationship with related concepts is explored in a semi-formal way.
Their informal definitions formed the basis of our original work \cite{Veeningen2010Modelingidentity} on representing knowledge of personal information.
Other formal work on identity management has mainly focused on safety properties with respect to misbehaving attackers, rather than privacy properties with respect to insiders who follow the protocol specification.
In this context, unlinkability \cite{Li2009VerifyingAnonymousCredential,Suriadi2010Strengtheningandformally} and undetectability \cite{Camenisch2010formalmodelof} properties have been considered for Identity Mixer and related anonymous credential schemes.
For SAML \cite{SAML20}, a standard for the exchange of identity information between identity and service providers used in the linking service model, secrecy properties have been considered \cite{Armando2008FormalAnalysisof}.
Our work differs from this latter category in two respects:
first, we define properties in a general setting, allowing comparisons between different systems;
and second, we distinguish between the roles of different insiders rather than considering one outsider, enabling us to express which (coalitions of) actors can associate or detect certain information, and which cannot.

In this work, we focus on minimising knowledge of personal information by technical means; other works address other aspects of privacy.
Landau et al.~\cite{Landau2009AchievingPrivacyin} argue that privacy protection can be achieved not just technically, but also by legal and policy means.
Hansen et al.~\cite{Hansen200435} argue that apart from ensuring data minimisation, privacy-enhancing IdM systems should also make the user aware of what information is exchanged about her and who can link it; and allow the user to control these aspects.
Bhargav-Spantzel et al. \cite{Bhargav-Spantzel2007PrivacyRequirementsin} stress the importance of trust between different parties in identity management, and in particular, trust of the user in other parties' handling of her personal information.
Our method can complement this demand for transparency by providing a precise view on how the choice of IdM system impacts privacy.
However, interestingly, recent research in behavioural economics suggests that offering transparency to users might actually reduce their privacy by inducing them to release more information \cite{Brandimarte2010MisplacedConfidences:Privacy}.

\section{Conclusions \& Future Work}\label{sec:conclusion}

In this work, we have presented a general formal framework to compare communication protocols with respect to privacy by data minimisation.
Requirements relevant in a given setting are formalised independently from any particular communication protocol in terms of the knowledge of (coalitions of) actors in a three-layer model of personal information.
These requirements are then verified automatically for particular protocols by computing this knowledge from a description of their communication.
Using this formal approach, we obtain results that are precise and verifiable, yet provide enough detail to obtain real insight into privacy differences.
In contrast to existing methods, our framework allows for the automated verification of a wide range of privacy requirements in one single model.

Our framework may be generalised and extended along several directions.
First, the model of personal information can be made more expressive.
For instance, to analyse privacy in application domains where the number and timing of transactions is relevant, the model can be extended to take these aspects into account.
Other relevant extensions include pieces of information that refer to multiple data subjects (see~\cite{Veeningen2013SymbolicPrivacyAnalysis}); or more flexible reasoning about attribute properties.
Second, the model of cryptographic primitives can be made more general.
Our current model is based on two assumptions (structural equivalence and visible failure) that limit the number of cryptographic primitives that can be modelled.
We are exploring how these limitations can be overcome by modelling cryptographic primitives using an equational theory.
Finally, our model depends on the choice of a particular scenario in which requirements are verified; we refer to~\cite{Veeningen2013SymbolicPrivacyAnalysis} for a generalisation of our model that is independent from a particular scenario.

We have demonstrated our framework by performing a privacy comparison of identity management systems.
In the process, we have defined a comprehensive and detailed set of privacy requirements; to the best of our knowledge, no such set of requirements was available before.
We have modelled 4 representative IdM systems, and verified which of the 11 requirements hold for which systems, giving 44 checks in total.
It is worth noting that only 17 of the 44 checks are mentioned as (parts of) requirements in the design of the respective IdM systems.
In one instance, we found such a requirement not to hold (a problem which  is also mentioned by the authors of the system themselves).
In another instance, we clarified the exact setting in which a requirement holds, which may a solution that is unrealistic for performance or accountability reasons.
The remaining 27 of the 44 checks do not correspond to requirements explicitly stated by the designers of the IdM systems.
In this work, we have established whether they hold or not, leading to a more comprehensive analysis and comparison  of IdM systems. 
Interesting extensions to the case study would be to consider requirements for IdM systems based on the extensions mentioned above (e.g.~requirements on knowledge about the number of transactions); and additional IdM systems like U-Prove \cite{Paquin2010U-ProveCTPWhite} and the STORK Platform (\url{https://www.eid-stork.eu/}) as well as other variants of the systems we considered.

\paragraph{Acknowledgements}
We thank the anonymous reviewers for their useful comments.
We thank Berry Schoenmakers for useful technical feedback.
This work is partially supported by STW through project 'Identity Management on Mobile Devices' (10522).

\bibliographystyle{spmpsci}
{\small\bibliography{references}}

\appendix

\section{Trace Validity}\label{sec:trace-validity}

In this appendix, we introduce ``trace validity'' as a way of verifying that all knowledge required for a trace has been modelled.
Our framework takes as input a trace, together with the initial knowledge of the actors.
However, there are no guarantees that the trace and initial knowledge provided by the analyst are correctly specified.
This is fundamental for the analysis, 
because the initial knowledge also determines whether an actor can link the information he has observed to information he already has.
The concept of ``trace validity'' checks whether the initial knowledge and trace correspond to a valid scenario (i.e.,~a scenario in that can actually occur), and hence serves as a ``sanity check'' for the model.

To define trace validity, we need to model whether a context item has occurred in communication before.
When an actor $a$ initiates a protocol instance $\pi$ in state $\{\mathcal{C}_x\}_{x\in\mathcal{A}}$, no communication in the protocol instance has taken place yet, so the state does not contain context items with domain $\pi$.
Hence, to check whether $a$ can send message $\mathsf{m}\ctxm{}{\pi}{}$, we cannot just verify if $\mathcal{C}_a\vdash\mathsf{m}\ctxm{}{\pi}{}$.
Instead, we need to model that the actor ``instantiates'' the context items in $\mathsf{m}\ctxm{}{\pi}{}$ by items from other domains.
On the other hand, if actor $b$ wants to reply to message $\mathsf{m}\ctxm{}{\pi}{}$, then he no longer has this freedom to instantiate context items because contents of the context items from $\mathsf{m}\ctxm{}{\pi}{}$ he uses in his reply should corresponds to their contents in $\mathsf{m}\ctxm{}{\pi}{}$ itself.
In the former case, we call the context items \emph{undetermined}; in the latter case, we call them \emph{determined}:

\begin{definition}
 Let $\{\mathcal{C}_x\}_{x\in\mathcal{A}}$ be a state.
 We say that $\mathsf{p}\in\mathsf{P}^c$ is \emph{determined} in $\{\mathcal{C}_x\}_{x\in\mathcal{A}}$ if, for some $a\in\mathcal{A}$ and $\mathsf{m}\in\mathcal{C}_a$, $\mathsf{p}$ occurs in $\mathsf{m}$; or if $\mathsf{p}$ is a property $\psi_i(\mathsf{q})$ of some $\mathsf{q}$ occurring in $\mathsf{m}$.
 Otherwise, $\mathsf{p}$ is \emph{undetermined}.
\end{definition}

We now formalise when an actor has sufficient knowledge in a certain state to send a certain message $\mathsf{m}\ctxm{}{\pi}{}$.
The actor can instantiate any undetermined items in $\mathsf{m}\ctxm{}{\pi}{}$, but needs to respect the existing instantiation of determined items in $\mathsf{m}\ctxm{}{\pi}{}$.
We capture this by requiring that the actor can derive a message $\mathsf{n}$ that is equal to $\mathsf{m}$, except that undetermined items are replaced by items of his choice.
Intuitively, the actor having sufficient knowledge to send $\mathsf{m}\ctxm{}{\pi}{}$ means that, when the message $\mathsf{m}$ is added to his knowledge base, he does not gain any new knowledge from this.
For instance, if the actor can associate personal information from message $\mathsf{m}\ctxm{}{\pi}{}$ to information in his knowledge base, then he should be able to make the same associations using the corresponding item in $\mathsf{n}$.
The restrictions on $\mathsf{n}$ in the definition below guarantee that this is indeed the case:

\begin{definition}\label{def:determinability}
 Let $\{\mathcal{C}_x\}_{x\in\mathcal{A}}$ be a state, and $a\in\mathcal{A}$ an actor.
 Context message $\mathsf{m}$ is \emph{determinable} by $a$ in $\{\mathcal{C}_x\}_{x\in\mathcal{A}}$ if there exists a context message $\mathsf{n}\equiv\mathsf{m}$ such that $\mathcal{C}_a\vdash\mathsf{n}$, and the following conditions hold:
\begin{enumerate}
 \item Whenever $\mathsf{m}@z$ is determined, then $\mathsf{m}@z=\mathsf{n}@z$;
 \item Whenever $\mathsf{m}@z_1=\mathsf{m}@z_2$, then $\mathsf{n}@z_1=\mathsf{n}@z_2$;
 \item If $\mathsf{m}@z=\ctxm{d}{\kappa}{k}$ ($k\ne\cdot$) and some $\ctxm{e}{\eta}{k}\in\mathsf{I}^c\cup\mathsf{D}^c$ is determined, then 
$\linkable{\mathsf{n}@z}{\ctxm{e}{\eta}{k}}{a}$;
\item If $\mathsf{m}@z_1=\ctxm{d}{\kappa}{k}$, $\mathsf{m}@z_2=\ctxm{d'}{\kappa}{k}$ ($k\ne\cdot$), and no $\ctxm{e}{\eta}{k}\in\mathsf{I}^c\cup\mathsf{D}^c$ is determined, then $\linkable{\mathsf{n}@z_1}{\mathsf{n}@z_2}{a}$. 
\end{enumerate}

\end{definition}
Condition 1 states that the actor cannot replace determined items; condition 2 states that he should replace items consistently.
Conditions 3 and 4 make sure that actors cannot learn new associations by using $\mathsf{n}$ as $\mathsf{m}$: condition 3 applies to contexts already used in previous communication, and condition 4 applies to previously unused contexts.
For determined messages, determinability and detectability coincide.

The following example demonstrates determinability:

\begin{example}\label{exa:determinability}
 Consider the state $\{\mathcal{C}^0_x\}_{x\in\mathcal{A}}$ from Example~\ref{exa:trace}.
 The client's message $\mathsf{m}=E'_{\ctxm{shkey}{}{\cdot}}(\ctxm{id}{}{su})\ctxm{}{\pi}{}$ is determinable by $cli$ in this state.
 Namely, take $\mathsf{n}=E'_{\ctxm{skey}{\cdot}{\cdot}}(\ctxm{id}{ab}{4})$.
 Then $\mathsf{m}\equiv\mathsf{n}$, and this message trivially satisfies conditions 1--4 of the definition.

 Also, the server's reply to this message is determinable.
 Namely, consider the state $\{\mathcal{C}^1_x\}_{x\in\mathcal{A}}$ that $\{\mathcal{C}^0_x\}_{x\in\mathcal{A}}$ evolves into.
 The server's knowledge base is 
$$\mathcal{C}^1_{srv}=\mathcal{C}^0_{srv}\cup \{\ctxm{ip}{\cdot}{cli}, \ctxm{ip}{\cdot}{srv},E'_{\ctxm{shkey}{}{\cdot}}(\ctxm{id}{}{su})\ctxm{}{\pi}{}\},$$
and the server's reply is
$$\mathsf{m}=E'_{\ctxm{shkey}{}{\cdot}}(\{\ctxm{age}{}{su},\ctxm{n}{}{\cdot},S_{\ctxm{k^-}{}{srv}}(\{\ctxm{age}{}{su},\ctxm{n}{}{\cdot}\})\})\ctxm{}{\pi}{}.$$
 Indeed, one can verify that
$$\mathsf{n}=E'_{\ctxm{shkey}{\pi}{\cdot}}(\{\ctxm{col1}{db}{1},\ctxm{n}{\cdot}{\cdot},S_{\ctxm{k^-}{\pi}{srv}}(\{\ctxm{col1}{db}{1},\ctxm{n}{\cdot}{\cdot}\})\})$$
satisfies the conditions from the above definition.
Namely, no determined items from $\mathsf{m}$ have been replaced in $\mathsf{n}$ (condition 1); both occurrences of $\ctxm{age}{\pi}{su}$ have been replaced by the same item, and similarly for $\ctxm{n}{\pi}{\cdot}$ (condition 2); and $\ctxm{col1}{db}{1}\leftrightarrow_{srv}\ctxm{id}{\pi}{su}$, i.e., the message contains only associations known by $srv$ (condition 3).
Condition 4 holds trivially because there are no two context items satisfying the given condition.
\qed
\end{example}

Trace validity is defined step-by-step from the validity of its message transmissions.
A message transmission consists of identifiers $\mathsf{a},\mathsf{b}$ of the communication parties and communicated message $\mathsf{m}$.
For validity, we require determinability both of the message, and of the communication identifiers.
This way, we check that both the knowledge required to send the message, and the knowledge of where to send the message to, have been modelled.
Formally, for a basic message transmission $\mathsf{a}\to\mathsf{b}:\mathsf{m}$, this means determinability by the sender of the context message $\{\mathsf{m},\mathsf{a},\mathsf{b}\}$.
For the other two types of the form $\transmission{\mathsf{a}}{\mathsf{b}}{\mathsf{m}}$ modelling cryptographic protocols, both actors contribute information: the initiator of the protocol should determine the sender and receiver addresses $\mathsf{a}$, $\mathsf{b}$, and both parties contribute parts of $\mathsf{m}$:

\begin{definition}\label{def:validity}
\begin{table}[tb]
\small\hspace{-0.7cm}\begin{tabular}[b]{l|ll}
  $\mathfrak{t}=$ & \textbf{Determinable by $a$} & \textbf{Determinable by $b$} \tabularnewline\hline
  $~\mathsf{a}\to\mathsf{b}:\mathsf{m}$&$\{\mathsf{a},\mathsf{b},\mathsf{m}\}$&$\emptyset$ \tabularnewline
  $~\transmission{\mathsf{a}}{\mathsf{b}}{\textrm{ZK}(\mathsf{m}_1;\mathsf{m}_2;\mathsf{m}_3;\{\mathsf{n}_a,\mathsf{n}_b\})}$&
        $\{\mathsf{a},\mathsf{b},\mathsf{m}_1,\mathsf{n}_a\}$&$\mathsf{n}_b$\tabularnewline
  $~\transmission{\mathsf{a}}{\mathsf{b}}{\mbox{ICred}_{\mathsf{m}_2}^{\mathsf{m}_1}(\mathsf{m}_3;\{\mathsf{n}_j\}_{j=1}^7)}$ &
        $\{\mathsf{a},\mathsf{b},\mathsf{pk}(\mathsf{m}_2),\mathsf{m}_1,\mathsf{n}_1,\mathsf{n}_2,\mathsf{n}_3,\mathsf{n}_7\}$&$\{\mathsf{pk}(\mathsf{m}_2),\mathsf{m}_2,\mathsf{m}_3,\mathsf{n}_4,\mathsf{n}_5,\mathsf{n}_6\}$
 \end{tabular}
 \caption{\label{tbl:transmission-validity}Determinability requirements for the different types of message transmissions}
\end{table}
Let $\{\mathcal{C}_x\}_{x\in\mathcal{A}}$ be a state, and $\mathfrak{t}$ a message transmission.
Let $\mathfrak{t}=\mathsf{a}\to\mathsf{b}:\mathsf{m}$ or $\transmission{\mathsf{a}}{\mathsf{b}}{\mathsf{m}}$, and let $a,b\in\mathcal{A}$ be the actors such that $a\leftrightarrow\sigma(\mathsf{a})$, $b\leftrightarrow\sigma(\mathsf{b})$.
We say that $\mathfrak{t}$ is \emph{valid} in $\{\mathcal{C}_x\}_{x\in\mathcal{A}}$ if the messages indicated in Table~\ref{tbl:transmission-validity} are determinable by $a$ and $b$, respectively.
Trace $\mathfrak{t}_1;\cdots;\mathfrak{t}_k$ is \emph{valid} in state $\{\mathcal{C}^0_x\}_{x\in\mathcal{A}}$ if, in the evolution
$$\{\mathcal{C}^0_x\}_{x\in\mathcal{A}} \stackrel{\mathfrak{t}_1}{\rightarrow} \{\mathcal{C}^1_x\}_{x\in\mathcal{A}} \stackrel{\mathfrak{t}_2}{\rightarrow} \cdots \stackrel{\mathfrak{t}_n}{\rightarrow} \{\mathcal{C}^n_x\}_{x\in\mathcal{A}},$$
each message transmission $\mathfrak{t}_i$ is valid in respective state $\{\mathcal{C}^{i-1}_x\}_{x\in\mathcal{A}}$.
\end{definition}
For ZK proofs, the prover needs to know the private information for the proof, and both parties contribute randomness.
Note that to participate in the protocol, the verifier does not need to know the public information or the properties to be proven; however, he does need to know this information to be able to interpret the proof (i.e., to apply the testing rule).
For credential issuing, the user needs to know her secret identifier $\mathsf{m}_1$, randomness, and the issuer's public key; the issuer needs to know his private/public key pair, the attributes to be signed, and additional randomness.

The following example highlights validity of message transmissions and traces.
\begin{example}
Consider the trace given in Example~\ref{exa:trace}.
In Example~\ref{exa:determinability}, we showed determinability of the two messages transmitted in the trace; this argument can be easily extended to conclude determinability of the messages $\{\mathsf{a},\mathsf{b},\mathsf{m}\}$ from Definition~\ref{def:validity}, and hence validity of the two message transmissions.
We conclude that the trace is valid.\qed
\end{example}

Trace validity is implemented in the tool supporting our framework.
We briefly discuss the implementation.
The main task in implementing trace validity is to check for determinability of a message $\mathsf{m}$; that is, to find a derivable message $\mathsf{n}$ that is equivalent to $\mathsf{m}$ and satisfies properties (1) to (4) from Definition~\ref{def:determinability}.
Properties (1) and (2) place restrictions on the form of the message, which can be expressed in terms of free variables in a Prolog query to the deductive system.
For properties (3) and (4) we check associability as in Section~\ref{subsec:prolog}.

\section{Inference Rules for Zero-Knowledge Proofs and Credential Issuing}\label{appendix}

In this appendix, we show how our models of ZK proofs and the credential issuing protocol are derived.

\subsection{Zero-Knowledge Proofs}\label{subsec:app-zero-knowledge}

ZK proofs allow a prover to prove to a verifier that he knows some secret information satisfying certain properties with respect to some public information, without revealing any information about the secret.
For instance, consider a large group of prime order $n$ generated by a group element $g$.
Note that given value $h$, it is infeasible to determine the discrete logarithm $x=\log_g h$; this property can be exploited to build a public key cryptosystem in which values of $h$ are public keys, and the corresponding values of $x$ are private keys.
A \emph{prover} who knows $x$ as well as $n$, $g$, and $h$ can engage in a ZK proof protocol with a \emph{verifier} who just knows $n$, $g$, and $h$; when the protocol has finished successfully, the verifier is convinced that the prover knows the value of $x$, without learning anything about its value.

The general definition of ZK proofs leaves open different kinds of implementations; we model a particular kind of ZK proof called $\Sigma$-protocols \cite{Cramer1997ModularDesignof}.
$\Sigma$-protocols are \emph{three-move} protocols in which the prover first sends a \emph{commitment}; the verifier responds with a randomly generated \emph{challenge}; and finally the prover sends a \emph{response}.
The ZK proofs used in the systems analysed \cite{Bangerter2004CryptographicFrameworkControlled,Camenisch2003signatureschemewith,Camenisch2004SignatureSchemesand,Fujisaki1997StatisticalZeroKnowledge} are of this kind.

\begin{figure}[tb]
\small\centering
\subfigure[Schnorr proof of knowledge\label{fig:schnorr}]{
$$
\xymatrixrowsep{0in}
\xymatrix{
\txt{$\mathsf{Prover}$\\$(x=\log_g h)$}&&\txt{$\mathsf{Verifier}$\\$(h)$} \\
\txt{$u\in_R \mathbb{Z}_n$\\$a\leftarrow g^u$} && \\
\ar[rr]^{a}&& \\
&&c\in_R\mathbb{Z}_n \\
&&\ar[ll]_c \\
r\leftarrow u+cx \\
\ar[rr]^r && \\
&& g^r\stackrel{?}{=} ah^c
}
$$
}
\subfigure[Formal model of Schnorr proof\label{fig:schnorr-formal}]{
\parbox{\linewidth}{
$$\transmission{\mathsf{p}}{\mathsf{v}}{\textrm{ZK}(\mathsf{k}^-;\mathsf{pk}(\mathsf{k}^-);\emptyset;\{\mathsf{n}_p,\mathsf{n}_v\})}~~~~~~~~~~~~~~~~~~~~$$
$$\mbox{~~~~~~~~~~~~~~~~~~~~(where $\mathsf{k}^-=x$, $\mathsf{pk}(\mathsf{k}^-)=h$, $\mathsf{n}_p=u$, $\mathsf{n}_v=c$)}$$
}
}
\caption{Schnorr proof of knowledge and its formal model\label{fig:schnorrs}}
\end{figure}
An example $\Sigma$-protocol is the protocol proposed by Schnorr to prove knowledge of $x=\log_g h$ in the setting given above
(Figure~\ref{fig:schnorr}).
The prover computes a random $u$ and sends a commitment $g^u$ to the verifier.
The verifier responds with a random challenge $c$.
The prover calculates response $r=u+cx$.
The verifier convinces himself that the prover indeed knows the secret $x$ by checking that $g^r = ah^c$ using the response, commitment and public information.
The prover can only calculate a valid response if he knows the secret; also, the response does not reveal any information about $x$~\cite{Schnorr1989EfficientIdentificationand}.

We formally model ZK proofs at a high level using the primitive $\textrm{ZK}(\mathsf{m}_1;\mathsf{m}_2;\mathsf{m}_3;\mathsf{n})$.
The secret information $\mathsf{m}_1$ and public information $\mathsf{m}_2$ are described in terms of messages; the ZK proof proves that the public information has a certain message structure with respect to the secret information.
In addition, the proof can show that context data items $\mathsf{d}$ occurring in $\mathsf{m}_1$ satisfy properties $\psi_k(\mathsf{d})$, listed in $\mathsf{m}_3$.
Finally, $\mathsf{n}$ represents randomness; in $\Sigma$-protocols, $\mathsf{n}=\{\mathsf{n}_p,\mathsf{n}_v\}$, representing the provers' randomness   $\mathsf{n}_p$ for the commitment and the verifier's randomness  $\mathsf{n}_v$ for the challenge.
For instance, $\mathsf{ZK}(\mathsf{k}^-;\mathsf{pk}(\mathsf{k}^-);\emptyset;\{\mathsf{n}_p,\mathsf{n}_v\})$ is a proof of knowledge of the private key $\mathsf{k}^-$ corresponding to public key $\mathsf{pk}(\mathsf{k}^-)$ with no properties and contributed randomness $\mathsf{n}_p,\mathsf{n}_v$.
From this high-level description in terms of structure of messages, the low-level description follows implicitly.
For instance, in a setting where public/private key pairs are of the form $(h,x=\log_g h)$, the proof $\mathsf{ZK}(\mathsf{k}^-;\mathsf{pk}(\mathsf{k}^-);$ $\emptyset;\{\mathsf{n}_p,\mathsf{n}_v\})$ corresponds to a proof of knowledge of the discrete logarithm $x=\log_g h$ of $h$ like the Schnorr protocol.
Figure~\ref{fig:schnorrs} shows the Schnorr protocol and its formal model in this setting.

\begin{figure}[tb!]
\fbox{
\parbox{0.96\columnwidth}{
\begin{centering}
\small
~~~
$\axvdashc{EZ$_1$'}{\mathcal{C}_a\vdash\textrm{ZK}(\mathsf{m}_1;\mathsf{m}_2;\mathsf{m_3};\{\mathsf{n}_p{,}\mathsf{n}_v\})}{\mathcal{C}_a\vdash\mathsf{m}_3}{}$
~~~
$\axvdashc{EZ$_2$'}{\mathcal{C}_a\vdash\textrm{ZK}(\mathsf{m}_1;\mathsf{m}_2;\mathsf{m_3};\{\mathsf{n}_p{,}\mathsf{n}_v\})}{\mathcal{C}_a\vdash\mathsf{n}_v}{}$
~~~
$\axvdashc{EZ$_3$'}{\mathcal{C}_a\vdash\textrm{ZK}(\mathsf{m}_1;\mathsf{m}_2;\mathsf{m_3};\{\mathsf{n}_p{,}\mathsf{n}_v\})}{\mathcal{C}_a\vdash\mathsf{m}_2}{}$
~~~
$\axvdashc{EZ$_4$'}{\mathcal{C}_a\vdash\textrm{ZK}(\mathsf{m}_1;\mathsf{m}_2;\mathsf{m_3};\{\mathsf{n}_p{,}\mathsf{n}_v\}),\mathcal{C}_a\vdash\mathsf{n}_p}{\mathcal{C}_a\vdash\mathsf{m}_1}{}$
~~~
$\axvdashc{EZ$_5$'}{\mathcal{C}_a\vdash\textrm{ZK}(\mathsf{m}_1;\mathsf{m}_2;\mathsf{m_3};\{\mathsf{n}_p{,}\mathsf{n}_v\}),\mathcal{C}_a\vdash\mathsf{m}_1}{\mathcal{C}_a\vdash\mathsf{n}_p}{}$
~~~
$\axvdashc{TZ$_1$'}{\mathcal{C}_a\vdash\textrm{ZK}(\mathsf{m}_1;\mathsf{m}_2;\mathsf{m_3};\{\mathsf{n}_p{,}\mathsf{n}_v\}),\mathcal{C}_a\vdash\mathsf{n}'_p}{\mathcal{C}_a\vdash\mathsf{n}_p}{(\mathsf{n}'_p\doteq\mathsf{n}_p)~~}$
~~~
$\axvdashc{CZ'}{\mathcal{C}_a\vdash\{\mathsf{m}_1{,}...{,}\mathsf{m}_j{,}\mathsf{n}_1{,}...{,}\mathsf{n}_k{,}\mathsf{p}_1{,}...{,}\mathsf{p}_l{,}\mathsf{n}_p{,}\mathsf{n}_v\}}
              {\mathcal{C}_a\vdash\textrm{ZK}(\{\mathsf{m}_1{,}...{,}\mathsf{m}_j\};\{\mathsf{n}_1{,}...{,}\mathsf{n}_k\};\{\mathsf{p}_1,...\mathsf{p}_l\};\{\mathsf{n}_p;\mathsf{n}_v\})}
              {}$
\par\end{centering}
}}
\caption{
Complete set of inference and rules for ZK
($\mathcal{C}_a$ a set of context messages;
$\mathsf{m}_*$, $\mathsf{n}_*$ context messages;
$\mathsf{p}_i$ properties of $\mathsf{m}_k$, i.e., every $\mathsf{p}_i=\psi_j(\mathsf{m}_k)\in\mathsf{D}^c$ for some $j$, $k$)
)
\label{fig:rules-zk}}
\end{figure}
In Figure~\ref{fig:rules-zk}, we present a set of inference rules for the ZK primitive.
We first explain them, and then argue that for privacy purposes and under certain assumptions, it suffices to consider the smaller set of rules presented in Figure~\ref{fig:deductive-system}.
We first discuss what messages can be derived from a ZK transcript $\textrm{ZK}(\mathsf{m}_1;\mathsf{m}_2;\mathsf{m}_3;\{\mathsf{n}_p,\mathsf{n}_v\})$ using elimination and testing rules.
The property proven by a ZK proof determines the format of the messages in the ZK proof protocol.
Hence, we allow any actor to derive the properties $\mathsf{m}_3$ from the transcript $\avdashc{EZ$_1$'}$.
(Because different properties may have identically-looking ZK proof protocols, this is an over-approximation of knowledge.)
The verifier randomness $\mathsf{n}_v$ is transmitted as challenge, and so can be derived from the transcript $\avdashc{EZ$_2$'}$.
Because both parties are assumed to know $\mathsf{m}_2$ before the start of a ZK proof, it does not need to follow from the transcript.
However, depending on the protocol, it may be possible to derive $\mathsf{m}_2$.
E.g., in the Schnorr example, $h=(g^r a^{-1})^{-c}$.
Hence, as a possible over-approximation, we allow any observer to derive the public information $\mathsf{m}_2$ $\avdashc{EZ$_3$'}$.

The fact that the protocol is zero-knowledge means that a verifier (who knows $\mathsf{m}_2$, $\mathsf{m}_3$ and $\mathsf{n}_v$) should not be able to learn anything about $\mathsf{m}_1$.
In fact, if there are several possible secrets $\mathsf{m}_1$ corresponding to public information $\mathsf{m}_2$, then the probability distribution for protocol transcripts is required to be independent from $\mathsf{m}_1$.
Thus, it is impossible to test $\mathsf{m}_1$ from the transcript.
(Of course, if $\mathsf{m}_2$ determines $\mathsf{m}_1$, e.g., if they are a public/private key pair, then $\mathsf{m}_1$ can be derived using $\mathsf{m}_2$, but this is not due to the ZK proof.)
Because the verifier, who knows all components of the ZK proof except $\mathsf{m}_1$ and $\mathsf{n}_p$, cannot deduce anything about the secret $\mathsf{m}_1$, any inference rule to derive it needs to have $\mathsf{n}_p$ as a prerequisite.
By a similar line of reasoning, if $\mathsf{m}_1$ can be derived from $\mathsf{n}_p$, then an inference rule for $\mathsf{n}_p$ needs $\mathsf{m}_1$, or it needs to be a testing rule.
In fact, in the Schnorr proof, in $\Sigma$-protocols all these inferences can be made: $\mathsf{m}_1$ can be derived directly from $\mathsf{n}_p$ $\avdashc{EZ$_4$'}$ and vice versa $\avdashc{EZ$_5$'}$, and $\mathsf{n}_p$ can be tested $\avdashc{TZ$_2$'}$.

To generate a transcript $\textrm{ZK}(\mathsf{m}_1;\mathsf{m}_2;\mathsf{m}_3;\{\mathsf{n}_p,\mathsf{n}_v\})$ of a $\Sigma$-protocol, an actor needs $\mathsf{n}_p$ for the commitment; $\mathsf{n}_v$ for the challenge; and both pieces of randomness and the private information for the response $\mathsf{n}_p$ $\avdashc{CZ'}$.
(Technically, the public information is not needed.)
Similarly, for determinability of the message transmission $\transmission{\mathsf{a}}{\mathsf{b}}{\mbox{ZK}(\mathsf{m}_1{;}\mathsf{m}_2{;}\mathsf{m}_3{;}}$ $\{\mathsf{n}_p{,}\mathsf{n}_v\})$, the prover needs 
$\{\mathsf{m}_1,\mathsf{n}_p\}$ in addition to the communication addresses $\{\mathsf{a},\mathsf{b}\}$; the verifier needs $\mathsf{n}_v$.

There are two aspects the above model does not take into account.
First, from two ZK proofs using the same prover randomness, the secret can be derived: in case of the Schnorr proof, by computing $(r-r')/(c-c')$ from transcripts $(a,c,r)$ and $(a,c',r')$.
This is a general property of $\Sigma$-protocols called \emph{special soundness}.
However, if the prover always honestly generates his randomness, then this is very unlikely and we can safely ignore it.
Second, an actor can also ``simulate'' a ZK proof transcript without knowing the secret information by first generating the challenge and response and from that determining the commitment.
Such a simulation has the exact same form as a ZK proof, but because the randomness in the commitment is unknown, it cannot be used to derive a secret corresponding to the public information.
Such simulations are very unlikely to correspond to ZK proofs that really took place, so they are not relevant for knowledge analysis.

To express privacy requirements, the knowledge of randomness is not directly relevant.
In addition, assuming that the randomness of the ZK proof is freshly generated and not reused elsewhere, it is clear that it cannot help to derive information indirectly: $\avdashc{EZ$_4$'}$ is the only rule to derive personal information (namely, $\mathsf{m}_1$) using randomness, and it has knowledge of $\mathsf{n}_p$ as prerequisite, which can only be derived when $\mathsf{m}_1$ is already known.
Ignoring rules $\avdashc{EZ$_2$'}$, $\avdashc{EZ$_5$'}$, we obtain the inference rules given in Figure~\ref{fig:deductive-system}, and determinability requirements in Table~\ref{tbl:transmission-validity}.

\subsection{Anonymous Credentials and Issuing}\label{subsec:app-anoncred}

In an anonymous credential system, credentials $\textrm{cred}^{M_1}_{\mathsf{k}^-}(M_2;M_3)$ assert the link between a user's identifier $M_1$ and her attributes $M_2$ using secret key $\mathsf{k}^-$, and such credentials are issued and shown anonymously \cite{Camenisch2003signatureschemewith}.
Anonymous issuing means the issuer of the credential does not learn the user's identifier $M_1$ (in particular, this means he cannot issue credentials containing the identifier without the user's involvement).
We model the issuing protocol by the $\textrm{ICred}^{M_1}_{\mathsf{k}^-}(M_2;M'_3)$ primitive.
The randomness $M'_3$ used in the issuing protocol determines the randomness $M_3$ in the credential.
Anonymous showing means that it is possible to perform ZK proofs of ownership of a credential proving certain properties.
This is captured by our $\textrm{ZK}$ primitive.

\begin{figure}[tb]
\fbox{
\parbox{0.96\columnwidth}{
\small\centering
$\axvdashc{CS$^0$}{\mathcal{C}_a\vdash\mathsf{pk}(\mathsf{k}^-),\mathcal{C}_a\vdash\mathsf{m}_1,\mathcal{C}_a\vdash\mathsf{n}_a}{\mathcal{C}_a\vdash S^0_{\mathsf{k}^-}(\mathsf{m}_1,\mathsf{n}_a)}{}$
~~~
$\axvdashc{CS$^0$'}{\mathcal{C}_a\vdash S^0_{\mathsf{k}^-}(\mathsf{m}_1{,}\mathsf{n}_a),\mathcal{C}_a\vdash\{\mathsf{k}^-{,}\mathsf{m}_2{,}\mathsf{n}_b\}}{\mathcal{C}_a\vdash S_{\mathsf{k}^-}(\mathsf{m}_1,\mathsf{m}_2,\mathsf{n}_a,\mathsf{n}_b)}{}$
}}
\caption{Inference rules for signature scheme with signatures on committed values ($\mathcal{C}_a$ a set of context messages; $\mathsf{k}^-$, $\mathsf{m}_*$, $\mathsf{n}_*$ context messages)\label{fig:inference-sigcomm}}
\end{figure}

\begin{figure}[tb]
\small
\subfigure[Issuing protocol for anonymous credentials\label{fig:credential-issuing}]{
\parbox{\linewidth}{
\begin{eqnarray*}
 \mathsf{a}\rightarrow\mathsf{b}&:&S^0_{\mathsf{k}^-}(\mathsf{m}_1,\mathsf{n}_2); \\
 \transmissionP{\mathsf{a}}{\mathsf{b}}&:&\textrm{ZK}(\mathsf{m}_1,\mathsf{n}_1,\mathsf{n}_2;\mathsf{pk}(\mathsf{k}^-),\mathcal{H}(\mathsf{m}_1,\mathsf{n}_1),S^0_{\mathsf{k}^-}(\mathsf{m}_1,\mathsf{n}_2);\emptyset;\{\mathsf{n}_3,\mathsf{n}_4\}); \\
 \mathsf{b}\rightarrow\mathsf{a}&:&\{S_{\mathsf{k}^-}(\mathsf{m}_1,\mathsf{m}_2,\mathsf{n}_2,\mathsf{n}_5),\mathsf{n}_5\}; \\
 \transmissionP{\mathsf{b}}{\mathsf{a}}&:&\textrm{ZK}(\mathsf{k}^-;\mathsf{pk}(\mathsf{k}^-),S^0_{\mathsf{k}^-}(\mathsf{m}_1,\mathsf{n}_2),\mathsf{m}_2,\mathsf{n}_5,\\&&S_{\mathsf{k}^-}(\mathsf{m}_1,\mathsf{m}_2,\mathsf{n}_2,\mathsf{n}_5);\emptyset;\{\mathsf{n}_6,\mathsf{n}_7\})
\end{eqnarray*}
\centering Credential obtained: $\{S_{\mathsf{k}^-}(\mathsf{m}_1,\mathsf{m}_2,\mathsf{n}_2,\mathsf{n}_5),\mathsf{n}_2,\mathsf{n}_5\}$
}
}
\subfigure[Formal model of anonymous credential issuing protocol \label{fig:credential-issuing-formal}]{
\parbox{\linewidth}{
$$\transmission{\mathsf{a}}{\mathsf{b}}{}\mbox{ICred}_{\mathsf{k}^-}^{\mathsf{m}_1}(\mathsf{m}_2;\{\mathsf{n}_i\}_{i=1}^7)$$
\centering Credential obtained:
$\textrm{cred}^{\mathsf{m}_1}_{\mathsf{k}^-}(\mathsf{m}_2;\{\mathsf{n}_2,\mathsf{n}_5\})$
}
}

\caption{Anonymous credentials from signature scheme with signatures on committed values\label{fig:anoncred-from-sig}}
\end{figure}
We model anonymous credential systems constructed from signature schemes \cite{Camenisch2003signatureschemewith,Camenisch2004SignatureSchemesand} as used in the Identity Mixer system \cite{Bangerter2004CryptographicFrameworkControlled}.
In general, this construction is possible if the signature scheme allows for issuing of signatures on committed values (Figure~\ref{fig:inference-sigcomm}).
That is, a commitment $S^0_{\mathsf{k}^-}(\mathsf{m}_1,\mathsf{n}_a)$ to message $\mathsf{m}_1$ using randomness $\mathsf{n}_a$ is constructed using public key $\mathsf{pk}(\mathsf{k}^-)$ $\avdashc{CS$^0$}$;
this commitment is turned into signature $S_{\mathsf{k}^-}(\mathsf{m}_1,\mathsf{m}_2,\mathsf{n}_a,\mathsf{n}_b)$ using private key $\mathsf{k}^-$, message $\mathsf{m}_2$ and randomness $\mathsf{n}_b$,  $\avdashc{CS$^0$'}$.
Based on such a scheme, an anonymous credential $\textrm{cred}^{\mathsf{m}_1}_{\mathsf{k}^-}(\mathsf{m}_2;\{\mathsf{n}_a,\mathsf{n}_b\})$ is simply a randomised signature (containing secret identifier $\mathsf{m}_1$ and attributes $\mathsf{m}_2$) along with its used randomness.
In the Identity Mixer system, two such signature schemes can be used: SRSA-CL signatures \cite{Camenisch2003signatureschemewith} and BM-CL signatures \cite{Camenisch2004SignatureSchemesand}.
There are slight technical differences between the two; we discuss SRSA-CL signatures and briefly outline the differences later.

The anonymous credential issuing protocol can be modelled as a trace in terms of the signature scheme (Figure~\ref{fig:credential-issuing}).
It involves a user $\mathsf{a}$ and an issuer $\mathsf{b}$.
As before, $\mathsf{a}$ is assumed to have sent a commitment $\mathcal{H}(\mathsf{m}_1,\mathsf{n}_1)$ to her secret identifier to $\mathsf{b}$ prior to initiating the protocol.
(Unlike the commitment $S^0_{\mathsf{k}^-}(\mathsf{m}_1,\mathsf{n}_2)$ for the signature, $\mathcal{H}(\mathsf{m}_1,\mathsf{n}_1)$ does not depend on $\mathsf{k}^-$ and can thus be shared with other issuing or showing protocols for credentials having a different key.)
In the first two messages, actor $\mathsf{a}$ provides her commitment for the signature, and then proves that it is formed correctly; that is, it indeed contains the identifier corresponding to the one in $\mathcal{H}(\mathsf{m}_1,\mathsf{n}_1)$.
Actor $\mathsf{b}$ uses the commitment to construct a signature on $\{\mathsf{m}_1,\mathsf{m}_2,\mathsf{n}_2,\mathsf{n}_5\}$, and sends the signature along with his randomness to $\mathsf{a}$.
At this point, $\mathsf{a}$ knows the signature and the two pieces of randomness used in it: these three components together form the anonymous credential, as shown in the figure.
(Note that $\mathsf{b}$ does not know $\mathsf{n}_2$, so he does not have the complete credential.)
In the last step, the signer $\mathsf{b}$ proves that $S_{\mathsf{k}^-}(\mathsf{m}_1,\mathsf{m}_2,\mathsf{n}_2,\mathsf{n}_5)$ is valid; when using the SRSA-CL signature scheme, this step is technically needed to ensure the security of the signature~\cite{Bangerter2004CryptographicFrameworkControlled}.
Figure~\ref{fig:credential-issuing-formal} displays our high-level model of the issuing protocol and the credential obtained from it.

The high-level inference rules (Figure~\ref{fig:deductive-system}) and determinability relation (Table~\ref{tbl:transmission-validity}) for $\textrm{cred}$ and $\textrm{ICred}$ follow from the lower-level model in Figure~\ref{fig:credential-issuing}.
The credential's signature can be verified using messages $\{\mathsf{pk}(\mathsf{k}^-),\mathsf{m}_1,\mathsf{m}_2\}$, and a credential can be constructed from its components $\avdashc{CR}$.
Although randomness can be inferred from the credential, we do not model these inferences in the high-level model because they are not relevant for knowledge of personal information.

From the issuing protocol, the user can infer the credential using the randomness from the credential \avdashc{EI$_1$}.
We check the messages of the trace for further possible inferences.
For the two ZK proofs, $\avdashc{EZ$_1$}$ does not apply because there are no proofs of properties.
The $\avdashc{EZ$_2$}$ rule can be applied to both ZK proofs occurring in the issuing protocol; this translates to rules $\avdashc{EI$_2$}$ and $\avdashc{EI$_3$}$.
We also consider the derivation of the nonces $\mathsf{n}_1$, $\mathsf{n}_2$ $\avdashc{EI$_2$}$: $\mathsf{n}_1$ is generated outside of the issuing protocol, so its derivation may be of interest; $\mathsf{n}_2$ is a prerequisite for $\avdashc{EZ$_2$}$.
Rule $\avdashc{EZ$_3$}$ gives $\avdashc{EI$_4$}$.
We do not add a rule to derive $S^0_{\mathsf{k}^-}(\mathsf{m}_1,\mathsf{n}_2)$ from the transcript because its knowledge is not relevant from a privacy point of view.
Also, this message does not allow the derivation of any information that was not already derivable from the zero-knowledge proofs.
However, it does give testing rule $\avdashc{TI$_2$}$.
Testing rule rules $\avdashc{TI$_1$}$ and $\avdashc{TI$_3$}$ follow from the first message transmission.
The other testing rules $\avdashc{TI$_4$}$, $\avdashc{TI$_5$}$ follow from the corresponding testing rule $\avdashc{TZ$_1$}$ for zero-knowledge proofs.

Finally, consider $\mbox{ICred}_{\mathsf{k}^-}^{\mathsf{m}_1}(\mathsf{m}_2;\{\mathsf{n}_i\}_{i=1}^7)$'s determinability requirements.
Assuming fresh nonces, determinability of $\{\mathsf{a},\mathsf{b},\mathsf{pk}(\mathsf{k}^-),\mathsf{m}_1,\mathsf{n}_2\}$ by $\mathsf{a}$ is required for the first message transmission.
For the first ZK proof, determinability by $\mathsf{a}$ of $\mathsf{n}_1$ and $\mathsf{n}_3$ is required; and determinability by $\mathsf{b}$ of $\mathsf{n}_4$.
The next message means determinability of $\{\mathsf{k}^-,\mathsf{m}_2,\mathsf{n}_5\}$ by $\mathsf{b}$.
The last ZK proof additionally means determinability of $\{\mathsf{pk}(\mathsf{k}^-),\mathsf{n}_6\}$ by $\mathsf{b}$, and $\mathsf{n}_7$ by $\mathsf{a}$.
We get the determinability requirements given in Table~\ref{tbl:transmission-validity}.
Note that technically, $\mathsf{a}$ does not need $\mathsf{m}_2$ to run the protocol, and $\mathsf{b}$ does not need $\mathcal{H}(\mathsf{m}_1,\mathsf{n}_1)$; however, in practice, they will check whether the data supplied matches their expectations using the checks expressed by the testing rules.

We mention two modelling details regarding the use of SRSA-CL signatures for anonymous credentials.
First, the last ZK proof in the issuing trace is technically not a proof of knowledge of the private key, but of the RSA inverse of part of the issuer's randomness.
However, in terms of knowledge this proof is equivalent because the private key can be determined from the RSA inverse and vice versa \cite{Boneh1999TwentyYearsof}.
Second, due to the structure of the signature, different choices for $\mathsf{n}_a$ and $\mathsf{n}_b$ can lead to content equivalent signatures.
However, assuming $\mathsf{n}_a$ and $\mathsf{n}_b$ are chosen at random, this happens with negligible probability.

Finally, an alternative signature scheme supporting signatures on committed values is the BM-CL scheme \cite{Camenisch2004SignatureSchemesand}.
There are two technical differences with the SRSA-CL-based system presented above.
First, BM-CL signatures have the additional property that they allow ``blinding'': a user can turn a valid credential $\textrm{cred}^{\mathsf{m}_1}_{\mathsf{k}^-}(\mathsf{m}_2;\{\mathsf{n}_a,\mathsf{n}_b\})$ into a different credential $\textrm{cred}^{\mathsf{m}_1}_{\mathsf{k}^-}(\mathsf{m}_2;\{\mathsf{n}'_a,\mathsf{n}_b\})$ (however, she is not able to change randomness $\mathsf{n}_b$).
Second, the final ZK proof in the issuing protocol of Figure~\ref{fig:anoncred-from-sig} is not necessary for a BM-CL-based scheme.
We chose the SRSA-CL-based signature scheme because the high-level model is simpler; however, in terms of privacy the choice of signature scheme does not matter.

\begin{table}[tb]
\newcommand{\cm}{\ding{51}}
\newcommand{\ob}{$\medsquare$}
\newcommand{\cb}{$\filledmedsquare$}
\small\centering
\rotatebox{90}{\begin{tabular}[b]{r||cccc||ccccccc|ccc|cccccc}
&\multicolumn{4}{c||}{Coalition of...}
&\multicolumn{7}{c|}{\cb: undetectable w.r.t. coalition}
&\multicolumn{3}{c|}{Involvement}
&\multicolumn{6}{c}{\cb: unassociable w.r.t. coalition}\\
Requi-
&\multicolumn{4}{c||}{}
&\multicolumn{7}{c|}{\ob: detectable w.r.t. coalition}
&\multicolumn{3}{c|}{unknown}
&\multicolumn{6}{c}{\ob: associable w.r.t. coalition}\\
rement   & $bs$ & $ii$ & $is$ & $ttp$ & $d_1$ & $d_2$ & $d_2{>}60$ & $d_3$ & $d_5$ & $d_6$ & $d_7$ & $ii$ & $is$ & $bs$ & $\kappa,\mu$ & $\kappa,\zeta$ & $\kappa,\xi$ & $\mu,\zeta$ & $\mu,\xi$ & $\zeta,\xi$\\\hline
AX & \cm  &      &      &       &  \ob  &       &   \ob      &       &       &  \ob  &       &&&& \\
SID& \cm  &      &      &       &       &       &            &  \cb  &  \cb  &       &       &&&& \\
SPD& \cm  &      &      &       &       &  \cb  &            &       &       &       &       &&&& \\
ID &      & \cm  &      &       &       &       &            &       &  \cb  &  \cb  &       &&&& \\
ID &      &      &  \cm &       &  \cb  &  \cb  &   \cb      &  \cb  &       &       &       &&&& \\\hline
IM &  \cm &      &      &       &       &       &            &       &       &       &       &      & \cb &       &\\
IM &      &  \cm & \cm  &       &       &       &            &       &       &       &       & \cb  &     &       &\\
ISM&  \cm &      &      &       &       &       &            &       &       &       &       &      &     &  \cb  &\\
ISM&      &  \cm &      &       &       &       &            &       &       &       &       &      &     &  \cb  &\\\hline
AR & \cm & \cm  &  \cm &  \cm   &       &       &            &       &       &       &       &      &     &       &                &      \ob         &     \ob        &               &             &    \ob        \\
SL & \cm &      &      &        &       &       &            &       &       &       &       &      &     &       &                &                  &                &               &             &    \cb        \\
IL &     & \cm  &      &        &       &       &            &       &       &       &       &      &     &       &                &      \cb         &     \cb        &               &             &               \\
IL &     &      &  \cm &        &       &       &            &       &       &       &       &      &     &       &                &                  &                &     \cb       &    \cb      &               \\
IIL&     & \cm  &  \cm &        &       &       &            &       &       &       &       &      &     &       &    \cb         &                  &                &               &             &               \\
ISL& \cm & \cm  &  \cm &        &       &       &            &       &       &       &       &      &     &       &                &      \cb         &     \cb        &      \cb      &    \cb      &               \\
\end{tabular}}
\caption{Schematic overview of the requirements in Table~\ref{tbl:formalization}. Each row indicates that with respect to the given coalition of actors, (a) the given items should be (un)detectable; (b) the involvement of the given actors should be unknown; and (c) Alice's profiles in the given domains should be (un)associable\label{tbl:coverage}
}
\end{table}

\end{document}